\documentclass[a4paper,12pt]{article}

\usepackage{mathrsfs}
\usepackage{amsmath,amssymb,latexsym}
\usepackage{tikz}
\usepackage{xspace}
\usepackage{slashed}
\usepackage{bbm} 

\usepackage{jheppub}                   
\makeatletter
\gdef\@fpheader{\ }                    
\makeatother

\usetikzlibrary{arrows,shapes,positioning}
\usetikzlibrary{decorations.markings}
\usetikzlibrary{calc,decorations.pathmorphing,patterns}
\usetikzlibrary{shapes,snakes}
\tikzstyle arrowstyle=[scale=1]
\tikzstyle directed=[postaction={decorate,decoration={markings,
    mark=at position .65 with {\arrow[arrowstyle]{stealth}}}}]
\tikzstyle reverse directed=[postaction={decorate,decoration={markings,
    mark=at position .65 with {\arrowreversed[arrowstyle]{stealth};}}}]

\newcommand{\dd}{\mathrm{d}}
\newcommand{\rme}{\mathrm{e}}

\newcommand{\der}{\partial}

\newcommand{\bbR}{\mathbb{R}}

\DeclareMathOperator{\SU}{\mathit{SU}}
\DeclareMathOperator{\SO}{\mathit{SO}}
\DeclareMathOperator{\ISO}{\mathit{ISO}}
\DeclareMathOperator{\CSO}{\mathit{CSO}}

\DeclareMathOperator{\SL}{\mathit{SL}}
\DeclareMathOperator{\GL}{\mathit{GL}}

\newcommand{\rep}[1]{\mathbf{#1}}

\newcommand{\id}{\mathbbm{1}}

\DeclareMathOperator{\vol}{vol}

\newcommand{\Lgen}{{L}}

\DeclareMathOperator{\adj}{ad}

\newcommand{\E}[1]{\mathit{E}_{#1(#1)}}

\newcommand{\be}{\begin{equation}}
\newcommand{\ee}{\end{equation}}
\newcommand{\bea}{\begin{eqnarray}}
\newcommand{\eea}{\end{eqnarray}}
\newcommand{\nn}{\nonumber}

\newcommand{\bpm}{\begin{pmatrix}}
\newcommand{\epm}{\end{pmatrix}}

\newcommand{\diff}{\mathrm{d}}

\newcommand{\calB}{\mathcal{B}}

\newcommand{\calE}{\mathcal{E}}

\newcommand{\calL}{\mathcal{L}}

\newcommand{\rg}[1]{\stackrel{\scriptscriptstyle{\circ}}{#1}}

\newcommand{\tend}[1]{\hat{#1}}

\newcommand{\din}{{\rm d^{in}}}

\newcommand{\xN}{\otimes_{N'}\!}
\newcommand{\eqs}{\stackrel{*}{=}}

\newcommand{\hE}{\hat{E}}

\newcommand{\hs}[1]{\hspace{#1}}

\newcommand{\tB}{{\tilde{B}}}
\newcommand{\tH}{{\tilde{H}}}

\newcommand{\ra}{\rightarrow}

\title{Exceptional generalised geometry for massive~IIA and consistent reductions}

\author[a]{Davide Cassani,}
\emailAdd{davide.cassani@lpthe.jussieu.fr}
\author[a]{Oscar de Felice,}
\emailAdd{odefelice@lpthe.jussieu.fr}
\author[a]{Michela Petrini,}
\emailAdd{petrini@lpthe.jussieu.fr}
\author[b,c]{\hbox{Charles Strickland-Constable,}}
\emailAdd{charles.strickland-constable@cea.fr}
\author[d]{and Daniel Waldram}
\emailAdd{d.waldram@imperial.ac.uk}

\affiliation[a]{LPTHE, Sorbonne Universit\'es UPMC Paris 06, CNRS,\\
4 place Jussieu, F-75005, Paris, France}

\affiliation[b]{Institut de physique th\'eorique, Universit\'e Paris Saclay, CEA, CNRS,\\ Orme des Merisiers, F-91191 Gif-sur-Yvette, France}

\affiliation[c]{Institut des Hautes \'Etudes Scientifiques,\\ Le Bois-Marie, 35 route de Chartres, F-91440 Bures-sur-Yvette, France}

\affiliation[d]{Department of Physics, Imperial College London,\\
Prince Consort Road, London, SW7 2AZ, UK}

\abstract{
We develop an exceptional generalised geometry formalism for massive type IIA supergravity. In particular, we construct a deformation of the generalised Lie derivative, which generates the type IIA gauge transformations as modified by the Romans mass. We apply this new framework to consistent Kaluza--Klein reductions preserving maximal supersymmetry. We find a generalised parallelisation of the exceptional tangent bundle on $S^6$, and from this reproduce the consistent truncation ansatz and embedding tensor leading to dyonically gauged $\ISO(7)$ supergravity in four dimensions. We also discuss closely related hyperboloid reductions, yielding a dyonic $\ISO(p,7-p)$ gauging. Finally, while for vanishing Romans mass we find a generalised parallelisation on $S^d$, $d=4,3,2$, leading to a maximally supersymmetric reduction with gauge group $\SO(d+1)$ (or larger), we provide evidence that an analogous reduction does not exist in the massive theory.
}

\begin{document}
\maketitle

\section{Introduction}

Consistent Kaluza--Klein truncations establish an exact map between supergravity theories in different dimensions  and have been used to embed into string theory interesting solutions such AdS vacua, black holes, domain walls, and Lifshitz or Schr\"odinger non-relativistic backgrounds, to mention only a few.  Consistency requires that the dependence of the higher-dimensional fields on the internal manifold factorises out once the truncation ansatz is plugged into the equations of motion. This is a highly non-trivial condition, which --- aside from the cases where it is ensured by some symmetry --- makes consistent truncations rare and hard to construct. Dimensional reductions on certain spheres  provide prominent examples of consistent truncations that are not the consequence of a manifest symmetry, the best known instances being eleven-dimensional supergravity on $S^7$ \cite{deWit:1986oxb} or $S^4$ \cite{Nastase:1999kf}, and type IIB supergravity on $S^5$ \cite{Cvetic:2000nc}. Recently, there has been progress in clarifying the systematics of these reductions by working with reformulations (and extensions) of higher-dimensional supergravity theories that make some form of U-duality symmetry manifest, see~e.g.~\cite{Riccioni:2007au,Berman:2012uy,Aldazabal:2013mya,Godazgar:2013dma,spheres,Hohm:2014qga,Ciceri:2014wya,Baguet:2015xha,
Baguet:2015sma,Malek:2015hma}.

An important aspect in dimensional reductions is to establish the precise relation between the gauge symmetry of the lower-dimensional theory and its higher-dimensional origin. This becomes even more crucial for compactifications preserving maximal supersymmetry, since specifying which subgroup of the U-duality group is gauged suffices to completely determine the truncated theory. Generically, a subgroup of the gauge group originates from the Killing symmetries of the internal manifold, as is standard in Kaluza--Klein reductions. However, higher-dimensional supergravities come with $p$-form potentials, which carry their own gauge symmetry and also contribute to the gauging of the truncated theory.
The problem is therefore studied more effectively if one can treat diffeomorphisms and  $p$-form gauge transformations in a unified fashion. 
A formalism that accomplishes this and at the same time has $E_{n(n)}$  manifest as a structure group is provided by  exceptional generalised geometry.

In exceptional generalised geometry~\cite{Hull:2007zu, Pacheco:2008ps},  given a $d$-dimensional internal manifold $M_d$, one studies geometric structures defined on a certain generalised tangent bundle, which extends the ordinary tangent bundle. While ordinary vectors generate diffeomorphisms, sections of this generalised bundle also encode all the gauge parameters of the supergravity theory on $M_d$, and --- if one starts from type II supergravity --- naturally transform under the U-duality group $E_{d+1(d+1)}\times \mathbb{R}^+$.
The full set of internal diffeomorphisms and $p$-form gauge transformations is generated by an extension of the usual Lie derivative. This is denoted by $L$ and is called {\it generalised Lie derivative} (or {\it Dorfman derivative}). This operator is a key tool to study the gauge symmetry of a compactification.  

In  \cite{spheres}  it was observed that consistent truncations with maximal supersymmetry are related to the existence of a {\it generalised Leibniz parallelisation}, namely a globally-defined frame $\{\hat E_A\}$ for the generalised tangent bundle that also satisfies the property
\be\label{frame_condition}
L_{\hat E_A}\hat E_{B} \,=\, X_{AB}{}^C \hat E_C\ ,
\ee
with {\it constant} coefficients $X_{AB}{}^C$. A frame satisfying~\eqref{frame_condition}  defines a Leibniz algebra, hence the qualification ``Leibniz'' attributed to the parallelisation. Starting from a generalised Leibniz parallelisation, one can define a {\it generalised Scherk--Schwarz reduction}. As the name suggests, this is a generalisation of conventional Scherk--Schwarz reductions on local group manifolds~\cite{Scherk:1979zr} to a larger class of manifolds, which preserves the same amount of supersymmetry as the original higher-dimensional theory. The constants $(X_A)_{B}{}^C$ correspond to the generators of the lower-dimensional gauge group, and are tantamount to the embedding tensor that fully determines the gauged maximal supergravity. The truncation defined by the generalised Scherk--Schwarz procedure is conjectured to be consistent. Although it has not been proved in full generality, this expectation is supported by a number of examples.   

Examples of spaces that are parallelisable in the generalised sense but not in the ordinary sense are provided by spheres.
It was argued in~\cite{spheres} that all known sphere consistent truncations can be understood as generalised Scherk--Schwarz reductions. It was shown there that all $d$-dimensional spheres $S^d$ admit a generalised Leibniz parallelisation trivialising the $\GL(d+1)$ bundle $T \oplus \Lambda^{d-2}T^*$ and satisfying the $\SO(d+1)$ algebra.
For eleven-dimensional supergravity on $S^4$ or $S^7$, or type IIB supergravity on $S^5$, this parallelisation was extended to a Leibniz parallelisation of the full exceptional generalised tangent bundle, which contains the $\GL(d+1)$ bundle. It was also shown how applying a Scherk--Schwarz procedure to the generalised frame,  instead of the ordinary one,  determines the consistent truncation ansatz for all of the lower-dimensional scalar fields. 

A similar approach has been adopted for studying generalised Scherk--Schwarz reductions using exceptional field theory, see~e.g.~\cite{Hohm:2014qga,Baguet:2015xha,Baguet:2015sma}. In particular the generalised parallelisation has been used to define, in addition, the gauge and higher-tensor fields in the truncation. Formally, under the section condition, the equations of exceptional field theory and exceptional generalised geometry are the same. In this paper, we will use the latter formulation, so that our perspective is that the geometric degrees of freedom are precisely those of ten-dimensional supergravity, and we will not consider any enlargement of the physical space-time.

In this paper, we use the formalism of generalised geometry to study consistent truncations of type IIA supergravity preserving maximal supersymmetry.
In the first part, we  develop further the formulation of generalised geometry relevant for type IIA supergravity, originally introduced in~\cite{Hull:2007zu,Grana:2009im}, 
and we extend it to the case where the Romans mass is switched on.
As first step, we specify the Dorfman derivative for massless type IIA supergravity; this is easily done by reducing the M-theory derivative given in~\cite{Coimbra:2011ky}. Then we extend it to the massive case. This step is non-trivial for the following reason. In generalised geometry, fluxes of the supergravity field strengths are incorporated via a twisting of the generalised tangent bundle by the respective potentials. However, this cannot accommodate the Romans mass, since being a zero-form flux it has no associated potential. We overcome this limitation by recalling that,  while introducing the Romans mass does not modify the basic degrees of freedom of IIA supergravity, it does affect the gauge transformations. This implies that the Dorfman derivative needs be modified, so that it generates the correct massive gauge transformations. We construct the operator accomplishing this, that we call {\it massive Dorfman derivative}.
This also implies that the generalised tangent bundle is patched in a way different from the massless case.

In the second part of the paper, we apply the generalised geometry formalism to consistent truncations of type IIA supergravity preserving maximal supersymmetry. We show how starting from a generalised Leibniz parallelisation one can construct a generalised Scherk--Schwarz truncation ansatz for the scalar fields as well as for the bosonic fields with external legs, giving explicit expressions in terms of the type IIA fields. We also provide a partial proof of consistency of the truncation by studying the reduction of the higher-dimensional gauge transformations, thus setting the reduction on more solid ground.

Then we construct explicit parallelisations of the type IIA $E_{d+1(d+1)}\times \mathbb{R}^+$ generalised tangent bundle on the $d$-dimensional spheres $S^d$, for $d=6,4,3,2$. For $S^6$, we recover the consistent truncation to the four-dimensional, dyonically gauged $\ISO(7)$ supergravity~\cite{Dall'Agata:2012bb,Dall'Agata:2014ita} recently worked out in~\cite{Guarino:2015jca,Guarino:2015qaa,Guarino:2015vca}. In particular, we reproduce the full bosonic truncation ansatz starting from the generalised parallelisation. In addition, we obtain the ansatz for the dual type IIA fields. The role of the Romans mass in this truncation is to introduce a magnetic gauging of the translational part of the group $\ISO(7)$. We also discuss closely related parallelisations where $S^6$ is replaced by one of the six-dimensional hyperboloids $H^{p,7-p}$; these yield dyonically gauged $\ISO(p,7-p)$  supergravity, where again the Romans mass entails a magnetic gauging of the translational symmetries. For vanishing Romans mass, these correspond to the $S^6$ and $H^{p,7-p}$ truncations identified in~\cite{Hull:1988jw}. Together they provide the uplift of all dyonic gaugings in the $\CSO(p,q,r)$ class \cite{Dall'Agata:2014ita}, aside for $\SO(8)$ (which, in contrast, has been shown not to have a locally geometric uplift~\cite{Lee:2015xga}). 

For massless type IIA on $S^4$, $S^3$ and $S^2$, we find generalised Leibniz parallelisations whose gauge group is $\SO(5)$, $\ISO(4)$ and $\SO(3)$, respectively. This matches previously known consistent truncations on such manifolds. We also find that when the Romans mass is switched on, these parallelising frames fail to satisfy an algebra. This is in contrast with the $S^6$ case, where the Romans mass modifies the gauge group generators $X_{AB}{}^C$ without spoiling the Leibniz property~\eqref{frame_condition} of the parallelising frame. We are thus led to investigate the existence of alternative frames. We analyse the $S^3$ case in detail, and prove a no-go result indicating the absence of a consistent truncation of massive type IIA supergravity down to a maximal supergravity in seven dimensions with $\SO(4)$ gauge group (or larger). We also comment on the $S^4$ and $S^2$ cases along similar lines.

The structure of the paper is as follows. In section~\ref{sec:gauge_IIA} we briefly summarise the features of massive type IIA supergravity that will be relevant for our construction. In section~\ref{sec:ExGenGeo_IIA} we present the type IIA generalised geometry for vanishing Romans mass. The deformation of the generalised Lie derivative accommodating for the latter is given in section~\ref{sec:massive_genLie}. In section~\ref{sec:GenParall} we illustrate the generalised Scherk--Schwarz reduction and in section~\ref{sec:examples} we apply it to various examples. Section~\ref{sec:conclusions} contains our conclusions.  Several technical details of our derivations are relegated to the appendices.

\section{Gauge symmetries of massive IIA}\label{sec:gauge_IIA}

In this section we give a brief summary of type IIA supergravity, focusing on its gauge transformations. We will work with the democratic formulation of~\cite{Bergshoeff:2001pv}, which is the natural framework for applying generalised geometry.

The NSNS sector of type II supergravity is given by the usual two-form potential $B$, with field strength $H = \dd B$, and a dual six-form potential $\tilde B$. 
The RR sector contains the odd potentials $C_1$, $C_3$, together with their duals $C_7$, $C_5$.  There is also a nine-form
$C_9$ which does not carry any degree of freedom and whose field strength is dual to the Romans mass $m$ \cite{Romans:1985tz}. In the 
democratic formulation, all RR potentials are treated on the same footing, and can be arranged in the poly-form
$C = \sum_{k=0}^4 C_{2k + 1}$. 
The field strength\footnote{\label{Abasis_foot} There exists another common choice for the RR potential, the  $A$-basis, which is related to the $C$-basis we use as
$A =  \rme^{-B} \wedge C  \, .$
In this basis the field strength \eqref{RRfieldstrength} reads 
$F =  \rme^{B}\wedge \left(\dd A + m \right) \,.$}
\be
\label{RRfieldstrength}
F  \,=\,  \dd C - H \wedge C + m\, \rme^B 
\ee
is invariant under an infinitesimal gauge transformation of the NSNS two-form and RR potentials\footnote{Our sign conventions for the gauge transformations are chosen so that they match the generalised geometry expressions to be introduced later.}
\begin{align}
\label{gtlin1}
\delta B \,&=\, -\dd \lambda \nn \ ,\\[1mm]
\delta C \,&=\, -\rme^B\wedge(\dd \omega - m  \lambda) \ , 
\end{align}
where $\lambda$ is a one-form and $\omega = \sum_{k=0}^4 \omega_{2k} $ a poly-form of even degree. When $m\neq 0$, the RR potential $C_1$ can be gauged away by a suitable choice of the NSNS gauge parameter $\lambda$. The field strength $F$  satisfies the Bianchi identity 
\be
\dd F \,=\, H \wedge F\ ,
\ee
together with the self-duality relation
\be\label{eq:selfdualityF}
*F \,=\, s(F)\ ,
\ee
where  $s(F_n) = (-1)^{[n/2]}F_n\,$ is the transposition operator, which reverses the order of the indices.  The Bianchi identity and the self-duality relation together imply the equations of motion for the RR fields.

The NSNS six-form potential $\tilde B$ is defined  by interpreting  the equation of motion for $B$,
\be
\label{eomH_3}
\dd \left( \rme^{-2\phi} * H \right) + \tfrac 12 \left[F \wedge * F \right]_8\ =\ 0\ ,
\ee
as the Bianchi identity for the dual seven-form field strength, defined as 
\be
\tH  = \,\rme^{-2\phi} * H\ .
\ee 
The  $[\ldots]_8$ in  \eqref{eomH_3}  denotes  the eight-form component of the poly-form in the bracket.
Using the self-duality relation~\eqref{eq:selfdualityF}, eq.~\eqref{eomH_3} can be written as
\begin{equation}\label{BianchiH7}
		\dd \big( \tH  +  \tfrac{1}{2} \left[ s(F) \wedge C + m \,\rme^{-B}\wedge C \right]_7 \big) \,=\, 0\ ,
\end{equation}
which is solved by
\be\label{eq:tildeH7}
\tilde H \,=\, \dd \tilde B - \tfrac{1}{2} \left[ s(F) \wedge C + m \,\rme^{-B}\wedge C\right]_7\ .
\ee
Requiring gauge invariance of the field strength $\tilde H$ fixes the linearised gauge transformation of $\tilde B$ as \cite{Bergshoeff:1997ak,Bergshoeff:2006qw}\footnote{The $\tB$ in the present paper
is related to the ones of~\cite{Bergshoeff:1997ak} and \cite{Bergshoeff:2006qw} by a field redefinition. The one in (B.7) of~\cite{Bergshoeff:1997ak} is related to ours as  
$\tB^{\rm there}= 
 (\tB + \frac12 C_1 \wedge C_5)^{\rm here}$.  
The one in  (2.29) of~\cite{Bergshoeff:2006qw} is $\tB^{\rm there} 
 = (\tB - \frac 12 C_1 \wedge C_5 + \frac 12 B\wedge C_1\wedge C_3)^{\rm here}$.}
\bea
\label{gtlin2}
\delta\tB \,=\, -(\dd \sigma +  m\, \omega_6) - \tfrac12 \left[\rme^{B} \wedge (\dd\omega - m \lambda) \wedge s(C)\right]_6 \ ,
\eea
where  $\sigma$ is  a five-form,  while   $\lambda$  and  $\omega_{2 k}$   are the parameters of the $B$-field and RR  gauge transformations \eqref{gtlin1}.
We thus find that in the massive theory, $\tB$ is no longer invariant under $B$ gauge transformations. Also notice that the $\omega_6$ gauge transformation can be used to set $\tB=0$ in the case where $m \neq 0$. 

From the infinitesimal gauge transformations above, we infer that the deformation due to the Romans mass can be summarised into the shifts
\bea\label{gauge_massless_to_massive}
\dd \omega_0 \ &\longrightarrow&\ \dd \omega_0 - m \lambda\ , \nn \\ [1mm]
\dd \sigma \ &\longrightarrow&\ \dd \sigma + m \,\omega_6\ .
\eea
This observation will guide us in the construction of generalised geometry for massive type IIA.

\section{Type IIA exceptional generalised geometry}
\label{sec:ExGenGeo_IIA}

In this section we give a detailed description of the exceptional generalised geometry for compactifications of massless type IIA supergravity. 
Generalised geometry  allows to  treat  on the same ground  diffeomorphisms  and  transformations of the gauge potentials of type II supergravity or M-theory.
This is achieved by constructing an extended tangent space over the compactification manifold  whose transition functions are combinations of standard $\GL(d)$ and gauge transformations of the supergravity potentials.

Generalised complex geometry, as originally proposed by Hitchin \cite{Hitchin:2004ut,Gualtieri:2003dx}, geometrises the NSNS sector of type II supergravity. Hitchin's generalised tangent bundle, is isomorphic to  the sum $T \oplus T^*$ of the tangent and cotangent bundle to the  $d$-dimensional compactification
manifold $M_d$,  and is patched by $\GL(d)$ transformations 
and gauge shifts of the NSNS two-form $B$. The structure group of this extended bundle is
 $O(d,d)$,  the T-duality group of the compactification on a $d$-dimensional torus. From a string theory perspective, $T$ and $T^*$ parameterise the quantum number of the string, 
that is momentum and winding charge.  
 Extending  this  construction to include the RR potentials in type II supergravity \cite{Hull:2007zu,Grana:2009im,Coimbra:2011nw}, or adapting it to M-theory compactifications \cite{Hull:2007zu, Pacheco:2008ps, Coimbra:2011ky,Coimbra:2012af}, leads to exceptional generalised geometry. In this case the structure group of the generalised tangent bundle is the U-duality group, and the bundle parameterises all the charges of the theory under study, that is momenta and winding, as well as NS- and D-brane (or M-brane) charges. 
 
While in $O(d,d)$ generalised geometry the structure of the generalised tangent bundle is the same  in type IIA and IIB  and does not depend on the dimension of the manifold $M_d$, 
 the exceptional tangent bundle takes a different form depending on whether one works in type IIA or type IIB supergravity, and depending on the dimension of $M_d$ its fibres transform in different representations of the structure group $E_{d+1(d+1)}\times \mathbb{R}^+$~\cite{Hull:2007zu,Grana:2009im,Coimbra:2011nw}.
 The exceptional geometry for IIA compactifications  was partly constructed in~\cite{Grana:2009im}.  In the following, we extend the analysis of~\cite{Grana:2009im} to the construction of the generalised Lie derivative, generalised
 split frame and generalised metric.   As we  show in appendix~\ref{fromMthToIIA},  a  
  straightforward way to obtain these objects is to reduce the corresponding M-theory ones 
 presented in \cite{Coimbra:2011ky,Coimbra:2012af}. 
 
For definiteness we will focus on structures defined on a six-dimensional manifold $M_6$. 
However, if one is interested in an internal space of dimension $d<6$, the relevant expressions are easily obtained from those given below by dropping all forms of degree higher than $d$.

\subsection{The generalised tangent bundle}

The exceptional generalised tangent bundle $E$ on $M_6$ is isomorphic (in a way that we will specify below) to the bundle \cite{Hull:2007zu,Grana:2009im}
 \begin{equation}
 \label{IIAtangentbundle}
	\check E \, =\, T \oplus T^* \oplus \Lambda^5 T^*\oplus (T^* \otimes \Lambda^6 T^*)
		\oplus \Lambda^{\rm even} T^*\, ,
\end{equation}
where $\Lambda^{\rm even}T^*=\mathbb{R} \oplus \Lambda^2 T^* \oplus \Lambda^4 T^* \oplus \Lambda^6 T^*$. 
Roughly speaking, the first two terms in~\eqref{IIAtangentbundle} are associated to the momentum and winding of string states, while 
$\Lambda^5 T^*$ and $(T^*\otimes \Lambda^6 T^*)$  can be seen as the NS five-brane and Kaluza--Klein monopole charges, respectively. 
Similarly  $\Lambda^{\rm even}T^*$ corresponds to charges of the IIA D-branes.

The structure group of $E$ is $E_{7(7)}\times \mathbb{R}^+$, and its fibres, also called {\it generalised vectors}, transform in the ${\bf 56_1}$ representation, where the subscript denotes the $\mathbb{R}^+$ weight.\footnote{One can construct the symplectic and quartic invariants characterising $E_{7(7)}$ and show that they are indeed preserved when the sections are patched as in~\eqref{patching} below.} The form \eqref{IIAtangentbundle} corresponds to the decomposition of this representation under the $\GL(6)$ structure group of $M_6$. According to this decomposition, a generalised vector can be written as
\begin{equation}		
	V \, =\, v + \lambda + \sigma + \tau + \omega\ ,
\end{equation}
where at any point on $M_6$, $v \in T$ is an ordinary vector, $\lambda \in T^*$ is a one-form, $\sigma \in \Lambda^5 T^*$ is a five-form, $\tau = \tau_1\otimes\tau_6 \in T^* \otimes \Lambda^6 T^*$ is the tensor product of a one-form and a six-form, and $\omega = \omega_0+\omega_2+\omega_4+\omega_6\in \Lambda^{\rm even}T^*$ is a  poly-form of even degree.

It is also useful to consider the decomposition under $\GL(6)$ of  the adjoint  bundle $\adj\subset E \otimes E^*$, 
\begin{equation}\label{decomp_adjoint}
	\adj \, =\, \bbR_\Delta \oplus \bbR_\phi 
		\oplus (T\otimes T^*) \oplus \Lambda^2 T \oplus \Lambda^2 T^*
		\oplus \Lambda^6 T \oplus \Lambda^6 T^*
		\oplus \Lambda^{\text{odd}} T \oplus \Lambda^{\text{odd}} T^*\, .
\end{equation}
Its sections $R$ transform in the ${\bf 133_0}+ {\bf 1_0}$ of $E_{7(7)}\times \mathbb{R}^+$, and decompose as
\begin{equation}\label{section_adjoint}
R \, = \, l + \varphi + r  + \beta  + B  +  \tilde \beta   +  \tilde B    + \Gamma  + C \, ,
\end{equation} 
where each term in the sum is a section of the corresponding sub-bundle in \eqref{decomp_adjoint}.
The adjoint bundle  encodes the transformations of the supergravity theory. Specifically,
$r \in End(T)$ corresponds to the $GL(6)$ action, while the scalars
 $l$ and $\varphi$ are related to the shifts of the warp factor and dilaton, respectively.  The forms $B$,  $\tilde B$ and $C = C_1 + C_3 + C_5$
 correspond to the internal components of the NSNS two-form, of its dual and of the RR potentials. 
The other elements are poly-vectors obtained by raising the indices of the forms, and do not have an immediate supergravity counterpart. 

There are two other objects that will appear later. The first is $N$, a sub-bundle of the symmetric product $S^2 E$, whose fibres transform in the ${\bf 133_2}$ representation of $E_{7(7)}\times \mathbb{R}^+$. It is given by
\be\label{Nbundle}
N \,\simeq\, \mathbb{R}  \oplus \Lambda^4T^*  \oplus \Lambda^{\rm odd}T^*  \oplus \Lambda^6T^*  \oplus (T^* \otimes \Lambda^5 T^*) \oplus (\Lambda^2T^*  \oplus \Lambda^6T^*  \oplus \Lambda^{\rm odd}T^*)\otimes \Lambda^6T^* .
\ee
The second is $K$, a sub-bundle of $E^*\otimes {\rm ad}$, whose fibres transform in the ${\bf 912_{-1}}$ representation. We will not give its $\GL(6)$ decomposition here, but it is simple to obtain it from the expression in \cite{Coimbra:2011ky}.

The NSNS and RR supergravity potentials do not need to be globally defined,
 and can give rise to fluxes threading non-trivial cycles of the internal manifold. This is encoded in generalised geometry by a {\it twist} of the generalised tangent bundle. If we start from the bundle $\check E$ in \eqref{IIAtangentbundle}, and denote by $\check V = \check v + \check \lambda + \check \sigma+ \check \tau + \check \omega$ its sections, then we define a section $V  = v + \lambda + \sigma+ \tau + \omega$ on the twisted bundle $E$ as
\be
\label{eq:twistC_short}
V \,=\, \rme^{\tilde B}\,\rme^{-B}\,\rme^C \cdot\check V\, ,
\ee  
where $\cdot$ denotes the adjoint action of the $E_{7(7)}\times \bbR^+$ algebra, whose explicit form is given in appendix~\ref{app:reductionIIA}.\footnote{In the $A$-basis of footnote~\ref{Abasis_foot}, the relation~\eqref{eq:twistC_short} between twisted and untwisted generalised vectors is expressed as
$V\, =\,  \rme^A\,\rme^{\tilde B}\, \rme^{-B}\cdot\check V ,
$
as it can be checked using the formula~\eqref{eq:exp_adj}. This is the form of the twist that was used in~\cite{Grana:2009im}.} It is this twist that specifies the isomorphism between $E$ and the untwisted bundle~$\check E$.
 Splitting \eqref{eq:twistC_short} in $\GL(6)$ representations yields 
\begin{align}\label{eq:twistC}
v \,&=\, \check v \ ,\nn \\[1mm]
\lambda \,&=\, \check\lambda + \iota_{\check v} B\ ,\nn \\[1mm]
\sigma \,&=\, \check \sigma + \iota_{\check v}\tilde B - \left[s(C)\wedge \big(\check\omega + \tfrac12 \iota_{\check v}C + \tfrac12\check \lambda \wedge C \big)\right]_5\ , \nn \\[1mm]
\tau \,&=\, \check\tau + jB\wedge \left[\check \sigma - s(C)\wedge \big(\check\omega + \tfrac12 \iota_{\check v}C + \tfrac12\check \lambda \wedge C \big)\right]_5 + j\tilde B\wedge (\check\lambda + \iota_{\check v}B) \nn \\[1mm]
\,&\ \quad - js(C)\wedge\big(\check\omega + \tfrac12\iota_{\check v}C + \tfrac12 \check\lambda \wedge C \big) \ ,\nn \\[1mm]
\omega \,&=\, \rme^{-B}\wedge \big( \check \omega + \iota_{\check v}C +  \check \lambda \wedge C \big) \ ,
\end{align}
where the ``$j$-notation'' is explained in appendix~\ref{app:notation}.
%
%

Given two coordinate patches $U_{\alpha}$ and $U_{\beta}$ on $M_6$, the patching condition for the generalised vector $V$ on the overlaps $U_{\alpha} \cap U_{\beta}$ includes $p$-form gauge transformations and reads
\be
\label{patching}
V_{(\alpha)} \,=\,  \rme^{\dd \tilde \Lambda_{(\alpha \beta)}} \,\rme^{\dd \Omega_{(\alpha \beta)}}\, \rme^{-\dd \Lambda_{(\alpha \beta)}} \cdot V_{(\beta)}  \,, 
\ee  
where $\Lambda_{(\alpha \beta)}$ is a one-form,  $\tilde\Lambda_{(\alpha \beta)}$ a five-form, and $\Omega_{(\alpha \beta)}$ a poly-form of even degree, all defined on $U_{\alpha} \cap U_{\beta}$.
Plugging~\eqref{eq:twistC_short} into \eqref{patching} and reorganising the exponentials on the right hand side, one obtains the  patching conditions for the supergravity potentials:
\begin{align}
\label{eq:gauge-field-patchingmassless}
	B_{(\alpha)} \,&=\, B_{(\beta)} + \dd \Lambda_{(\alpha\beta)} \ ,\nn\\[1mm]
	C_{(\alpha)} \,&=\, C_{(\beta)} + \rme^{B_{(\beta)} +\dd \Lambda_{(\alpha\beta)}} \wedge \dd \Omega_{(\alpha\beta)} \ , \nn\\[1mm]
	\tB_{(\alpha)} \,&=\, \tB_{(\beta)} + \dd \tilde \Lambda_{(\alpha\beta)} 
		  + \tfrac12 \left[ \dd \Omega_{(\alpha\beta)} \wedge \rme^{B_{(\beta)} + \dd \Lambda_{(\alpha\beta)}}\wedge s(C_{(\beta)}) \right]_6 \,.
\end{align}
As we clarify in appendix \ref{app:gauge-patching}, these do indeed correspond to the finite supergravity gauge transformations between patches (here given for vanishing Romans mass, $m=0$). This construction generalises the standard definition of a gerbe connection.

While here we constructed the twisted bundle $E$ by postulating the patching~\eqref{patching} and showing that with the twist \eqref{eq:twistC_short} it leads to the appropriate supergravity transformations, in appendix~\ref{app:gauge-patching} we take a converse viewpoint and illustrate how one can instead start from the supergravity transformations and from these derive the patching conditions \eqref{patching}. 

Formally the twisted bundle $E$ is described as a series of extensions 
\begin{align}
\label{eq:twistE}
   0 \longrightarrow T^* \longrightarrow \; & E'
      \longrightarrow T \longrightarrow 0 \,, \nn\\[1mm]
   0 \longrightarrow \Lambda^{\rm even} T^* \longrightarrow \; &E''  
      \longrightarrow E' \longrightarrow 0 \,, \nn\\[1mm]
   0 \longrightarrow \Lambda^5T^* \longrightarrow \; &E''' 
      \longrightarrow E'' \longrightarrow 0 \,, \nn\\[1mm]
   0 \longrightarrow T^* \otimes \Lambda^6T^* 
      \longrightarrow \; &E 
      \longrightarrow E''' \longrightarrow 0 \,,
\end{align}
where, as above, the exact sequences are split by the supergravity potentials, which provide an isomorphism $E\cong \check{E}$. Note that these sequences define a natural surjective mapping $\pi : E \ra T$ known as the anchor map. One can additionally view $E$ as an extension of Hitchin's generalised tangent space $E'$~\cite{Hitchin:2004ut,Gualtieri:2003dx} by $O(d,d)\times\bbR^+$ tensor bundles, as we describe in appendix~\ref{app:Odd}.

When discussing the generalised Scherk--Schwarz truncations we will need sub-bundles of  $E$ and $N$ that do not fully span a representation of $E_{7(7)}\times \bbR^+$
and are obtained projecting out some of the components of the original bundle. 
A first example is the bundle $E'''$   in \eqref{eq:twistE}, which corresponds to projecting out the dual graviton term $\tau \in T^* \otimes \Lambda^6T^*$ using the natural map in the last line of \eqref{eq:twistE}.
Hence the sections of $E'''$  are simply given by  $ v + \lambda + \sigma + \omega$. 
We will also need a  bundle $N'$, given by
\be\label{NprimeBundle}
N' \,\simeq \, \mathbb{R} \oplus \Lambda^4T^* \oplus \Lambda^{\rm odd} T^* \, , 
\ee
which is obtained from the bundle $N$ defined in \eqref{Nbundle} with analogous projections (see appendix~\ref{fromMthToIIA}). 
Sections of $N'$ can be constructed pairing two generalised vectors $V$ and $V'$ into the product
\begin{align}\label{N'prod_IIA}
V \otimes_{N'} V' \,&=\, v\,\lrcorner\, \lambda' + v' \lrcorner\, \lambda\nn\\[1mm]
\,&\ +  v \,\lrcorner \,\sigma' + v' \,\lrcorner \,\sigma + [\omega \wedge s(\omega')]_4 \nn\\[1mm]
\,&\ + v \,\lrcorner\, \omega' + \lambda\wedge \omega' + v' \lrcorner\, \omega + \lambda'\wedge \omega\,.
\end{align}


\subsection{The (massless) generalised Lie derivative}

On the generalised bundle one can define a {\it generalised Lie derivative} (or {\it Dorfman derivative}). This is an extension of the ordinary Lie derivative that 
generates the infinitesimal generalised diffeomorphisms, namely the ordinary diffeomorphisms together with the NSNS and RR gauge transformations~\cite{Coimbra:2011ky,Coimbra:2012af}. 
Given two vectors $v,v'\in \Gamma(TM_6)$, the ordinary Lie derivative $\mathcal{L}_vv'$ can be written in components as a $\mathfrak{gl}(6)$ action 
\be
( \mathcal{L}_v v')^m \,=\,  v^n \partial_n v'^{\,m} - (\partial \times v)^m{}_n v'^{\,n}  \, ,
\ee
where the symbol $\times$ is the product of the fundamental and dual representation of $\GL(6)$,  $(\partial \times v)^m{}_n = \partial_n v^m$. 
The Dorfman derivative $L$ is defined in an analogous way, with ordinary vectors replaced by generalised vectors of the twisted bundle $E$.  Namely, using an index $M$ to denote the components of a generalised vector $V$ in a standard coordinate basis, 
\be
V^M \,=\, \{v^m , \, \lambda_m,\, \sigma_{m_1\ldots m_5},\, \tau_{m,m_1\ldots m_6} , \, \omega,\, \omega_{m_1m_2},\,\omega_{m_1\ldots m_4},\, \omega_{m_1\ldots m_6}\}\,,
\ee
and embedding the standard derivative operator 
as a section of the dual generalised tangent bundle $E^*$, $\partial_M =  (\partial_m, 0, \ldots,0)$, 
 the Dorfman derivative is defined as~\cite{Coimbra:2011ky}
\be 
\label{eq:Liedefg}
(L_V V')^M \,=\,  V^N \partial_N  V'^{\,M} - (\partial \times_{\rm ad} V)^M{}_N V'^{\,N}\, , 
\ee
where $ \times_{\rm ad}$ is the projection onto the adjoint bundle
\be
 \times_{\rm ad} \, : \, E^* \otimes E \rightarrow  {\rm ad} \, . 
\ee
This gives 
\be
\partial \times_{\rm ad} V \,=\, \partial \times v  -\dd \lambda + \dd \sigma + \dd \omega\ .
\ee
The derivative \eqref{eq:Liedefg} satisfies the 
Leibniz property
\begin{equation}
\label{eq:Leibniz}
L_V (L_{V'} V'') \,=\, 	L_{V'} (L_V V'') + L_{L_V V'} V'' \, ,
\end{equation}
but in general is not antisymmetric, $L_V V' \neq - L_{V'}V$.  In the $\GL(6)$  decomposition, \eqref{eq:Liedefg} takes the form
\bea
\label{dorf6}
L_V V' &=& \mathcal{L}_v v' + \left(\mathcal{L}_v \lambda' - \iota_{v^\prime} \mathrm{d}\lambda\right)  + \left( \mathcal{L}_v \sigma' - \iota_{v^\prime}\mathrm{d}\sigma + [s(\omega^\prime) \wedge \mathrm{d}\omega]_5\right) \nn \\ [1mm]
&& + \left(\mathcal{L}_v \tau' + j \sigma' \wedge \mathrm{d}\lambda + \lambda^\prime \otimes \mathrm{d}\sigma + j s(\omega^\prime) \wedge \mathrm{d}\omega \right) \nn \\ [1mm]
&& + \left(\calL_v \omega' +  \dd \lambda \wedge \omega' - (\iota_{v'}+  \lambda' \wedge)\dd\omega\right) \ .
\eea
This can be derived from \eqref{eq:Liedefg} computing the adjoint action or, as we show in appendix \ref{fromMthToIIA}, by reducing the M-theory Dorfman derivative. It can also be written in terms of natural derivative operators in $O(d,d)$ generalised geometry (see appendix~\ref{app:Odd}).

One can also construct the action of the generalised Lie derivative on the untwisted bundle $\check{E}$ defined in~\eqref{IIAtangentbundle}. The new operator can be denoted by $\check{L}$ and is defined as:
\be
\label{eq:fromLtotildeL}
 \check L_{\check V} {\check V'} \,  = \, \rme^{-C} \, \rme^{B}  \,  \rme^{- \tilde B} \cdot  L_V V'  \,,
\ee
where we used the relation \eqref{eq:twistC_short}  between twisted and untwisted generalised vectors.  $\check L$ may be dubbed {\it twisted} Dorfman derivative.\footnote{When the generalised tangent bundle is untwisted, the Dorfman derivative is twisted, and vice-versa.}  Expanding in $\GL(6)$ components one can
show that the twisted Dorfman derivative has the same expression as \eqref{dorf6},  where all twisted components are replaced by the untwisted ones and with the substitutions 
\begin{align}\label{fluxterms_massless_Dorfman}
\dd \check\lambda \ &\to \ \dd \check\lambda - \iota_{\check v}H \, ,\nn\\[1mm]
\dd \check\sigma \ &\to \ \dd \check\sigma -  [s(\check \omega)\wedge F]_6 \, ,\nn\\[1mm]
\dd \check \omega \ &\to \ (\dd-H\wedge) \check \omega - (\iota_{\check v} + \check \lambda\wedge)F\, ,
\end{align}
where $H$ is the supergravity NSNS three-form on $M_6$ while $F = F_2 + F_4 + F_6$ are the RR fluxes, sitting in a spinor representation of $O(6,6)$, and both are contained in the bundle $K$ whose fibres transform in the ${\bf 912_{-1}}$ representation of $E_{7(7)}\times \bbR^+$~\cite{Grana:2009im}. 
Applying the projection ${\bf 56_{1}} \otimes {\bf 912_{-1}} \ra {\bf 133_0}$ to the generalised vector $\check V$ and the flux part of the ${\bf 912_{-1}}$ we obtain an element of the adjoint 
\begin{equation}\label{FluxesAdjoint}
R =  - \iota_{\check v} H + \check \omega \wedge H    - ( \iota_{ \check v} F +   \check \lambda \wedge F)  +   \check \omega \wedge F
\end{equation}
 and it is the action of this on $\check V'$ which gives the flux terms \eqref{fluxterms_massless_Dorfman} in the twisted Dorfman derivative.

The twisted Dorfman derivative is often more useful than  \eqref{dorf6} in concrete computations, as it contains the gauge-invariant NSNS and RR field strengths and not the
potentials.

Since only the gauge-invariant field strengths appear, it is clear that the twisted derivative $\check L_{\check V} {\check V'}$ yields a well-definite section of the untwisted bundle $\check E$. Hence~\eqref{eq:fromLtotildeL} proves that the twisted vector $L_V V'$ transforms as a section of the twisted bundle $E$.

In view  of the extension to massive IIA, it is useful to stress again that the Dorfman derivative generates the infinitesimal generalised diffeomorphisms on the internal manifold $M_d$. Interpreting a generalised vector $V$ as a gauge parameter, the infinitesimal gauge transformation of any field is given by
\be
\delta_V \,=\, L_V\ .
\ee
The Leibniz property~\eqref{eq:Leibniz} then just expresses the gauge algebra $[\delta_V, \delta_{V'}] \,=\, \delta_{L_V V'}$.

\subsection{Generalised frame and metric}
\label{gen_frame_metric}

In generalised geometry the physical fields of a given supergravity theory, dilaton, metric and gauge potentials, are encoded in the generalised metric $G$. 
In the same way as the ordinary metric on the manifold $M_6$ can be seen as an $O(6)$ structure on $T M_6$ parameterising the coset $\GL(6)/\SO(6)$, the generalised metric can be seen as an $\SU(8) / \mathbb{Z}_2$ structure on the generalised tangent bundle, and parameterises the coset $\E7/(\SU(8) / \mathbb{Z}_2)$. 

The construction of the generalised metric is a natural extension of what is done for the more familiar metric $g$.
For instance, $G$ can be defined by its action on two generalised twisted vectors $V$ and  $V'$ as
\begin{align}\label{GofVandV'}
G(V,V')\,&=\, \check{v} \,\lrcorner\, \check{v}' + \check{\lambda} \,\lrcorner\, \check{\lambda}' + \check{\sigma} \,\lrcorner\, \check{\sigma}' + \check{\tau} \,\lrcorner\, \check{\tau}'+ \sum_{k=0}^3\check{\omega}_{2k} \,\lrcorner\, \check{\omega}'_{2k} \nn\\[1mm]
\,&=\, \check{v}^m \check{v}'_m + \check{\lambda}^m  \check{\lambda}'_m + \tfrac{1}{5!}\check{\sigma}^{m_1\ldots m_5} \check{\sigma}'_{m_1\ldots m_5} + \tfrac{1}{6!}\check{\tau}^{m,m_1\ldots m_6} \check{\tau}'_{m,m_1\ldots m_6}\nn\\
\,&\, \quad + \sum_{k=0}^3\tfrac{1}{(2k)!}\,\check{\omega}^{m_1 \ldots m_{2k}} \check{\omega}'_{m_1 \ldots m_{2k}} \ ,
\end{align}
where the indices are lowered/raised using the ordinary metric $g_{mn}$ and its inverse $g^{mn}$.

One can also define a generalised frame  $\{\hat{E}_A\}$ on $E$ and then construct the inverse generalised metric as the tensor product of two such frames
\be
\label{genmetfor}
G^{-1} \,=\, \delta^{AB} \hat{E}_A \otimes \hat{E}_B \,. 
\ee
We will give  below  a precise definition of  this product.  
To construct the generalised frame,  we first consider  the  {\it untwisted} generalised tangent bundle $\check E$.  Let
 $\hat e_a$, with $a=1,\ldots,6$, be an ordinary frame, namely a basis for the tangent space at a point of $M_6$, and let $e^a$ be the dual basis for the cotangent space.\footnote{We are using the hat symbol to distinguish frame vectors, $\hat{e}_a$, from co-frame one-forms, $e^a$. Similarly, the hat on $\hat E_A$ indicates that this is a generalised frame vector.}
 Then we can define a frame  $\check{\hat E}_A$ for the untwisted generalised tangent space as the  collection of bases for the subspaces  that compose it
\begin{equation}
\label{untframe}
\{  \check{\hat{E}}_A\} = \{\hat e_a\} \cup \{e^a\} \cup \{e^{a_1 \ldots a_5} \} \cup \{e^{a,a_1\ldots a_6}\} \cup \{1\} \cup \{e^{a_1a_2}\} \cup \{e^{a_1\ldots a_4}\} \cup \{e^{a_1 \ldots a_6}\}\, ,
\end{equation} 
where $e^{a_1\ldots a_p}= e^{a_1}\wedge \cdots \wedge e^{a_p}$ and $e^{a,a_1\ldots a_6} = e^a\otimes e^{a_1\ldots a_6}$. 
 A frame for the {\it twisted} generalised tangent space is obtained by twisting \eqref{untframe} by the local $E_{7(7)}\times \mathbb{R}^+$ transformation
\be
\label{twist_splitfr}
\hat{E}_A \ = \ \rme^{\tB}\rme^{-B}\rme^{C}\rme^{\Delta}\rme^{\phi}\cdot \check{\hat{E}}_A\ ,
\ee
where in addition to the twist \eqref{eq:twistC_short} we also include a rescaling by the dilaton $\phi$ and warp factor $\Delta$, acting as specified in~\eqref{IIAadjvecCompact}.  Because of the rescaling by $\Delta$ the frame
\eqref{twist_splitfr} was called  {\it conformal split frame} in~\cite{Coimbra:2011ky}.   Note that  \eqref{twist_splitfr}  is just a particular choice of frame, not the most
general one. Any other frame can be obtained from \eqref{twist_splitfr}  acting with an $E_{7(7)} \times \mathbb{R}^+$ transformation.  

We denote the components of $\hat{E}_A$ carrying different flat indices as
\begin{equation}
\{\hat E_A\} = \{\hat{\calE}_a\} \cup \{\calE^a\} \cup \{\calE^{a_1 \ldots a_5} \} \cup \{\calE^{a,a_1\ldots a_6}\} \cup \{\calE\} \cup \{\calE^{a_1a_2}\} \cup \{\calE^{a_1\ldots a_4}\} \cup \{\calE^{a_1 \ldots a_6}\}\, . 
\end{equation} 
Explicit expressions for each of these terms are given in appendix~\ref{splitfr_MtoIIA}.

Once we have the generalised frame, we can derive the expression for the inverse generalised metric $G^{-1}$.  Expanded in $\GL(6)$ components,
the product  \eqref{genmetfor} becomes
\begin{equation}
\label{genmetr}
\begin{split}
G^{-1} &=\, \delta^{aa'} \hat{\mathcal{E}}_a \otimes \hat{\mathcal{E}}_{a'} + \delta_{aa'} \mathcal{E}^a \otimes \mathcal{E}^{a'} + \mathcal{E} \otimes \mathcal{E} + \tfrac{1}{2}\delta_{a_1a'_1}\delta_{a_2a'_2} \mathcal{E}^{a_1a_2} \otimes \mathcal{E}^{a'_1a'_2}  \\[1mm]
&\,\quad + \tfrac{1}{4!}\delta_{a_1a'_1}\cdots\delta_{a_4a'_4} \mathcal{E}^{a_1\ldots a_4} \otimes \mathcal{E}^{a'_1\ldots a'_4} + \tfrac{1}{5!}\delta_{a_1a'_1}\cdots\delta_{a_5a'_5} \mathcal{E}^{a_1\ldots a_5} \otimes \mathcal{E}^{a'_1\ldots a'_5} \\[1mm]
&\,\quad +  \tfrac{1}{6!}\delta_{a_1a'_1}\cdots\delta_{a_6 a'_6} \mathcal{E}^{a_1\ldots a_6} \otimes \mathcal{E}^{a'_1\ldots a'_6} + \tfrac{1}{6!}\delta_{a a'}\delta_{a_1a'_1}\cdots\delta_{a_6a'_6} \mathcal{E}^{a,a_1\ldots a_6} \otimes \mathcal{E}^{a',a'_1\ldots a'_6}\, .
\end{split}
\end{equation}
The full expression for $G^{-1}$ is long and ugly, so we only give the terms that will be relevant for the 
next section. Arranging them according to their 
curved index structure, we have
\begin{align}\label{invG_comp_1}
(G^{-1})^{mn} \,&=\, \rme^{2\Delta}g^{mn}\, ,\nn\\[1mm]
(G^{-1})^{m} \,&=\,  \rme^{2\Delta}g^{mn} C_n \, ,\nn\\[1mm]
(G^{-1})^{m}_{\phantom{m}n} \,&=\, - \rme^{2\Delta}g^{mp}B_{pn}\, , \nn\\[1mm]
(G^{-1})^{m}_{\phantom{m}np} \,&=\,  \rme^{2\Delta}g^{mq}\left(C_{qnp} - C_{q}B_{np} \right) \, , \nn\\[1mm]
(G^{-1})^m{}_{npqr} \,&=\, {\rme}^{2\Delta}g^{ms}\left(C_{snpqr} -C_{s[np}B_{qr]} + \tfrac{1}{2}C_s B_{[np}B_{qr]}\right) \, , \nn\\[1mm]
(G^{-1}) \,&=\, \rme^{2\Delta}\left( \rme^{-2\phi} + g^{mn}C_m C_n\right)\, .
\end{align}
These terms are sufficient to read off all the supergravity physical fields from the generalised metric (we are omitting the formula determining $\tilde B_{m_1\ldots m_6}$).
Some other components of $G^{-1}$ are
\begin{align}\label{invG_comp_2}
(G^{-1})_{m} \,&=\,  \rme^{2\Delta}g^{np}C_n B_{pm} \, , \nn\\[1mm]
(G^{-1})_{(mn)} \,&=\, \rme^{2\Delta}\left( g_{mn} + g^{pq}B_{pm}B_{qn}\right)\, , \nn\\[1mm]
(G^{-1})_{[mn]} \,&=\, - \rme^{2\Delta}\left( \rme^{-2\phi}B_{mn} - g^{pq}C_q\left( C_{pmn} - C_{p}B_{mn} \right) \right) \, , \nn\\[1mm]
(G^{-1})_{m,np} \,&=\, -\rme^{2\Delta}\left(g_{m[n}C_{p]} + g^{qr}B_{qm}\left( C_{rnp} - C_{r}B_{np} \right) \right) \, .
\end{align}

There is also a density associated to the generalised metric which trivialises the $\mathbb{R}^+$ factor of the $E_{d+1(d+1)} \times\bbR^+$ structure group. In terms of the field content of type IIA it is given by
\begin{equation}\label{gen_density}
	\Phi = (\det G)^{-(9-d)/ (\mathrm{dim} E)} = {g}^{1/2}\, \rme^{-2\phi} \rme^{(8-d)\Delta}\,,
\end{equation}
as can be seen by decomposing the corresponding M-theory  density~\cite{Coimbra:2011ky}. This equation provides an easy way to solve relations such as~\eqref{invG_comp_1} explicitly for the supergravity fields. For example, to solve the first, second and last of equations in~\eqref{invG_comp_1}, one can begin by setting
\begin{equation}\label{fields_from_G_first}
	(M^{-1})^{mn} := (G^{-1})^{mn} \,=\, \rme^{2\Delta}g^{mn}\,.
\end{equation}
The second of equations~\eqref{invG_comp_1} then becomes
\begin{equation}
	C_m = M_{mn} (G^{-1})^{n}\,,
\end{equation}
which can be substituted into the last equation in~\eqref{invG_comp_1} to give
\begin{equation}
	\rme^{2\Delta} \rme^{-2\phi} = (G^{-1}) - M_{mn} (G^{-1})^{m} (G^{-1})^{n} :=  Q\,.
\end{equation}
One then easily obtains the expressions for $g_{mn}, C_m, \rme^\Delta$ and $\rme^{-2\phi}$ as
\begin{equation}\label{fields_from_G_last}
\begin{aligned}
	\rme^\Delta = \bigg( \frac{\Phi}{Q \sqrt{\det M}} \bigg)^{1/6}\,,
	\hs{30pt}
	&\rme^{-2\phi} = \bigg( \frac{Q^4 \sqrt{\det M}}{\Phi} \bigg)^{1/3}\,, \\[1mm]
	g_{mn} = M_{mn} \bigg( \frac{\Phi}{Q \sqrt{\det M}} \bigg)^{1/3}\,,
	\hs{30pt}
	&C_m = M_{mn} (G^{-1})^n\,,
\end{aligned}
\end{equation}
where $M_{mn},Q$ and $\Phi$ are given in terms of the generalised metric as above. In particular, we have expressions for $\rme^\Delta$ and $g_{mn}$, so that solving the remaining relations in~\eqref{invG_comp_1} becomes straightforward.

The above method to compute the warp factor from an arbitrary generalised metric involves evaluating $\det G$, which is in general a slightly difficult computation. A simpler way to attain the same result is to evaluate the determinant of a subset of the components of the generalised metric, denoted $\mathcal{H}$, corresponding to the degrees of freedom in the coset
\begin{equation}
	\mathcal{H} \in \frac{\SO(d,d)\times\bbR^+}{\SO(d)\times\SO(d)}\,.
\end{equation}
Explicitly, we construct $\mathcal{H}^{-1}$ in components via
\begin{equation}
\label{eq:GB-metric}
	\mathcal{H}^{-1} \,=\, \begin{pmatrix} (G^{-1})^{mn} & (G^{-1})^m{}_n \\[1mm]
		 (G^{-1})_m{}^n & (G^{-1})_{mn} \end{pmatrix}
		 \,=\, \rme^{2\Delta} \begin{pmatrix} g^{mn} & -(g^{-1} B)^m{}_n \\[1mm]
		 (B g^{-1})_m{}^n & (g - B g^{-1} B)_{mn} \end{pmatrix}
\end{equation}
where in the second equality we have used~\eqref{invG_comp_1} and~\eqref{invG_comp_2}. We recognise the last matrix as the components of (the inverse of) the $O(d,d)$ generalised metric of~\cite{Gualtieri:2003dx}, which has unit determinant. Therefore we can immediately write
\begin{equation}
	\rme^{\Delta}\, =\, (\det \mathcal{H})^{-1/4d}\,.
\end{equation}
We comment on the appearance of the $O(d,d)$ generalised metric in appendix~\ref{app:Odd}.

\section{The massive generalised Lie derivative}\label{sec:massive_genLie}

One of the main goals of this paper is to give a generalised geometric description of massive IIA supergravity and apply it to consistent truncations where
the Romans mass contributes to the gauging of the lower-dimensional theory.  
The difficulty in incorporating the mass $m$ in
this  formalism is that the construction of the generalised tangent bundle encodes the fluxes as derivatives of the potentials which untwist the bundle, while the zero-form flux $m = F_0$ is not expressible as the derivative of a potential.  This means that it is not possible to introduce the mass term as an additional twist of the generalised bundle $E$.

The key point in solving this problem is to look at the way the gauge transformations of the NSNS and RR potentials are realised in exceptional generalised
geometry. 
We saw in section \ref{sec:gauge_IIA} how the mass affects the gauge transformations of type IIA supergravity. Since the gauge transformations of the supergravity potentials are encoded in the way the twisted generalised vectors patch, the introduction of the Romans mass requires a modification of the patching conditions~\eqref{patching}.  Following a similar reasoning as in the massless case, we find that new patching conditions of the form
 \be
\label{patching_m}
V_{(\alpha)} \,=\,  \rme^{\dd \tilde \Lambda_{(\alpha \beta)}} \,\rme^{\dd \Omega_{(\alpha \beta)} + m\,\Omega_{6(\alpha\beta)} }\, \rme^{-\dd \Lambda_{(\alpha \beta)}- m\,\Lambda_{(\alpha\beta)}} \cdot V_{(\beta)}   
\ee 
reproduce the massive supergravity gauge transformations on overlapping patches $U_\alpha \cap U_\beta$. A first-principles derivation of this is also given in appendix~\ref{app:gauge-patching}. 

Although the structure of the exact sequences~\eqref{eq:twistE} is left intact by this deformation, the precise details of the twisting~\eqref{patching_m} do change.\footnote{A consequence of this is the following. In massless IIA we can project a generalised vector onto its vector and zero-form parts $v+\omega_0$, giving a well-defined section of a bundle with seven-dimensional fibre. This is the dimensional reduction of the M-theory tangent bundle $TM_7$. However, with the massive IIA patching rules~\eqref{patching_m}, this projection would no longer give a section of a bundle with seven-dimensional fibre. Hence, the massive patching rules do not arise from a seven-dimensional geometry.}
 An important feature of massive type IIA is that by virtue of the Bianchi identity we have (globally)
\begin{equation}
\label{eq:H-exact}
	H_3 \,=\, \tfrac{1}{m}\, \dd F_2
\end{equation}
so that for $m\neq 0$ $H_3$ is trivial in cohomology. The first extension in~\eqref{eq:twistE} is thus naturally equivalent to the trivial one.

Also, a pure NSNS gauge transformation no longer acts in the $O(d,d)$ subgroup of $E_{d+1(d+1)}\times\bbR^+$, simply because it also generates a $C_1$ RR potential. As such, there is no massive version of Hitchin's $O(d,d)$ generalised geometry.\footnote{Though see~\cite{Hohm:2011cp} for a double field theory approach to this, where the $F_0$ flux is generated by introducing a linear dependence on the additional non-geometric coordinates dual to the winding modes of the string.}

The modification \eqref{patching_m} of the patching condition also requires a deformation of the Dorfman derivative. Recall that the latter generates the infinitesimal gauge transformations, and that these are affected by the Romans mass via the shifts  \eqref{gauge_massless_to_massive}. It follows that the massive form of the Dorfman derivative is obtained implementing the same shift in the massless expression~\eqref{dorf6}:
\begin{align}
\label{dorf6m}
L_V V' \,&=\, \mathcal{L}_v v' + \left(\mathcal{L}_v \lambda' - \iota_{v^\prime} \mathrm{d}\lambda\right)  + \left( \mathcal{L}_v \sigma' - \iota_{v^\prime}(\dd\sigma+m\omega_6) + [s(\omega^\prime) \wedge (\dd\omega-m\lambda)]_5\right) \nn \\[1mm]
\,&\, + \left(\mathcal{L}_v \tau' + j \sigma' \wedge \mathrm{d}\lambda + \lambda^\prime \otimes (\dd\sigma+m\omega_6) + j s(\omega^\prime) \wedge (\dd\omega-m\lambda) \right) \nn \\[1mm]
\,&\, + \left(\calL_v \omega' +  \dd \lambda \wedge \omega' - (\iota_{v'}+  \lambda' \wedge)(\dd\omega-m\lambda)\right)\, ,
\end{align}
which contains the mass as a deformation parameter.  
 More formally, \eqref{dorf6m} is related to the massless Dorfman derivative~\eqref{eq:Liedefg} (here denoted by $L^{(m=0)}$) as
\be
\label{mdefDorf}
\Lgen_V V' = \Lgen^{(m=0)}_V V' + \underline{m}(V) \cdot V'\,,
\ee
where, given a generalised vector $V$, we define the map $\underline{m}$ such that
\be
\underline{m}(V) \, = \, m\,\lambda - m\,\omega_6 \, 
\ee
is an object that acts in the adjoint of $E_{7(7)}$ (see~\eqref{IIAadjvecCompact}) as
\begin{align}\label{madj}
\underline{m}(V)\cdot V' \,=\, m \left( - \iota_{v'} \omega_6 - \lambda \wedge \omega_4' +  \lambda' \otimes \omega_6  -  \lambda \otimes \omega'_6 + \iota_{v'} \lambda + \lambda' \wedge \lambda\right)\,.
\end{align}
It is a tedious but straightforward computation to verify that  \eqref{dorf6m} satisfies the Leibniz property 
\eqref{eq:Leibniz}.\footnote{A very subtle point is that neither of the terms on the RHS of \eqref{mdefDorf} transforms correctly as a generalised vector under \eqref{patching_m}, and as a consequence $\underline{m}(V)$ does not transform as a section of the adjoint bundle. However, overall $L_V V'$ defines a good section of $E$.}

To justify further our definition, we rewrite the massive Dorfman derivative in the untwisted picture.
Using \eqref{eq:fromLtotildeL} we find 
\begin{align}
\label{eq:twistedDorfmanmassive}
\check L_{\check V} {\check V'}  &=\mathcal{L}_{\check v} \check v' + (\mathcal{L}_{\check v}\check  \lambda' - \iota_{\check v^\prime} \mathrm{d}\check \lambda + \iota_{\check v'} \iota_{\check v} H ) \nn \\[1mm]
& +  \mathcal{L}_{\check v} \check \sigma'\! - \iota_{\check v^\prime}\mathrm{d}\check \sigma  +\! \left[ \iota_{\check v'}(s(\check\omega) \wedge F)+ s(\check\omega^\prime) \!\wedge\! \big(\mathrm{d} \check\omega - H\wedge  \check\omega - (\iota_{\check v}+ \check \lambda \wedge)F\big) \right]_5  \nn \\[1mm]
& +   \mathcal{L}_{\check v} \check\tau' + j \check\sigma' \wedge (\dd \check\lambda - \iota_{\check v}H) + \check\lambda' \otimes \big( \dd \check\sigma  - [s(\check \omega)\wedge F]_6 \big) \nn \\[1mm] 
\,&\ \quad  + j s(\check\omega') \wedge \big( \dd \check \omega -  H\wedge\check \omega - (\iota_{\check v} + \check \lambda\wedge)F \big) \nn \\[1mm] 
& +  \mathcal{L}_{\check v}\check\omega' +  (\mathrm{d}\check \lambda - \iota_{\check v} H) \wedge\check\omega' - (\iota_{\check v^\prime}+ \check \lambda' \wedge) \big(\dd\check \omega - H \wedge\check \omega -(\iota_{\check v} +\check\lambda\wedge)F \big)\ ,
\end{align}
where $F = F_0+F_2 +F_4+F_6$ is now the complete $O(6,6)$ spinor with $m\neq 0$, as in \eqref{RRfieldstrength}. So the twisted version of the massive Dorfman derivative produces precisely the expected flux terms including the Romans mass. Again, these are given by the action of \eqref{FluxesAdjoint}, now with $m\neq 0$. 

Note that of all the flux terms in~\eqref{eq:twistedDorfmanmassive}, the mass term is the only one which is diffeomorphism-invariant. It is also the only true deformation of the generalised Lie derivative, since it cannot be removed by twisting the generalised tangent bundle.

\section{Generalised parallelisations and \hbox{consistent reductions} }\label{sec:GenParall}

In this section we apply the formalism developed above to dimensional reductions of type IIA supergravity on spheres and hyperboloids. We build on ideas put forward in~\cite{spheres}, where certain sphere consistent truncations were understood as {\it generalised Scherk--Schwarz reductions}. This gave evidence that generalised geometry sheds light on the hidden structure of a class of dimensional reductions whose consistency relies on the conspirancy between different terms, which seems ``miracolous'' from an ordinary Kaluza--Klein viewpoint. 

In addition to the truncation ansatz for the lower-dimensional scalar fields already given in~\cite{spheres}, we  provide the complete ansatz for the fields with one or two legs in the external space-time. We mainly have in mind reductions on six-dimensional manifolds, however the expressions we obtain also apply to reductions on spaces of dimension $d\leq 6$, after truncating away all forms of degree larger than $d$.
Moreover, while in this paper we focus on type IIA supergravity, it is straightforward to adapt the procedure to other higher-dimensional supergravity theories, such as type IIB and eleven-dimensional supergravity.

\subsection{Ordinary Scherk--Schwarz reductions}

Before coming to the generalised Scherk--Schwarz reduction, it may be useful to briefly recall how a conventional Scherk--Schwarz reduction~\cite{Scherk:1979zr} is defined. In this case, the internal manifold is chosen to be a $d$-dimensional Lie group, $M_d = G$. It follows that $M_d$ is parallelisable, namely there exists a global frame $\{\hat{e}_a\}$,  $a=1,\ldots,d$, trivialising the frame bundle and thus the tangent bundle $TM_d$. This frame is constructed by considering a basis of vectors that are invariant under the (say) left-action of the group $G$ on itself. Under the Lie derivative, the left-invariant frame satisfies the algebra
\be
\mathcal{L}_{\hat{e}_a}\hat{e}_b \,=\, f_{ab}{}^c \, \hat{e}_c\ ,
\ee
where $f_{ab}{}^c$ are the structure constants of $G$. These vectors generate the right-isometries of the bi-invariant metric on the group manifold. A truncation ansatz for the internal metric is defined by ``twisting'' the original frame on $M_d$ by a $\GL(d)$ matrix $U_a{}^b$ depending on the external spacetime coordinates $x^\mu$,
\be
\hat{e}'_a{}^m(x,z) \,=\, U_a{}^b(x)\, \hat{e}_b{}^m(z)\ ,
\ee
and setting
\be
g^{mn}(x,z) \,=\, \delta^{ab}\, \hat{e}'_a{}^m(x,z) \,\hat{e}'_b{}^n(x,z) \,=\, \mathcal{M}^{ab}(x) \,\hat{e}_a{}^m(z) \,\hat{e}_b{}^n(z)\ ,
\ee
where $\mathcal{M}^{ab} = \delta^{cd}\,U_c{}^a U_d{}^b\,$. As we are free to redefine the frame by $x$-dependent $\SO(d)$ transformations, the $\mathcal{M}^{ab}$ matrix parameterises the coset $\GL(d)/\SO(d)$; hence it defines $\frac{1}{2}d(d+1)$ scalars on the external spacetime. It follows that $g_{mn} =\mathcal{M}_{ab} e_m{}^a e_n{}^b$, where $\mathcal{M}_{ab}$ is the inverse of $\mathcal{M}^{ab}$, and, as before, the one-forms $e^a$ are dual to the vectors $\hat{e}_a$. The full ten-dimensional metric is given by
\be
\dd \tend{s}^2\, = \, g_{\mu\nu}\dd{x}^\mu\dd{x}^\nu + \mathcal{M}_{ab}  (e^a - \mathcal{A}^a)(e^b - \mathcal{A}^b)\,.
\ee
The $d$ one-forms $\mathcal{A}^a=\mathcal{A}_\mu{}^a(x)\dd{x}^\mu$ gauge the right-isometries on the group manifold, and are therefore $G$ gauge fields on the external spacetime.
For the RR one-form one takes
\be
\tend{C}_1(x) \,=\, C_\mu(x) \dd{x}^\mu + C_a(x) (e^a - \mathcal{A}^a) + \rg{C_1}\ ,
\ee
where $\rg{C_1}$ is the potential for a background, left-invariant two-form flux. This gives an additional one-form and $d$ more scalars. A similar ansatz is taken for the other form potentials.
 
The reduction defined in this way is consistent by symmetry reasons: the dependence of the type IIA fields on the internal coordinates is fully encoded in the left-invariant tensors $\hat{e}_a$ and $e^a$, and there is no way the singlet modes can source the truncated non-singlet modes in the equations of motion.
The gauge group of the lower-dimensional, truncated theory arises from the interplay between the right-Killing symmetries generated by the left-invariant vectors $\hat{e}_a$ and the gauge transformations of the form potentials with flux, and corresponds to a semi-direct product of $G$ with a non-compact factor. The full supersymmetry of the original theory is preserved in the truncation. 

We refer to e.g.~\cite{Kaloper:1999yr,Dall'Agata:2005ff,D'Auria:2005er,Hull:2005hk,Hull:2006tp} for a detailed account of conventional Scherk--Schwarz reductions in a context related to the one of this paper.

\subsection{Generalised Scherk--Schwarz reductions}\label{sec:genSS}

Extensions of conventional Scherk--Schwarz reductions to reformulations (or extensions) of high-dimensional supergravity theories with larger structure groups have been considered by several authors, see~e.g.~\cite{Riccioni:2007au,Berman:2012uy,Aldazabal:2013mya,Godazgar:2013dma,spheres,Hohm:2014qga,Ciceri:2014wya,Baguet:2015xha,
Baguet:2015sma,Malek:2015hma}. Here we will follow~\cite{spheres} and define a {\it generalised Scherk--Schwarz reduction} on a $d$-dimensional manifold $M_d$ (not necessarily a Lie group) as the direct analogue of an ordinary Scherk--Schwarz reduction, with the ordinary frame on the tangent bundle replaced by a frame on the generalised tangent bundle. In particular this will allow us to derive an explicit ansatz for the fields with one or two external legs for type IIA (in analogy to the exceptional field theory expressions for eleven-dimensional and type IIB supergravity given in~\cite{Hohm:2014qga,Baguet:2015xha,Baguet:2015sma}). 

As in any Kaluza--Klein reduction, we start by decomposing the type IIA fields according to the $\SO(1,9) \to \SO(1,9-d)\times \SO(d)$ splitting of the  Lorentz group. We will use coordinates $x^\mu$, $\mu = 0,\ldots, 9-d$ for the external spacetime and $z^m$, $m=1,\ldots,d$ for the internal manifold $M_d$, of dimension $d \leq 6$. Then the ten-dimensional metric can be written as
\be\label{KK_decomp_metr}
\tend{g} \,= \, \rme^{2\Delta}g_{\mu\nu}\dd{x}^\mu\dd{x}^\nu + g_{mn} Dz^m Dz^n\ ,
\ee
where 
\be
Dz^m \,=\, \dd{z}^m - h_\mu{}^m \dd{x}^\mu \ ,
\ee 
and the scalar $\Delta$ is the warp factor of the external metric $g_{\mu\nu}$. In this section the symbol hat  denotes the original ten-dimensional fields. The form fields are decomposed as
\begin{align}\label{expand_10dfields}
\tend{B}  &=  \tfrac{1}{2} B_{m_1m_2} Dz^{m_1m_2} + \overline{B}_{\mu m} \dd{x}^\mu \wedge Dz^m + \tfrac{1}{2}\overline{B}_{\mu\nu} \dd{x}^{\mu\nu} \, ,\nn\\[1mm]
\tend{\tilde{B}} &= \tfrac{1}{6!} \tilde{B}_{m_1\ldots m_6} Dz^{m_1\ldots m_6} + \tfrac{1}{5!} \overline{\tilde B}_{\mu m_1\ldots m_5} \dd{x}^\mu \!\wedge\! Dz^{m_1\ldots m_5} +  \tfrac{1}{2\cdot 4!} \overline{\tilde B}_{\mu\nu m_1\ldots m_4} \dd{x}^{\mu\nu} \!\wedge\! Dz^{m_1\ldots m_4} + \ldots  ,\nn\\[1mm]
\tend{C}_1 &= C_m Dz^m + \overline{C}_{\mu,0} \,\dd{x}^\mu  \, ,\nn\\[1mm]
\tend{C}_3 &= \tfrac{1}{3!} C_{m_1m_2m_3} D{z}^{m_1m_2m_3} + \tfrac{1}{2}\overline{C}_{\mu m_1m_2} \dd{x}^\mu\wedge D{z}^{m_1m_2} + \tfrac{1}{2}\overline{C}_{\mu\nu m}\dd{x}^{\mu\nu} \wedge Dz^m + \ldots \, ,\nn\\[1mm]
\tend{C}_5 &= \tfrac{1}{5!} C_{m_1\ldots m_5} D{z}^{m_1\ldots m_5} + \tfrac{1}{4!}\overline{C}_{\mu m_1\ldots m_4} \dd{x}^\mu\!\wedge\! D{z}^{m_1\ldots m_4} +\tfrac{1}{2\cdot 3!} \overline{C}_{\mu\nu m_1m_2m_3}\dd{x}^{\mu\nu} \!\wedge\! Dz^{m_1m_2m_3}+\ldots ,\nn\\[1mm]
\tend{C}_7 &= \tfrac{1}{6!}\overline{C}_{\mu m_1\ldots m_6} \dd{x}^\mu\wedge D{z}^{m_1\ldots m_6} +\tfrac{1}{2\cdot 5!} \overline{C}_{\mu\nu m_1\ldots m_5}\dd{x}^{\mu\nu} \wedge Dz^{m_1 \ldots m_5} + \ldots \, ,
\end{align}
where $\dd{x}^{\mu\nu} = \dd x^\mu\wedge \dd x^\nu$ and $Dz^{m_1\ldots m_p} = Dz^{m_1}\wedge \cdots \wedge Dz^{m_p}$.
The ellipsis denote forms with more than two external indices, that we will not need. 
The expansion in $Dz$ instead of $\dd z$ is standard in Kaluza--Klein reductions, and ensures that the components transform covariantly under internal diffeomorphisms. We stress that at this stage the field components still depend on all the coordinates $\{x^\mu,z^m\}$:  we are decomposing the various tensors according to their external or internal legs but we have not specified their dependence on the internal space yet. The only exception is the external metric, which is assumed to depend just on the external coordinates: $g_{\mu\nu} = g_{\mu\nu}(x)$.

The barred fields  appearing in~\eqref{expand_10dfields} can also be identified by introducing the vector 
\be
\partial_\mu + h_\mu \,=\, \frac{\partial}{\partial x^\mu} + h_\mu{}^m \frac{\partial}{\partial z^m}\,,
\ee 
which satisfies $\iota_{(\partial_\mu + h_\mu)}Dz^m =0$. For the the fields with one external leg  we have 
\begin{align}
\overline{B}_{\mu} \,&=\, \iota_{(\partial_\mu + h_\mu)} \hat B \, \big| \,,\nn\\[1mm]
\overline{\tilde B}_{\mu} \,&=\, \iota_{(\partial_\mu + h_\mu)} \hat{\tilde{B}} \, \big| \,,\nn\\[1mm]
\overline{C}_{\mu} \,&=\, \iota_{(\partial_\mu + h_\mu)} \hat C \, \big| \,,
\end{align}
where by the symbol ``$|$'' we mean that after having taken the contraction $\iota_{(\partial_\mu + h_\mu)}$, the forms on the right hand side are restricted to have just internal legs. In other words, we set $\dd x \equiv 0$. Similarly, for the fields with two external legs we find
\begin{align}\label{redef_one_forms_2}
\overline{B}_{\mu\nu} \,&=\, \iota_{(\partial_\nu + h_\nu)} \iota_{(\partial_\mu + h_\mu)} \tend{B}  \, ,\nn\\[1mm]
\overline{\tilde B}_{\mu\nu} \,&=\, \iota_{(\partial_\nu + h_\nu)} \iota_{(\partial_\mu + h_\mu)} \tend{\tilde B}  \, \big|\, ,\nn\\[1mm]
\overline{C}_{\mu\nu} \,&=\,  \iota_{(\partial_\nu + h_\nu)}\iota_{(\partial_\mu + h_\mu)} \tend{C} \, \big| \, .
\end{align}
Moreover, we are arranging the RR potentials in the poly-forms 
\begin{align}
\overline{C}_\mu \,&=\, \overline{C}_{\mu,0} + \overline{C}_{\mu,2} + \overline{C}_{\mu,4} + \overline{C}_{\mu,6}\,,\nn\\[1mm]
\overline{C}_{\mu\nu} \,&=\, \overline{C}_{\mu\nu,1} + \overline{C}_{\mu\nu,3} + \overline{C}_{\mu\nu,5}\,.
\end{align}
These barred fields need a field redefinition. This can be seen by decomposing the gauge transformations of the ten-dimensional fields and imposing that they are covariant under the generalised diffeomorphisms so that they will eventually reproduce the gauge transformation of the lower-dimensional supergravity theory after the truncation is done. Here we just provide the correct redefinitions, postponing their full justification to the next section. 
We introduce the new fields
\begin{align}\label{redef_one_forms}
B_{\mu} \,&=\, \overline{B}_\mu \, ,\nn\\[1mm]
C_\mu \,&=\, \rme^{-B}\wedge \overline{C}_\mu 
\, ,\nn\\[1mm]
\tilde{B}_\mu \,&=\, \overline{\tilde{B}}_\mu - \tfrac{1}{2} [\,\overline{C}_\mu \wedge s(C)\,]_5 
\,,
 \end{align}
where $B$, $C$ are just internal, and
\begin{align}\label{redef_two_forms}
B_{\mu\nu}\,&=\, \overline{B}_{\mu\nu} + \iota_{h_{[\mu}}B_{\nu]} 
 \,,\nn\\[1mm]
\tilde B_{\mu\nu}\,&=\, \overline{\tilde B}_{\mu\nu} - \tfrac{1}{2} \big[ \,\overline{C}_{\mu\nu} \wedge s(C)\, \big]_4  + \iota_{h_{[\mu}} \tilde{B}_{\nu]}  \,,\nn\\[1mm]
C_{\mu\nu} \,&=\, \rme^{-B} \wedge \overline{C}_{\mu\nu} + \iota_{h_{[\mu}} C_{\nu]} + B_{[\mu}\wedge C_{\nu]} \,.
\end{align}
Note that we are using a notation where the various tensors are treated as differential forms on the internal manifold, while we explicitly display their external indices.

Having decomposed the higher-dimensional fields in a suitable way, we are now ready to construct our truncation ansatz. 
As a first thing we rearrange the type IIA fields with zero, one or two external indices in terms of generalised geometry objects. The fields with purely internal legs, i.e. 
\be
\{g_{mn},\, B_{m_1m_2},\, \tilde B_{m_1\ldots m_6} ,\, C_m,\, C_{m_1m_2m_3},\, C_{m_1 \ldots m_5} \}\ ,
\ee
together with the warp factor $\Delta$ and the dilaton $\phi$, parameterise a generalised metric $G^{MN}$.
The (redefined) fields with one external index are collected in the generalised vector
\be\label{def_calA_mu^M}
\mathcal{A}_\mu{}^M \,=\, \{ h_\mu{}^m ,\, B_{\mu m} ,\, \tilde B_{\mu m_1\ldots m_5},\, \tilde{g}_{\mu m_1\ldots,m_6,m} ,\, C_{\mu,0},\, C_{\mu m_1m_2},\, C_{\mu m_1\ldots m_4},\, C_{\mu m_1\ldots m_6} \}\ .
\ee
Here, $\tilde{g}$ is a tensor belonging to $ \Lambda^7T^*M_{10}\otimes T^*M_{10}$, related to the dual graviton. This is not part of type IIA supergravity in its standard form and we will thus ignore it by projecting $\mathcal{A}_\mu$ on the $E'''$ bundle introduced in \eqref{eq:twistE},
\be
\mathcal{A}_\mu{}^M \,\eqs\, \{ h_\mu{}^m ,\, B_{\mu m} ,\, \tilde B_{\mu m_1\ldots m_5},\, C_{\mu,0},\, C_{\mu m_1m_2},\, C_{\mu m_1\ldots m_4},\, C_{\mu m_1\ldots m_6} \}\ .
\ee
Here and below, the $\eqs$ symbol in an equation involving generalised vectors means that the equality holds after projecting on the bundle $E'''$ using the natural mappings~\eqref{eq:twistE}, namely after dropping the $T^*\otimes \Lambda^{6}T^*$ component.

The fields with $\mu\nu$ indices defined in \eqref{redef_two_forms} are components of a generalised tensor $\mathcal{B}_{\mu\nu}{}^{MN}$, which is  a two-form in the external spacetime and a section of the bundle $N$ on $M_6$ defined in~\eqref{Nbundle}. They actually correspond to the components of this object living on the bundle $N'$ given in~\eqref{NprimeBundle}, that is
\be
\mathcal{B}_{\mu\nu}{}^{MN} \,\eqs\, \{ B_{\mu\nu},\, \tilde{B}_{\mu\nu m_1\ldots m_4} ,\, C_{\mu\nu m},\, C_{\mu\nu m_1m_2m_3} ,\, C_{\mu\nu m_1\ldots m_5}\}\,.
\ee
For the equations involving sections of the bundle $N$, by the $\eqs$ symbol we mean that the equality holds after having projected on the bundle $N'$.

Suppressing the internal indices, the objects introduced above read
\begin{align}
\mathcal{A}_\mu \,&\eqs\,  h_\mu + B_{\mu} + \tilde B_{\mu} + C_{\mu,0}+  C_{\mu,2}+ C_{\mu ,4} + C_{\mu,6} \ ,\nn\\[1mm]
\mathcal{B}_{\mu\nu} \,&\eqs\, B_{\mu\nu} + \tilde{B}_{\mu\nu} + C_{\mu\nu,1}+ C_{\mu\nu ,3} + C_{\mu\nu,5}\,.
\end{align}

The construction of a (bosonic) truncation ansatz leading to a $(10-d)$-dimensional theory preserving maximal supersymmetry is then specified by the following steps:

\medskip

{\it Step 1.} One should find a generalised parallelisation $\{\hat E_A\}$, namely a globally-defined frame for the $E_{d+1(d+1)}\times \mathbb{R^+}$ generalised tangent bundle on $M_d$.  This means that the frame $\{\hat E_A\}$ must be  an $E_{d+1(d+1)}$ frame, namely that it is given by an $E_{d+1(d+1)}$ transformation of the coordinate frame.\footnote{By coordinate frame we mean 
 \be
 \{  \check{\hat{E}}_A\} = \{\partial_m\} \cup \{\dd x^m\} \cup \{\dd x^{m_1 \ldots m_5} \} \cup  
 \{\dd x^{m,m_1\ldots m_6}\} \cup \{1\} \cup \{\dd x^{m_1m_2}\} \cup \{\dd x^{m_1\ldots m_4}\} \cup \{\dd x^{m_1 \ldots ,_6}\} \, .  \nn 
 \ee}
We will see how this condition applies in the examples below. 
In addition, the frame must satisfy the algebra 
\begin{equation}\label{LeibnizParall}
L_{\hat E_A}\hat E_B \, =\, X_{AB}{}^C \hat E_C\ ,
\end{equation}
with {\it constant} coefficients $X_{AB}{}^C$. Following~\cite{spheres}, we call this a {\it generalised Leibniz parallelisation}; the name is due to the fact that since the generalised Lie derivative $L$ is not antisymmetric, the frame algebra defined by \eqref{LeibnizParall} is a Leibniz algebra and not necessarily a Lie algebra. The constants $X_{AB}{}^C$ correspond to the generators of the gauge group: in gauged supergravity they are defined by contracting the {\it embedding tensor}  $\Theta_A{}^\alpha$ encoding the gauging of the theory with the generators $(t_\alpha)_B{}^C$ of the U-duality group, $X_{AB}{}^C = \Theta_A{}^\alpha (t_\alpha)_B{}^C$ (we refer to e.g.~\cite{Samtleben:2008pe} for a review of the embedding tensor formalism). 
Using the Leibniz property~\eqref{eq:Leibniz} of the Dorfman derivative together with~\eqref{LeibnizParall}, we see that indeed the constants $X_{AB}{}^C$ realise the gauge algebra
\be
\label{eq:gauge-alg-X}
[X_A,X_B] \ = \ -X_{AB}{}^C X_C\ .
\ee
We emphasise that, provided the dimensional reduction goes through consistently, the knowledge of $X_{AB}{}^C$ alone is sufficient to completely determine the resulting gauged maximal supergravity.

\medskip

{\it Step 2.} One twists the parallelising frame by an $E_{d+1(d+1)}$ matrix $U_A{}^B$ depending on the external spacetime coordinates $x^\mu$:
\be
\hat E'_A{}^M(x,z) \,=\, U_A{}^B(x)\hat E_B{}^M(z)\ ,
\ee
and use this to construct a generalised inverse metric:
\be\label{invG_from_parall}
G^{MN}(x,z) \,=\, \delta^{AB}\hat E'_A{}^M (x,z)  \hat E'_B{}^N (x,z) \,=\, \mathcal{M}^{AB}(x) \hat E_A{}^M(z) \hat E_B{}^N(z)\ .
\ee
The matrix 
\be
\mathcal{M}^{AB} \,=\, \delta^{CD}U_C{}^A U_D{}^B
\ee 
parameterises the coset $E_{d+1(d+1)}/K$, where $K$ is the maximal compact subgroup of $E_{d+1(d+1)}$ (indeed, we are free to redefine the generalised frame by $x$-dependent $K$ transformations). Hence it accommodates all the scalars of the lower-dimensional theory.

Now one equates~\eqref{invG_from_parall} to the generic form of the generalised inverse metric $G^{-1}$ introduced in section~\ref{gen_frame_metric}, whose relevant components are given in \eqref{invG_comp_1} and  \eqref{invG_comp_2}. In this way we obtain the truncation ansatz for the full set of higher-dimensional degrees of freedom with purely internal components, which gives the scalar fields in the lower-dimensional theory. This also provides the expression for the warp factor $\Delta$. Concretely, these can be extracted following eqs.~\eqref{fields_from_G_first}--\eqref{fields_from_G_last}. Note that, since  the generalised density $\Phi$ appearing in~\eqref{fields_from_G_last} is independent of the twist matrix $U_A{}^B$,  it can be advantageously computed at the origin of the scalar manifold, where $\mathcal{M}^{AB}=\delta^{AB}$. So at any point on the scalar manifold the density is given by
\begin{equation}\label{gen_density_background}
	\Phi \,= \,\ \rg{g}{}^{\!1/2}\, \rme^{-2\rg\phi} \, \rme^{(8-d)\rg\Delta}\, ,
\end{equation}
where the {\large${\rg{\,}}$} symbol denotes the ``reference'' values of the corresponding fields, namely the values for trivial twist matrix.

\medskip

{\it Step 3.} The full set of vector fields in the lower-dimensional theory is specified by taking the following ansatz for the generalised vector $\mathcal{A}_\mu{}^M$ introduced in \eqref{def_calA_mu^M}
\be\label{trunc_ansatz_vec}
\mathcal{A}_\mu{}^M(x,z)  \,=\, \mathcal{A}_\mu{}^A(x) \hat{E}_A{}^M(z) \ .
\ee
The ansatz for the two-forms is
\be\label{ansatz_two-forms}
\mathcal{B}_{\mu\nu}{}^{MN}(x,z) \,\eqs\, \tfrac{1}{2}\,\mathcal{B}_{\mu\nu}{}^{AB}(x) (\hat{E}_A \otimes_{N'} \!\hat{E}_B)^{MN}(z)\,, 
\ee
where $\mathcal{B}_{\mu\nu}{}^{AB} = \mathcal{B}_{\mu\nu}{}^{(AB)}$, and the product $\otimes_{N'}$ is defined in~\eqref{N'prod_IIA}.

\medskip

A few comments are in order. Although the conditions in {\it Step~1} above are definitely non-trivial to satisfy, they are not as constraining as requiring that $M_d$ is a Lie group as needed in ordinary Scherk--Schwarz reductions. In fact, one can see that a necessary condition for the existence of a generalised parallelisation satisfying~\eqref{LeibnizParall} is that $M_d$ is a coset manifold, $M_d = G/H$ for some $G$ and $H\subset G$ \cite{spheres}.
 
 In the particular case that $M_d$ is a Lie group, a generalised Scherk--Schwarz reduction coincides with an ordinary Scherk--Schwarz reduction if the chosen generalised parallelisation uses just left-invariant tensors.\footnote{See~\cite[app.$\:$C]{spheres} for a discussion. In this case, adopting a generalised geometry approach still has some advantage in that~\eqref{LeibnizParall} directly provides the full embedding tensor.}
 However, when reducing the NSNS sector, it is possible to obtain a generalised parallelisation which realises a $G\times G$ gauge group rather than just $G$~\cite{Baguet:2015iou}. In the next section we will provide a frame for the full type IIA exceptional generalised geometry on $S^3$ which gives rise to an $\SU(2)\times\SU(2)$ gauging (this has also appeared in~\cite{Malek:2015hma}).

The spheres $S^d = \SO(d+1)/SO(d)$ provide examples of generalised parallelisations that are not based on Lie groups.
 In~\cite{spheres}, the ideas above were applied to give evidence that the sphere consistent truncations based on eleven-dimensional supergravity on $S^7$ \cite{deWit:1986oxb}, eleven-dimensional supergravity on $S^4$ \cite{Nastase:1999kf}, type IIB supergravity on $S^5$ and the NSNS sector of type II supergravity on $S^3$, can be interpreted as generalised Scherk--Schwarz reductions. 
In section~\ref{sec:examples} we will provide additional examples.

\subsection{Consistent reduction of gauge transformations}

We now provide a partial proof of the consistency of our generalised Scherk--Schwarz truncation ansatz  by showing that the internal diffeomorphisms together with the NSNS and RR gauge transformations  consistently reduce to the appropriate gauge variations in lower-dimensional maximal supergravity.\footnote{A more thorough proof would require studying the reduction of the supersymmetry variations or the equations of motion.} This will also justify the field redefinitions performed in \eqref{redef_one_forms} and  \eqref{redef_two_forms}. The reader not interested in the details of this computation, which is rather technical, can safely skip to the next section.

The gauge transformations of the ten-dimensional fields were given in section~\ref{sec:gauge_IIA}. Including also the diffeomorphisms, they read
\begin{align}
\label{gauge_var_full}
\delta\tend{g} \,&=\, \mathcal{L}_{\tend{v}} \tend{g} \,,\nn\\[1mm]
\delta \tend{B} \,&=\, \mathcal{L}_{\tend{v}} \tend{B} - \dd \tend{\lambda} \nn \ ,\nn\\[1mm]
\delta \tend{C} \,&=\, \mathcal{L}_{\tend{v}} \tend{C} -\rme^{\tend{B}}\wedge(\dd \tend{\omega} - m  \tend{\lambda}) \ ,\nn \\[1mm]
\delta \tend{\tilde{B}} \,&=\, \mathcal{L}_{\tend{v}} \tend{\tilde{B}} - (\dd \tend{\sigma} +  m\, \tend{\omega}_6) - \tfrac12 [\rme^{\tend{B}} \wedge (\dd\tend{\omega} - m \tend{\lambda}) \wedge s(\tend{C})]_6 \ .
\end{align}

We can immediately see why the redefinition of the RR potentials in~\eqref{redef_one_forms} is needed: for the gauge transformation of $C_\mu$ to start with $\partial_\mu \omega$ (as required for a gauge field in supergravity), we need to remove the $B$-terms with internal legs appearing in front of $\dd \omega$. The same argument determines the redefinition of the six-form NSNS potential in~\eqref{redef_one_forms}.

In order to decompose the gauge transformations, we express the gauge parameters as
\begin{align}\label{10d_gauge_param}
{\tend{v}} \,&=\, v \,=\, v^m \frac{\partial}{\partial z^m}\,,\nn\\[1mm]
\tend{\lambda} \,&=\, \lambda + \overline{\lambda}_\mu \,=\,  \lambda_m \dd{z}^m + \overline{\lambda}_\mu \dd{x}^\mu \, ,\nn\\[1mm]
\hat\sigma  \,&=\, \sigma + \overline{\sigma}_\mu + \overline{\sigma}_{\mu\nu} \, =\,  \tfrac{1}{5!}\sigma_{m_1\ldots m_5}\dd z^{m_1\ldots m_5} + \tfrac{1}{4!}\,\overline{\sigma}_{\mu m_1\ldots m_4}\dd x^\mu\wedge \dd z^{m_1\ldots m_4}
\nn\\[1mm]
\,&\qquad \qquad \qquad  \qquad\,\  + \tfrac{1}{2\cdot3!}\,\overline{\sigma}_{\mu\nu m_1\ldots m_3}\dd x^{\mu\nu}\wedge \dd z^{m_1\ldots m_3} + \ldots \, ,
\end{align}
where the ellipsis denote terms with more than two external indices, that we will ignore. Note that the vector $\hat v$ is purely internal, that is the diffeomorphisms we consider are just the internal ones. Similarly for the RR poly-form gauge parameter
we find
\begin{align}
\hat\omega \,=\, \omega + \overline{\omega}_\mu + \overline\omega_{\mu\nu} \,& =\,  (\omega_0+\omega_2+\omega_4+\omega_6) + (\overline{\omega}_{\mu,1}+ \overline{\omega}_{\mu,3}+\overline{\omega}_{\mu,5}) \nn\\[1mm]
& \,\quad + (\overline{\omega}_{\mu\nu,0}+\overline{\omega}_{\mu\nu,2}+\overline{\omega}_{\mu\nu,4}+\overline{\omega}_{\mu\nu,6}) + \ldots .
\end{align}
As in \eqref{expand_10dfields}, initially we impose no restriction on the dependence of the components of the gauge parameters on the coordinates $\{x^\mu,z^m\}$. However, differently from \eqref{expand_10dfields}, note that the expansion of the gauge parameters is made in $\dd{z}^m$ and not in $Dz^m=\dd z^m - h_\mu{}^m \dd x^\mu$. The fields marked with a bar require a redefinition, which will be introduced below.

The gauge transformations of the fields with purely internal legs maintain precisely the same form as in~\eqref{gauge_var_full}. 
As for the fields with one external leg, redefined as in~\eqref{redef_one_forms},  
after some computation we find that their variations are
\begin{align}\label{var_one_ext_index}
\delta h_\mu \,&=\, -\partial_\mu v + \mathcal{L}_v h_\mu \,,\nn\\[1mm]
\delta B_\mu \,&=\, -\partial_\mu \lambda + \din \overline{\lambda}_\mu + \mathcal{L}_v B_\mu - \iota_{h_\mu} \din\lambda\,,\nn \\[1mm]
\delta \tilde B_\mu \,&=\, -\partial_\mu \sigma + \din{\overline{\sigma}_{\mu}} - m\,\overline{\omega}_{\mu,5} + \mathcal{L}_v \tilde{B}_{\mu} - \iota_{h_\mu}(\din\sigma+m\omega_6) + \left[ C_\mu \wedge s(\din \omega - m\lambda) \right]_5 \,,\nn\\[1mm]
\delta C_\mu \,&=\, - \partial_\mu \omega + \din \overline{\omega}_\mu + m \overline{\lambda}_\mu + \mathcal{L}_v C_\mu + C_{\mu} \wedge \din \lambda - (\iota_{h_\mu}+ B_\mu \wedge)(\din\omega -m\lambda)\,,
\end{align}
where the exterior derivative $\din := \dd z^m \partial_m$  and  $\mathcal{L}$  act on the internal coordinates only.
The fields with two external legs have the following gauge variations
\begin{align}\label{transf_2form_BtildeB}
\delta B_{\mu\nu} \,&=\, -2 \partial_{[\mu} \overline{\lambda}_{\nu]} +\iota_{h_{[\mu}}\partial_{\nu]} \lambda - \iota_{h_{[\mu}}\din \overline{\lambda}_{\nu]} + \mathcal{L}_v B_{\mu\nu} - \iota_{\partial_{[\mu}v} B_{\nu]}\, ,\nn\\[2mm]
\delta \tilde B_{\mu\nu}  \,&=\,  - 2 \partial_{[\mu}\overline{\sigma}_{\nu]} -\din \overline{\sigma}_{\mu\nu} -m\, \overline{\omega}_{\mu\nu,4} + \iota_{h_{[\mu}}\big(\partial_{\nu]}\sigma - \din \overline{\sigma}_{\nu]} + m \,\overline{\omega}_{\nu],5} \big) + \mathcal{L}_v \tilde B_{\mu\nu}- \iota_{\partial_{[\mu}v}\tilde{B}_{\nu]} \nn\\[1mm]
\,&\,\quad   + \big[C_{\mu\nu} \wedge s(\din\omega-m\lambda)  +  ( -\partial_{[\mu}\omega + \din \overline{\omega}_{[\mu} + m\overline{\lambda}_{[\mu} ) \wedge s(C_{\nu]}) \big]_4
\end{align}
and (we give the transformations for the barred fields, as those of the unbarred field $C_{\mu\nu}$ are more cumbersome)
\begin{align}\label{Cbar_transf}
\delta (\rme^{-B} \wedge \overline{C}_{\mu\nu}) \,&= -\, 2\partial_{[\mu} \overline{\omega}_{\nu]} - 2\iota_{h_{[\mu}}\din \overline{\omega}_{\nu]} +2\iota_{h_{[\mu}}\partial_{\nu]}\omega -\iota_{h_{\nu}} \iota_{h_\mu} \din \omega - \din \overline{\omega}_{\mu\nu} \nn \\[1mm]
\, &\,\quad + \mathcal{L}_v (\rme^{-B}\wedge\overline{C}_{\mu\nu}) +\din\lambda \wedge (\rme^{-B}\wedge\overline{C}_{\mu\nu})  - \overline{B}_{\mu\nu} (\din \omega -m\lambda)
\nn \\[1mm]
\, &\,\quad  + 2 B_{[\mu} \wedge \iota_{h_{\nu]}}(\din \omega -m\lambda)  + 2 B_{[\mu}\wedge ( \partial_{\nu]} \omega - \din\overline{\omega}_{\nu]} -m\overline{\lambda}_{\nu]}) \nn\\[1mm]
\, &\,\quad  + B_{\mu} \wedge B_\nu \wedge \left(\din\omega - m \lambda\right) \,.
\end{align}

The gauge parameters with purely internal indices can be arranged into a generalised vector with the $T^*\otimes \Lambda^6T^*$ component projected out,
\be
\Lambda^M \,\eqs\, \{ v^m,\, \lambda_m,\, \sigma_{m_1\ldots m_5},\, \omega_0,\, \omega_{m_1m_2} ,\,\omega_{m_1\ldots m_4},\,\omega_{m_1\ldots m_6} \}\, ,
\ee
while the gauge parameters with one external leg form a section of the bundle $N'$,
\be
\overline{\Xi}_\mu^{\;(MN)} \,\eqs\, \{ \overline{\lambda}_\mu ,\, \overline{\sigma}_{\mu n_1\ldots n_4} ,\, \overline{\omega}_{\mu n},\, \overline{\omega}_{\mu n_1n_2n_3} ,\, \overline{\omega}_{\mu n_1\ldots n_5} \}\, .
\ee
The transformations for the fields with two external legs will be discussed below.
 
The gauge transformation of the fields with purely internal indices is given by the compact expression
\be\label{gauge_var_G}
\delta_\Lambda G^{-1} \,=\, L_\Lambda G^{-1}\,,
\ee
where $L_\Lambda$ is the massive Dorfman derivative~\eqref{dorf6m}.
The gauge variation \eqref{var_one_ext_index} of fields with one external leg can be repackaged into
\be\label{Avaria0}
\delta \mathcal{A}_\mu \, \eqs\, -\partial_\mu \Lambda + L_\Lambda \mathcal{A}_\mu +  \dd_{\rm m} \overline{\Xi}_\mu\,,
\ee
where it is understood that the differentials in the generalised Lie derivative act on the internal coordinates only. The  operator $\dd_{\rm m}$ is defined on any element $W = W_0 + W_4 + W_{\rm odd}$ of the bundle $N'$ as
\be\label{m_twist_ext_der}
\dd_{\rm m} W \,=\, \dd W + m(W_0 - W_5)\,,
\ee
and can be seen as an exterior derivative twisted by the Romans mass. Then in the present case we have
\be
\dd_{\rm m} \overline{\Xi}_\mu \,=\,\din \overline{\Xi}_\mu + m (\overline{\lambda}_\mu - \overline{\omega}_{\mu,5})\,.
\ee
It is easy to verify that for $W = V \otimes_{N'} V' $,
\be\label{eq:symmetricL}
\dd_{\rm m} W \,\eqs\, L_V V' + L_{V'}V \,\,.
\ee
If we now redefine the gauge parameter with one external leg as
\be\label{redefXi}
\overline{\Xi}_\mu\, =\, \Xi_\mu - \mathcal{A}_\mu \xN \Lambda\,,
\ee
and use the property~\eqref{eq:symmetricL},  we obtain 
\be
\dd_{\rm m} \overline{\Xi}_\mu \,\eqs\, \dd_{\rm m} \Xi_\mu - L_{\mathcal{A}_\mu} \Lambda - L_\Lambda \mathcal{A}_\mu\, . 
\ee
This redefinition allows to cast \eqref{Avaria0}  in the form
\be \label{Avaria}
\delta \mathcal{A}_\mu \, \eqs\, -\partial_\mu \Lambda - L_{\mathcal{A}_\mu} \Lambda  +  \dd_{\rm m} \Xi_\mu \,,
\ee
where one recognise the derivative $(\partial_\mu + L_{\mathcal A_\mu})\Lambda$, covariant under generalised diffeomorphisms. This is the appropriate form for matching the gauged supergravity covariant derivative after Scherk--Schwarz reduction.

We need to express the gauge transformations  \eqref{transf_2form_BtildeB} and  \eqref{Cbar_transf} of the external two-form fields 
in generalised geometry terms. This requires a rather complicated redefinition of the gauge parameters 
 $\overline{\omega}_{\mu\nu}= \overline{\omega}_{\mu\nu,0}+\overline{\omega}_{\mu\nu,2}+\overline{\omega}_{\mu\nu,4}$ and $\overline{\sigma}_{\mu\nu}$:
\begin{align}\label{expr_2form_gauge_par}
\overline{\omega}_{\mu\nu} &= \omega_{\mu\nu} +  (\iota_v+\lambda\wedge) C_{\mu\nu}  - \omega\,B_{\mu\nu}  + (2\lambda_{[\mu} + \iota_{h_{[\mu}}\lambda + \iota_v B_{[\mu})C_{\nu]} \nn\\[1mm]
& \, \quad  + (\iota_{h_{[\mu}} + B_{[\mu}\wedge\, )( 2\omega_{\nu]}+\iota_v C_{\nu]} + \lambda\wedge C_{\nu]} + \iota_{h_{\nu]}}\omega + B_{\nu]}\wedge \omega) \nn\\[2mm]
\overline{\sigma}_{\mu\nu} &= \sigma_{\mu\nu} + 2\iota_{h_{[\mu}}\sigma_{\nu]} +\iota_{h_\mu}\iota_{h_\nu}\sigma + \iota_v(\tilde{B}_{\mu\nu} - \iota_{h_{[\mu}}\tilde{B}_{\nu]}) + \iota_v C_{[\mu,4}C_{\nu],0}  \nn\\[1mm]
&\,\quad  -\iota_v C_{[\mu,2}\wedge C_{\nu],2} + 2\lambda\wedge (C_{[\mu,2}C_{\nu],0})  \nn\\[1mm]
&\,  \quad  - \big[(C_{\mu\nu} - \iota_{h_{[\mu}}C_{\nu]} + B_{[\mu}\wedge C_{\nu]})\wedge s(\omega) - 2C_{[\mu}\wedge s(\omega_{\nu]})\big]_3\,.
\end{align}
We repackage the new parameters $\sigma_{\mu\nu}$ and $\omega_{\mu\nu} = \omega_{\mu\nu,0} +\omega_{\mu\nu,2}+\omega_{\mu\nu,4}$ into
\be
\Phi_{\mu\nu}\,=\, \sigma_{\mu\nu} + \omega_{\mu\nu}\,.
\ee
This object lives in a sub-bundle of a bundle transforming in the {\bf 912} representation of $E_{7(7)}$, and collects the gauge parameters of the potentials that are three-forms in the external spacetime. One can then show that,  with the identifications \eqref{expr_2form_gauge_par},  the gauge transformations  for $B_{\mu, \nu}$,
$\tilde B_{\mu \nu}$, \eqref{transf_2form_BtildeB},  and  $C_{\mu\nu}$ (these follow from \eqref{Cbar_transf} and the last in \eqref{redef_two_forms})
 can be expressed as
\begin{align}\label{delta_calB_munu}
\delta \mathcal{B}_{\mu\nu} \,&=\, -2\partial_{[\mu} \overline{\Xi}_{\nu]} - 2 L_{\mathcal A_{[\mu}} \overline{\Xi}_{\nu]} - \dd_{\rm m} \overline{\Xi}_{[\mu} \xN \mathcal{A}_{\nu]}  - \partial_{[\mu} \Lambda \xN \mathcal{A}_{\nu]} +  \dd_{\rm m} \mathcal{B}_{\mu\nu}\xN \Lambda \nn\\[1mm]
\,&\; \quad - Y_{\mu\nu} - \dd_{\rm m}\Phi_{\mu\nu}\,,
\end{align}
where the action of  $\dd_{\rm m}$ on an element of $N'$  is given in \eqref{m_twist_ext_der}, and we define  
\be\label{dmOnPhimunu}
\dd_{\rm m}\Phi_{\mu\nu} = \din (\sigma_{\mu\nu} + \omega_{\mu\nu}) + m \,\omega_{\mu\nu,4}\,.
\ee 
The tensor $Y_{\mu\nu}$ is given in terms of 
$W_\nu \equiv \mathcal{A}_\nu \xN \Lambda$
 by
\begin{align}
Y_{\mu\nu} \,&=\,  \dd \big( \iota_{h_{[\mu}} W_{\nu]} + B_{[\mu}\wedge W_{\nu],{\rm odd}} - C_{[\mu}W_{\nu],0} - C_{[\mu,0} W_{\nu],3} + C_{[\mu,2}W_{\nu],1} \big)  \nn\\[1mm]
&\ \quad  + m\big(\iota_{h_{[\mu}} W_{\nu],5} + B_{[\mu}\wedge W_{\nu],3} -C_{\mu,4}W_{\nu,0}  \big)  \, .
\end{align}
After some manipulations, this can be re-expressed as
\begin{align}
Y_{\mu\nu}
\,&=\,  L_{\mathcal{A_{[\mu}}}\mathcal{A}_{\nu]} \xN \Lambda + 2 L_{\mathcal{A_{[\mu}}}\Lambda \xN \mathcal{A}_{\nu]} + L_\Lambda \mathcal{A}_{[\mu} \xN \mathcal{A}_{\nu]}\,,
\end{align}
which in turn allows to rewrite \eqref{delta_calB_munu} as
\begin{align}
\delta \mathcal{B}_{\mu\nu} &= -2\partial_{[\mu} \overline{\Xi}_{\nu]} - 2 L_{\mathcal A_{[\mu}} \overline{\Xi}_{\nu]} - \dd_{\rm m} \overline{\Xi}_{[\mu} \xN \mathcal{A}_{\nu]}  - \partial_{[\mu} \Lambda \xN \mathcal{A}_{\nu]}  +  \dd_{\rm m} \mathcal{B}_{\mu\nu}\xN \Lambda \nn\\[1mm]
&\quad  - L_{\mathcal{A_{[\mu}}}\mathcal{A}_{\nu]} \xN \Lambda - 2 L_{\mathcal{A_{[\mu}}}\Lambda \xN \mathcal{A}_{\nu]} -  L_\Lambda \mathcal{A}_{[\mu} \xN \mathcal{A}_{\nu]}  - \dd_{\rm m}\Phi_{\mu\nu}  \,.
\end{align}
Introducing the gauge field strength
\be\label{eq:defHmunu}
\mathcal{H}_{\mu\nu} \,=\, 2 \partial_{[\mu}\mathcal{A}_{\nu]} + L_{\mathcal{A}_{[\mu}}\mathcal A_{\nu]} + \dd_{\rm m} \mathcal{B}_{\mu\nu}\,,
\ee
and recalling the expression for $\delta{\mathcal A}_\mu$ given in \eqref{Avaria} and the redefinition of the gauge parameter $\overline \Xi_\mu$ in~\eqref{redefXi}, 
the variation of $\mathcal{B}_{\mu\nu}$ eventually takes the compact form
\be\label{var_Bmunu}
\delta \mathcal{B}_{\mu\nu} \,=\, -2\partial_{[\mu} \Xi_{\nu]} - 2 L_{\mathcal A_{[\mu}} \Xi_{\nu]} + \Lambda\xN \mathcal{H}_{\mu\nu}  + \mathcal{A}_{[\mu} \xN \delta\mathcal{A}_{\nu]} - \dd_{\rm m}\Phi_{\mu\nu}\,.
\ee

We can now plug in our truncation ansatz and show that it reproduces the correct lower-dimensional gauge-transformations. For the gauge parameters we take an ansatz similar to the one for the physical fields, that is
\begin{align}
\Lambda^M(x,z) \,&=\, -\Lambda^A(x) \hat{E}_A{}^M(z)\ ,\nn\\[1mm]
\widetilde{\Xi}_\mu{}^{MN}(x,z) \,&=\, -\tfrac{1}{2}\,\widetilde{\Xi}_\mu{}^{AB}(x)\, (\hat{E}_A{} \otimes_N \!\hat{E}_B)^{MN}(z)\,.
\end{align}

Plugging the ansatz into the variation \eqref{gauge_var_G} of the generalised metric, and using the action~\eqref{LeibnizParall} of the generalised Lie derivative on the parallelisation, we obtain
\be
\delta_{\Lambda} \mathcal{M}^{AB} \,=\, -\Lambda^C (X_{CD}{}^A \mathcal{M}^{DB} + X_{CD}{}^B\mathcal{M}^{AD})\,,
\ee
which is the correct variation of the scalar fields in gauged maximal supergravity, see e.g.~\cite{Samtleben:2008pe}.

In order to write the variation of $\mathcal A_\mu$, let us first observe that the ansatz together with the property~\eqref{eq:symmetricL} implies
\be
\dd_{\rm m} \Xi_\mu \,\eqs\, -\tfrac{1}{2}(L_{\hat{E}_B}\hat{E}_C + L_{\hat{E}_C}\hat{E}_B)\, \Xi_\mu{}^{BC} \,=\, - Z^A{}_{BC}\, \Xi_\mu{}^{BC}\hat{E}_A\,,
\ee
where we introduced the symmetrised structure constants $Z^A{}_{BC}=X_{(BC)}{}^A$.
Then, interpreting the variation of $\mathcal A_\mu$ in eq.~\eqref{Avaria} as $(\delta \mathcal{A}_\mu{}^A) \hat{E}_A $ and plugging the ansatz in, we obtain
\begin{align}
\delta \mathcal{A}_\mu{}^A \,&=\, \partial_\mu \Lambda^A + \mathcal{A}_\mu^B X_{BC}{}^A \Lambda^C - Z^A{}_{BC}\, \Xi_\mu{}^{BC}\,.
\end{align}
This is the correct gauge variation of the gauge fields in maximal supergravity (see again~\cite{Samtleben:2008pe}).

Finally, we need to consider the transformation of $\calB_{\mu\nu}$.
Eq.~\eqref{eq:defHmunu} yields
\be
\mathcal{H}_{\mu\nu} \,=\, \mathcal{H}_{\mu\nu}^A \hat{E}_A \,,
\ee
with
\be\label{eq:defHmunu_comp}
\mathcal{H}_{\mu\nu}^{A} \,=\, 2 \partial_{[\mu} \mathcal{A}_{\nu]}{}^A + X_{BC}{}^A \mathcal{A}_{[\mu}{}^B\mathcal{A}_{\nu]}{}^C + Z^{A}{}_{BC} \, \mathcal{B}_{\mu\nu}{}^{BC}\,.
\ee
This is the expression for the covariant field strengths used in gauged supergravity.
We also obtain
\begin{align}
L_{\mathcal{A}_\mu} \Xi_\nu \,&=\, -\tfrac12\mathcal{A}_\mu{}^C\, \Xi_\nu{}^{AB} L_{\hat{E}_C}( \hat{E}_A\otimes_{N'}\!\hat{E}_B)\nn\\[1mm]
\,&=\, - \mathcal{A}_\mu{}^C \,\Xi_\nu{}^{(DA)} X_{CD}{}^B  \hat{E}_A\otimes_{N'}\!\hat{E}_B\,,
\end{align}
where to pass from the first to the second line we distributed the Lie derivative on the factors of the $\otimes_{N'}$ product and used the Leibniz property of the generalised frame. Therefore:
\be
-2\partial_{[\mu} \Xi_{\nu]} - 2 L_{\mathcal A_{[\mu}} \Xi_{\nu]} \,=\,  D_{[\mu}  \Xi_{\nu]}{}^{AB} \hat{E}_A \xN \hat{E}_B\,,
\ee
where
\be
D_{[\mu}  \Xi_{\nu]}{}^{AB} \,=\, \partial_{[\mu} \Xi_{\nu]}{}^{AB}  + 2\mathcal{A}_{[\mu}{}^C \,\Xi_{\nu]}{}^{(DA)} X_{CD}{}^B  \,.
\ee

Putting everything together, \eqref{var_Bmunu} eventually takes the appropriate form to describe the two-form gauge transformations in gauged supergravity:
\be
\delta \mathcal{B}_{\mu\nu}{}^{AB} \,=\, 2 \,D_{[\mu}  \Xi_{\nu]}{}^{AB} - 2\, \Lambda^{(A} \mathcal{H}_{\mu\nu}{}^{B)} + 2\,\mathcal{A}_{[\mu}{}^{(A} \delta\mathcal{A}_{\nu]}{}^{B)} + \ldots\,,
\ee
where $\mathcal{H}_{\mu\nu}^{A}$ was given in \eqref{eq:defHmunu_comp}. The ellipsis denote a term coming from expressing $\dd_{\rm m}\Phi_{\mu\nu}$ in~\eqref{var_Bmunu} by means of the parallelisation  that we will not discuss in detail. This eventually gives the two-form gauge parameters in the lower-dimensional supergravity theory, contracted with the gauge group generators $X$. In four-dimensional supergravity, this term drops from all relevant equations, because the two-forms $\mathcal{B}_{\mu\nu}{}^{AB}$ always appear contracted with the embedding tensor, namely as $Z^A{}_{BC}\mathcal{B}_{\mu\nu}{}^{BC}$ \cite{deWit:2007kvg}, which implies that the term in the ellipsis is projected out due to the quadratic constraint. From a generalised geometry perspective, the corresponding statement is that in a reduction to four dimensions \eqref{var_Bmunu} always appears under the action of the exterior derivative twisted by the Romans mass, $\dd_{\rm m}$; given the definitions \eqref{dmOnPhimunu} and \eqref{m_twist_ext_der}, it is immediate to check that $\dd_{\rm m}(\dd_{\rm m}\Phi_{\mu\nu})=0$, hence the gauge parameters with two external indices drop from all relevant equations. This is no longer the case in reductions to supergravities in dimension six or higher, where the tensor hierarchy stops at one form degree higher, so that the three-form gauge potentials, as well as their two-form gauge parameters, also play a role. 

In conclusion, we have shown that under the generalised Scherk--Schwarz ansatz, the (massive) type IIA gauge transformations consistently reduce to the correct gauge transformation in lower-dimensional supergravity.

\section{Examples}\label{sec:examples}

In this section, we apply the generalised Scherk--Schwarz procedure to study consistent reductions of massless and massive type IIA supergravity on the spheres $S^6$, $S^4$, $S^3$ and $S^2$, as well as on six-dimensional hyperboloids. 
While for the massless case it is always possible to find generalised parallelisations that reproduce the known reductions to maximal gauged supergravities in lower-dimensions, for the massive theory we could only find a suitable generalised parallelisation on $S^6$ and the six-dimensional hyperboloids. We propose a general argument of why this is the case.

\subsection{$S^6$ parallelisation and $D=4$, $\ISO(7)_m$ supergravity}

We start our series of examples by revisiting the consistent reduction of type IIA supergravity on the six-sphere $S^6$  down to $D=4$ maximal supergravity with $\ISO(7)$ gauge group that was recently studied in detail in~\cite{Guarino:2015jca,Guarino:2015qaa,Guarino:2015vca}. For vanishing Romans mass, this reduction can be  understood as a limit of the consistent truncation of eleven-dimensional supergravity on $S^7$ (or on a seven-dimensional hyperboloid), where the seven-dimensional manifold degenerates into the cylinder $S^6 \times \mathbb {R}$~\cite{Hull:1988jw,Boonstra:1998mp}. In that case the group $\ISO(7)$ is gauged purely electrically~\cite{Hull:1984yy}. This means that only the 28 electric vector fields participate in the gauging, while the 28 magnetic duals do not appear in the Lagrangian. When the Romans mass $m$ is switched on, the truncation ansatz remains consistent with no modifications required.  However one finds that the magnetic vectors now also enter in the gauge covariant derivatives~\cite{Guarino:2015vca}, thus providing a {\it dyonic} gauging. The resulting four-dimensional supergravity is not equivalent to the theory with purely electric $\ISO(7)$ gauging~\cite{Dall'Agata:2014ita}; for this reason, we will denote it as the $\ISO(7)_m$ theory. This is an example of {\it symplectic deformation} of maximal supergravity of the type first discovered for the $D=4$, $\SO(8)$ theory in~\cite{Dall'Agata:2012bb}. The  $\ISO(7)_m$ theory admits several supersymmetric and non-supersymmetric AdS$_4$ solutions \cite{DallAgata:2011aa,Gallerati:2014xra,Guarino:2015qaa}, which all disappear when the parameter $m$ is sent to zero.\footnote{Specific formulae uplifting these AdS$_4$ vacua to massive type IIA supergravity were given in~\cite{Guarino:2015jca,Guarino:2015vca,Varela:2015uca}. Three of them are $G_2$-invariant and also included in the truncation of massive IIA supergravity on $S^6\simeq G_2/\SU(3)$ of~\cite{Cassani:2009ck}.}
 The structure of the $\ISO(7)_m$ theory was analysed in detail in~\cite{Guarino:2015qaa}.

In the following, we introduce a parallelisation of the $E_{7(7)}\times \mathbb{R^+}$ tangent bundle on $S^6$. Then, evaluating our massive generalised Lie derivative on the frame we obtain precisely the embedding tensor characterising the dyonic $\ISO(7)_m$ gauging. We also re-derive the truncation ansatz for the four-dimensional bosonic fields from generalised geometry. 

A generalised parallelisation on $S^6$ is defined as follows. Let $y^i, i = 1, \dots , 7$, with $\delta_{ij} y^i y^j = 1$, be the constrained coordinates on $S^6$, describing its embedding in $\mathbb{R}^7$ (see appendix~\ref{app:constrcoo} for some useful details about spheres in constrained coordinates). Let $v_{ij}$ be the $\SO(7)$ Killing vectors and define the following forms
\begin{align}\label{deflof}
	\omega^{ij} \,&=\,  R^2\,\dd y^i \wedge \dd y^j  \hs{85pt} \in \Lambda^2 T^*\, ,\nn \\[1mm]
	\rho_{ij} \, &=\ \rg{*}(R^2\,\dd y_i \wedge \dd y_j) 
		\hs{68pt} \in \Lambda^4 T^*\, , \nn \\[1mm]
	\kappa_i \, &=\, -\rg{*}(R\,\dd y_i) 
	\hs{95pt} \in \Lambda^5 T^*\, , \nn \\[1mm]
	\tau^{ij} \, &=\, R\,(y^i \dd y^j - y^j \dd y^i)\otimes \rg{\rm vol}_6  \hs{25pt} \in  T^*\otimes \Lambda^6T^*\, .
\end{align}
Here and in the rest of this section, the symbol $\rg{}$  means that the corresponding quantity is computed using the reference round metric of radius $R$. The index on the coordinates $y^i$ is lowered with the $\mathbb{R}^7$ metric $\delta_{ij}$.
We also twist the generalised tangent bundle with a five-form RR potential $\rg{C_5}$ such that 
\be\label{F6onS6}
\rg{F_6} \,\ =\, \dd\! \rg{C_5} \,\ =\, \frac{5}{R} \rg{\rm vol}_6\ ,
\ee 
with all other $p$-form potentials vanishing; the reason for this choice will become clear soon.

The generalised frame can be split according to the decomposition 
\begin{align}\label{decomp_E7}
E_{7(7)} \;&\supset\; \SL(8,\mathbb R) \;\supset\; \SL(7,\mathbb R) \nn \\[1mm]
{\bf 56}\, &\to\, {\bf 28} + {\bf 28'} \to\, {\bf 21} + {\bf 7} + {\bf 21'}+ {\bf 7'} 
\end{align}
as
\be
\{\hat{E}_A\} \,=\, \{\hat{E}_{IJ}, \hat{E}^{IJ}\} \,=\, \{\hat{E}_{ij},\hat{E}_{i8},\hat{E}^{ij},\hat{E}^{i8}\}\ .
\ee
We will call ``electric'' the $\hat{E}_{IJ}$ frame elements, transforming in the ${\bf 28}$ of $\SL(8)$, and ``magnetic'' the $\hat{E}^{IJ}$, transforming in the ${\bf 28'}$.

A generalised parallelisation is given by
\be\label{s6frame}
\hat{E}_{A} \,=\, \begin{cases}
\ \ \hat{E}_{ij} \!\!\!&=\, v_{ij} + \rho_{ij} + \iota_{v_{ij}}\!\rg{C_5}\ ,  \\[1mm]
\ \ \hat{E}_{i8} \!\!\!&=\, y_i + \kappa_i - y_i \rg{C_5}\ ,  \\[1mm]
\ \ \hat{E}^{ij} \!\!\!&=\, -\omega^{ij} - \tau^{ij} + j \!\rg{C_5} \wedge \omega^{ij}\ ,  \\[1mm]
\ \ \hat{E}^{i8} \!\!\!&=\, R\,\dd y^i - y^i \rg{\rm vol}_6 + R\,\dd y^i \wedge \rg{C_5} \ .
\end{cases}
\ee
It is not hard to see that this is globally defined. For instance, $\hat{E}_{ij}$ is nowhere vanishing as the Killing vectors $v_{ij}$ vanish at $y_i = y_j =0$, while the four-forms $\rho_{ij}$ vanish at $y_i^2+y_j^2=1$. Moreover, $\hat{E}_{i8}$ never vanishes as the locus $\kappa_i = 0$ does not overlap with $y_i =0$; similar considerations hold for the magnetic part of the frame.
The frame is also orthonormal with respect to the generalised metric~\eqref{GofVandV'}. Indeed, invoking the contraction formulae in~\eqref{contractions_sphere}, we have
\begin{align}
G(\hat{E}_{ij} , \hat{E}_{kl}) \,&=\, v_{ij}\,\lrcorner\, v_{kl} + \rho_{ij} \,\lrcorner\, \rho_{kl} \,=\, \delta_{ik}\delta_{jl} - \delta_{il}\delta_{jk}\ , \nn \\[1mm]
G(\hat{E}_{i8} , \hat{E}_{k8}) \,&=\, y_i \, y_k + \kappa_i\,\lrcorner\, \kappa_k \,=\, \delta_{ik}\ , \nn \\[1mm]
G(\hat{E}^{ij} , \hat{E}^{kl}) \,&=\, \omega^{ij}\,\lrcorner\, \omega^{kl} + \tau^{ij} \,\lrcorner\, \tau^{kl} \,=\, \delta^{ik}\delta^{jl} - \delta^{il}\delta^{jk} \ ,\nn \\[1mm]
G(\hat{E}^{i8} , \hat{E}^{k8}) \,&=\, R^2 \dd y^i \,\lrcorner\,\dd y^k + y^iy^k \rg{\rm vol}_6\,\lrcorner\, \rg{\rm vol}_6 \,=\, \delta^{ik} \ ,
\end{align}
with all other pairings vanishing.

We now evaluate the massive Dorfman derivative~\eqref{dorf6m} between two arbitrary frame elements, making use of various properties of the round spheres given in appendix~\ref{app:constrcoo}. In particular, we need identity~\eqref{eq:contrvol}, which together with our choice~\eqref{F6onS6} for $\rg{C_5}$ implies
\be
\iota_{v_{ij}}\!\rg{F_6}\,\ =\, \dd \rho_{ij} \ .
\ee
We find that the electric-electric pairings give
\begin{align}
L_{\hat{E}_{ij}} \hat{E}_{kl} \,&=\, \tfrac{2}{R}\big( \delta_{i[k}\hat{E}_{l]j} - \delta_{j[k}\hat{E}_{l]i} \big) \ ,\nn \\[1mm]
L_{\hat{E}_{ij}}\hat{E}_{k8} \,&=\, -\tfrac{2}{R} \delta_{k[i}\hat{E}_{j]8} \ ,\nn\\[1mm]
L_{\hat{E}_{i8}} \hat{E}_{kl} \,&=\, \tfrac{2}{R} \delta_{i[k}\hat{E}_{l]8} \ ,\nn\\[1mm]
L_{\hat{E}_{i8}}\hat{E}_{k8} \,&=\, 0\ ,
\end{align}
while for the electric-magnetic ones we have
\begin{align}
L_{\hat{E}_{ij}}\hat{E}^{kl} \,&=\, \tfrac{4}{R} \delta_{[i}^{[k} \delta_{j]j'} \hat{E}^{l]j'}\ , \nn \\[1mm]
L_{\hat{E}_{ij}}\hat{E}^{k8} \,&=\,  - \tfrac{2}{R} \delta^k_{[i} \delta_{j]j'} \hat{E}^{j'8} \ ,\nn \\[1mm]
L_{\hat{E}_{i8}}\hat{E}^{kl} \,&=\,  0 \ ,\nn \\[1mm]
L_{\hat{E}_{i8}}\hat{E}^{k8} \,&=\,  -\tfrac{1}{R}\,\delta_{ij} \hat{E}^{jk}\ ,
\end{align}
for the magnetic-electric 
\begin{align}
L_{\hat{E}^{ij}}\hat{E}_{kl} \,&=\,  L_{\hat{E}^{ij}}\hat{E}_{k8} \,=\,  L_{\hat{E}^{i8}}\hat{E}_{k8}\, = \,  0\ , \nn \\[1mm]
L_{\hat{E}^{i8}}\hat{E}_{kl} \,&=\, -2\, m \,\delta^i_{[k} \hat{E}_{l]8}\ ,
\end{align}
and for the magnetic-magnetic
\begin{align}
L_{\hat{E}^{ij}}\hat{E}^{kl} \,&=\, L_{\hat{E}^{ij}}\hat{E}^{k8} \, =\, L_{\hat{E}^{i8}}\hat{E}^{kl} \, =\, 0 \ ,\nn \\[1mm]
L_{\hat{E}^{i8}}\hat{E}^{k8} \,&=\, m\, \hat{E}^{ik}\ .
\end{align}
We thus obtain that condition~\eqref{LeibnizParall} is satisfied, namely the frame defines a Leibniz algebra under the massive Dorfman derivative. The non-vanishing  constants $X_{AB}{}^C$ read in $\SL(8)$ indices
\begin{align}\label{ISO7m_emb_tens}
X_{[II'][JJ']}{}^{[KK']} \,&=\, - X_{[II']}{}^{[KK']}{}_{[JJ']} \,=\, \,8 \, \delta_{[I}^{[K} \, \theta_{I'][J} \,\delta_{J']}^{K']}\ ,\nn \\[2mm]
X^{[II']}{}_{[JJ']}{}^{[KK']} \,&=\, - X^{[II'][KK']}{}_{[JJ']} \,=\, \,8\, \delta^{[I}_{[J} \, \xi^{I'][K} \,\delta^{K']}_{J']}\ ,
\end{align}
with
\be
\theta_{IJ} \,=\, \frac{1}{2R} \,{\rm diag}\big(\underbrace{1,\ldots,1}_{7},0\big)\ ,\qquad \xi^{IJ} \,=\, \frac{m}{2} \,{\rm diag}\big(\underbrace{0,\ldots,0}_{7},1\big)\ .
\ee
%
These match precisely the embedding tensor given in~\cite{Guarino:2015qaa} (modulo renormalising the generators by a $-1/2$ factor, see appendix C therein). The latter determines a dyonic $ISO(7)_m$ gauging of maximal $D=4$ supergravity, where the $\SO(7)$ rotations are gauged electrically while the seven translations are gauged dyonically. When $m=0$, we have $\xi^{IJ} =0$ and the $\ISO(7)$ gauging becomes purely electric.

Following the procedure for a generalised Scherk--Schwarz reduction described in the previous section, we can use our generalised parallelisation to deduce the truncation ansatz for the bosonic supergravity fields. 
 We start from the scalar ansatz. In four-dimensional maximal supergravity, the scalar matrix $\mathcal{M}^{AB}$ parameterises the coset $E_{7(7)}/\SU(8)$. Under the decomposition \eqref{decomp_E7}, this splits as 
\begin{align}
\mathcal{M}^{AB} &= \{ \mathcal{M}^{II',JJ'}, \, \mathcal{M}^{II'}{}_{JJ'} ,\,\mathcal{M}_{II'}{}^{JJ'} ,\,\mathcal{M}_{II',JJ'}\}  \nn \\
& = \{ \mathcal{M}^{ii',jj'}, \, \mathcal{M}^{ii',j8},\, \ldots ,\,\mathcal{M}_{i8,j8}\}\ .
\end{align}
Equating the components~\eqref{invG_comp_1} of the inverse generalised metric to those constructed from the parallelisation as in~\eqref{invG_from_parall}, we obtain
\begin{align}\label{scalar_ansatz_S6}
\rme^{2\Delta} g^{mn} \,&=\, \tfrac{1}{4}\mathcal{M}^{ii',jj'} v^m_{ii'}\, v^n_{jj'}\ , \nn \\[1mm]
\rme^{2\Delta} g^{mn}C_n \,&=\, \tfrac{1}{2}\mathcal{M}^{ii',j8}\, v^m_{ii'}\,y_{j}\ , \nn \\[1mm]
-\rme^{2\Delta} g^{mp}B_{pn} \,&=\, \tfrac{1}{2} \mathcal{M}^{ii'}{}_{j8}\, v^m_{ii'}\,R\, \partial_n y^{j}\ , \nn \\[1mm]
 \rme^{2\Delta}g^{mq}\left(C_{qnp} - C_{q}B_{np} \right)  \,&=\, - \tfrac{1}{4}\mathcal{M}^{ii'}{}_{jj'}\, v^{m}_{ii'}\, \omega^{jj'}_{np}\ , \nn\\[1mm]
\rme^{2\Delta}\left( \rme^{-2\phi} + g^{mn}C_m C_n\right)  \,&=\, \mathcal{M}^{i8,j8}\, y_i\, y_j   \ ,\nn \\[1mm]
{\rme}^{2\Delta}g^{ms}\big(C_{snpqr} - \rg{C}_{snpqr} -C_{s[np}B_{qr]} + \tfrac{1}{2}C_s B_{[np}B_{qr]}\big)  \,&=\,\tfrac{1}{4} \mathcal{M}^{ii',jj'} v^{m}_{ii'}\, (\rho_{jj'})_{npqr} \ ,
\end{align}
where we recall that the indices $i,i',j,j' =1\ldots, 7$ label the constrained coordinates while $m,n,\ldots=1,\ldots,6$ are curved indices on $S^6$. The scalar ansatz obtained in this way agrees with the formulae given in~\cite{Guarino:2015vca} (cf.\ eqs.~(3.14)--(3.18) therein). 
The additional relations appearing in eqs.~(3.19)--(3.22) of~\cite{Guarino:2015vca} can also be retrieved in the same way. The last equation in~\eqref{scalar_ansatz_S6} does not appear in~\cite{Guarino:2015vca}, and determines how the four-dimensional scalars enter in $C_{m_1\ldots m_5}$. Dualising its field strength $F_{m_1\ldots m_6}$ it should be possible to derive the expression of the Freund--Rubin term.

One can disentangle the different supergravity fields in~\eqref{scalar_ansatz_S6} by  following the procedure in eqs.~\eqref{fields_from_G_first}--\eqref{fields_from_G_last}. We recall that the generalised density $\Phi$ appearing in~\eqref{fields_from_G_last} can be computed at the origin of the scalar manifold, where $\mathcal{M}^{AB}=\delta^{AB}$, and is given by eq.~\eqref{gen_density_background}. Evaluating the first, second and second-last line of~\eqref{scalar_ansatz_S6} with $\mathcal{M}^{ii',jj'}=\delta^{i[j}\delta^{j']i'}$, $\mathcal{M}^{ii',j8}=0$, $\mathcal{M}^{i8,j8}=\delta^{ij}$, we find that $\rg{\Delta}\,=\,\rg{\phi}\,=\,0$.  Hence for the present truncation the generalised density is simply $\Phi \,=\ \rg{g}{}^{\!1/2}\,.$

We can also provide the ansatz for the vector fields as explained in section~\ref{sec:genSS}. Separating the components of eq.~\eqref{trunc_ansatz_vec},
we obtain
\begin{align}
h_\mu \,&=\, \tfrac{1}{2}\mathcal{A}_\mu{}^{ii'} v_{ii'} \ ,\nn \\[1mm]
B_{\mu}  \,&=\, \mathcal{A}_{\mu\,i8} \,R \,\dd y^i\ ,\nn \\[1mm]
C_{\mu,0}  \,&=\, \mathcal{A}_\mu{}^{i8} \,y_i \ ,\nn \\[1mm]
C_{\mu,2}  \,&=\, -\tfrac{1}{2}\mathcal{A}_{\mu\,ii'}\,R^2 \dd y^i \wedge \dd y^{i'} \, ,
\end{align}
which again agrees with~\cite{Guarino:2015vca}. Here,  $\mathcal{A}^{IJ} = \{\mathcal{A}^{ij}, \mathcal{A}^{i8} \}$ are the electric one-form fields in the four-dimensional theory while $\mathcal{A}_{IJ} = \{\mathcal{A}_{ij}, \mathcal{A}_{i8} \}$ are their magnetic duals.
We can also provide an ansatz for the type IIA dual fields with one external leg
\begin{align}
C_{\mu ,4}  \,&=\, \tfrac{1}{2}\mathcal{A}_\mu{}^{ii'} \big(\rho_{ii'} + \iota_{v_{ii'}}\! \rg{C_5} \!\big)\ ,\nn \\[1mm]
C_{\mu ,6}  \,&=\, \mathcal{A}_{\mu\,i8} \big(-y^i \rg{\rm vol}_6 + R\, \dd y^i\wedge \rg{C_5}\big) \ ,\nn \\[1mm]
\tilde{B}_{\mu}  \,&=\, \mathcal{A}_\mu{}^{i8} \, \big(\kappa_i -y_i \!\rg{C_5}\big)  \, .
\end{align}

Finally, the ansatz for the fields with two external legs follows from the general formula~\eqref{ansatz_two-forms} 
\begin{align}
B_{\mu\nu}\,&=\,  \mathcal{B}_{\mu\nu}{}^{ij}{}_{j8}\, y_i \,,\nn\\[1mm]
\tilde{B}_{\mu\nu} \,&=\, \tfrac{1}{8}\big(\tfrac 12 \mathcal{B}_{\mu\nu}{}^{i_1i_2,i_38}  y^j y_{[i_1}\epsilon_{i_2i_3]j k_1\ldots k_4}  - \mathcal{B}_{\mu\nu\,k_1k_2,k_3k_4}\big) R^4 \dd y^{k_1} \wedge \dd y^{k_2} \wedge\dd y^{k_3} \wedge\dd y^{k_4} \,,\nn\\[1mm]
C_{\mu\nu ,1}\,&=\, \big(\mathcal{B}_{\mu\nu\,ij}{}^{kj}+ \mathcal{B}_{\mu\nu\,i8}{}^{k8} \big) y_k R\, \dd y^i  \,,\nn\\[1mm]
C_{\mu\nu ,3}\,&=\,  \big(\tfrac{1}{12} \mathcal{B}_{\mu\nu}{}^{ii',jj'} y_{[i}\epsilon_{i']jj' k_1\ldots k_4} y^{k_4} - \tfrac12 \mathcal{B}_{\mu\nu\,k_1k_2,k_38} \big)R^3 \dd y^{k_1}\wedge \dd y^{k_2}\wedge \dd y^{k_3} \,,\nn\\[1mm]
C_{\mu\nu,5}\,&=\,  \mathcal{B}_{\mu\nu}{}^{ij}{}_{j8}\big(-\kappa_i + y_i\! \rg{C_5}\big) \,.
\end{align}

\subsection{Hyperboloids and $D=4$, $\ISO(p,7-p)_m$ supergravity}\label{Hpq}

The generalised Leibniz parallelisation on $S^6$ presented above  can be adapted to construct a similar  one on the six-dimensional hyperboloids $H^{p,7-p}$. This leads to a consistent truncation of massive type IIA supergravity to four-dimensional $\ISO(p,7-p)_m$ maximal supergravity. The existence of the purely electric gaugings goes back to~\cite{Hull:1984vg,Hull:1984qz}, while their relation to type IIA on $H^{p,7-p}$ was first given in~\cite{Hull:1988jw}. 

The hyperboloid $H^{p,q}$ is the homogeneous space
\begin{equation}
H^{p,q} \,=\, \frac{\SO(p,q)}{\SO(p-1,q)}\ ,
\end{equation}
and can be seen as the hypersurface in the Euclidean space $\mathbb{R}^{p+q}$ defined by the equation
\begin{equation}
\label{embH}
\eta_{ij}\, y^i y^j \,=\, 1\, ,
\end{equation}
where $i,j=1,\ldots,p+q$ and
\be
\eta_{ij} \,=\, {\rm diag}\big( \underbrace{+1,\ldots, +1}_{p}, \underbrace{-1,\ldots, -1}_{q}\big)\ .
\ee
Clearly, taking $q=0$ yields the sphere $S^{p-1}$.

Let us focus on the six-dimensional hyperboloids $H^{p,7-p}$, with $1\leq p < 7$.
A parallelisation on these manifolds can be introduced following the same path as for $S^6$, replacing the Kronecker $\delta_{ij}$ by $\eta_{ij}$ where appropriate.
In particular, the Killing vectors $v_{ij}$, that for the six-sphere satisfy the $\mathfrak{so}(7)$ algebra~\eqref{algebra_Killing_v}, now respect the  $\mathfrak{so}(p,7-p)$ algebra,
\begin{equation}
\mathcal{L}_{v_{ij}} v_{kl} \,=\, 2R^{-1}\left(\eta_{i[k}v_{l]j} - \eta_{j[k}v_{l]i}\right)\ .
\end{equation}
The equations~\eqref{Lie_on_y}--\eqref{Lie_on_omega} also need to be modified by replacing $\delta_{ij}$ with $\eta_{ij}$ everywhere. 
We can keep the definitions \eqref{deflof}, noting however that they now transform in representations of $\SO(p,7-p)$ instead of $\SO(7)$.
Then \eqref{s6frame} defines a generalised parallelisation on $H^{p,7-p}$. The Dorfman derivative between two frame elements satisfies~\eqref{LeibnizParall}, with the non-vanishing embedding tensor components being still given by \eqref{ISO7m_emb_tens}, where however now
\be
\theta_{IJ} \,=\, \frac{1}{2R} \,{\rm diag}\big(\underbrace{1,\ldots,1}_{p },\underbrace{-1,\ldots,-1}_{7-p },0\big)\ ,
\ee
while $\xi^{IJ}$ remains unchanged.
%
 This corresponds to an $\ISO(p,7-p)\simeq \CSO(p,7-p,1)$ frame algebra, where the seven translational symmetries are gauged dyonically.

The truncation ansatz remains formally the same as for the reduction on $S^6$. We thus infer that there exists a consistent truncation of massive IIA supergravity on the six-dimensional hyperboloids $H^{p,7-p}$, down to $\ISO(p,7-p)_m$ gauged supergravity. As above, the subscript $m$ emphasises that the translational isometries are gauged dyonically. Setting $m=0$, one recovers a truncation of massless type IIA supergravity on $H^{p,7-p}$ down to the $\ISO(p,7-p)$ theory with purely electric gauging~\cite{Hull:1988jw} (see also \cite{Hohm:2014qga}).

It was found in~\cite{Dall'Agata:2012bb} that the only gaugings of four-dimensional maximal supergravity in the $\CSO(p,q,r)$ class (with $r >0$) admitting a symplectic deformation are $\CSO(p,7-p,1)\simeq \ISO(p,7-p)$.\footnote{For these gaugings, the symplectic deformation is of on/off type: all non-zero values of the parameter controlling the magnetic gauging are equivalent.} Here we have established that all these symplectic deformations arise as consistent truncations of massive type IIA supergravity: while for $p=7$ the internal manifold is $S^6$, for $1\leq p < 7 $ the internal manifold is the hyperboloid $H^{p,7-p}$.

The same ideas could be applied to products of hyperboloids and tori, $H^{p,q}\times T^r$, with $p+q+r=7$. In this case, the parallelisation would satisfy the $\CSO(p,q,r+1)$ algebra.

\subsection{$S^4$ parallelisation with $m=0$ and $D=6$, $\SO(5)$ supergravity}

The U-duality group for type IIA on a four-dimensional manifold $M_4$ is $E_{5(5)}\simeq \SO(5,5)$ and the generalised tangent bundle is
\begin{equation}
E \ \simeq\ T \oplus T^* \oplus \mathbb{R} \oplus \Lambda^2 T^*\oplus \Lambda^4 T^*\ .
\end{equation}
A section of $E$
\begin{equation}
V \,=\, v + \lambda + \omega_0 + \omega_2 + \omega_4 
\end{equation}
transforms in the spinorial ${\bf 16^+}$ representation of $\SO(5,5)$.

We are interested in the case where $M_4$ is the four sphere $S^4$ and we describe it using constrained coordinates $y^i$ in $\mathbb{R}^5$.
It is then convenient to consider the decomposition of  the generalised frame $\hat{E}_A$, $A = 1\ldots,16$ under $\SL(5, \mathbb{R})$ 
 \begin{align}
\SO(5,5) \;&\supset\; \SL(5,\mathbb R)  \nn \\[1mm]
{\bf 16^+}\, &\to\, {\bf 10} + {\bf 5} + {\bf 1} \ ,
\end{align}
so that $\{\hat{E}_{A}\} \,=\, \{\hat{E}_{ij}\} \cup \{\hat{E}_{i}\} \cup \{\hat{E}\}$, with $i,j=1,\ldots,5$.

For {\it massless} type IIA supergravity on $S^4$, we take the frame
\be\label{parall_S4_meq0}
\hat{E}_{A} \,=\, \begin{cases} \ \ \hat{E}_{ij} \!\!\!&=\, v_{ij} + \rho_{ij} + \iota_{v_{ij}}\!\rg{C_3} \ ,\\[1mm]
\ \  \hat{E}_{i} \!\!\!&=\, R\,\dd y_i + y_i \! \rg{\rm vol}_4 + R\, \dd y_i\, \wedge\! \rg{C_3}\ ,\\[1mm]
\quad \hat{E} \!\!\!&=\, 1 \ ,
\end{cases} 
\ee
where $v_{ij}$ are the $\SO(4)$ Killing vectors and
\be
\rho_{ij} \, =\ \rg{*} (R^2 \dd y_i \wedge \dd y_j) \, =\, \frac{R^2}{2}\, \epsilon_{ijk_1k_2k_3}\, y^{k_1}\mathrm{d}y^{k_2}\wedge \mathrm{d}y^{k_3} \ .\ee
Note that we have twisted the frame by a background RR potential $\rg{C_3}$, that is the supergravity potential whose field strength threads the whole $S^4$.\footnote{The twist by $C_3$ acts on a vector $\check V$ of the untwisted generalised tangent bundle $\check E$ on $M_4$ as (cf.~eq.~\eqref{eq:twistC}):
\begin{equation*}
V \,=\, \rme^{C_3}\cdot \check  V \,=\, \check v + \check \lambda + \check \omega_0 + (\check \omega_2 + \iota_{\check v}C_3) + (\check \omega_4 + \check \lambda \wedge C_3)\ .
\end{equation*}
}
This is chosen such that
\be
\rg{F_4} \ = \dd \!\rg{C_3} \ = \frac{3}{R} \rg{\rm vol}_4\ , 
\ee
which, recalling~\eqref{eq:contrvol}, implies
\begin{equation}
\iota_{v_{ij}} \!\rg{F_4} \ =\, \mathrm{d}\rho_{ij}\ .
\end{equation}
We will not twist by $C_1$ or $B$ instead, as there are no two- or three-cycles on $S^4$.
Following similar reasoning as for $S^6$, it is easy to see that the frame above is globally defined and orthonormal with respect to the generalised metric~\eqref{GofVandV'}, thus it specifies a generalised parallelisation.

In four dimensions (or lower), the massive generalised Lie derivative simplifies considerably and reads
\begin{align}\label{dorf4m}
L_V V' \,=&\,\ \mathcal{L}_v v' + \left(\mathcal{L}_v \lambda' - \iota_{v^\prime} \mathrm{d}\lambda\right) + \left(\iota_v \mathrm{d}\omega_0' - \iota_{v^\prime} (\dd\omega_0 -  m\lambda) \right) \nn\\[1mm]
\,&\, + \left(\mathcal{L}_v \omega_2^\prime - \iota_{v^\prime}\mathrm{d}\omega_2 - \lambda' \wedge (\dd\omega_0 -  m\lambda) +  \omega_0' \mathrm{d}\lambda \right) \nn\\[1mm]
\,&\, + \left( \mathcal{L}_v \omega_4^\prime - \iota_{v^\prime}\mathrm{d}\omega_4 - \lambda' \wedge \mathrm{d}\omega_2 +  \omega_2^\prime \wedge \mathrm{d}\lambda\right) \ .
\end{align}
Using the relations in appendix~\ref{app:constrcoo}, we compute the {\it massless} Dorfman derivative (that is expression~\eqref{dorf4m} with $m=0$) between the frame elements. We find that the only non-vanishing pairings are
\begin{align}\label{Dorfman_par_S4}
L_{\hat{E}_{ij}}\hat{E}_{kl} \,&=\, 2R^{-1}\big(\delta_{i[k}\hat{E}_{l]j} - \delta_{j[k}\hat{E}_{l]i}  \big)\ ,\nn\\[1mm]
L_{\hat{E}_{ij}}\hat{E}_{k} \,&=\, -2R^{-1}\delta_{k[i}\hat{E}_{j]} \ .
\end{align}
This defines a Leibniz algebra since $L_{\hat{E}_{ij}}\hat{E}_{k} \neq -L_{\hat{E}_{k}}\hat{E}_{ij}=0$; the associated gauge algebra, following from~\eqref{eq:gauge-alg-X}, is the $\SO(5)$ algebra.

A consistent truncation of massless type IIA supergravity on $S^4$ preserving maximal supersymmetry has been constructed in~\cite{Cowdall:1998rs,Cvetic:2000ah} by simply reducing on a circle the seven-dimensional theory defined by eleven-dimensional supergravity on $S^4$. The gauge group of the resulting $\mathcal N =(2,2)$ six-dimensional theory is indeed~$\SO(5)$ (see also~\cite{Bergshoeff:2007ef} for a discussion of the gauging in six dimensions). This theory does not admit AdS$_6$ vacua: the most symmetric solution is a half-BPS domain-wall, originating from a circle reduction of the AdS$_7\times S^4$ vacuum of eleven-dimensional supergravity, and describing the near-horizon geometry of D4-branes.

Following the example of $S^6$, one might expect  that the same frame \eqref{parall_S4_meq0} would lead to a generalised parallelisation for $m\neq0$
with a modified gauge group in six-dimensions. However, it is easy to check by direct computation that with the massive Dorfman derivative the frame \eqref{parall_S4_meq0} 
does not satisfy a Leibniz algebra.  We will  further comment on this  in section \ref{massive_algebras}.

\subsection{$S^3$ parallelisation with $m=0$ and $D=7$, $\ISO(4)$ supergravity}\label{S3and7dsugra}

The U-duality group of type IIA supergravity on a three-dimensional manifold $M_3$ is $E_{4(4)} \simeq \SL(5,\mathbb{R})$, and the corresponding generalised tangent bundle is
\begin{equation}
E \ \simeq\ T \oplus T^* \oplus \mathbb{R} \oplus \Lambda^2 T^*\, , 
\end{equation}
with sections
\begin{equation}
V \,=\, v + \lambda + \omega_0 + \omega_2 
\end{equation}
transforming in the ${\bf 10}$ of $\SL(5,\mathbb{R})$.
A generalised frame $\{\hat{E}_A\}$, $A = 1\ldots,10$, can equivalently be denoted as $\{\hat{E}_{IJ}= \hat{E}_{[IJ]}\}$, with $I,J =1,\ldots, 5$.
We consider again  $M_3 = S^3$ in constrained coordinates $y^i$ in $\mathbb{R}^4$,  and we decompose the frame under $\SL(4, \mathbb{R})$ as
 \begin{align}
\SL(5,\mathbb{R}) \;&\supset\; \SL(4,\mathbb R)  \nn \\[1mm]
{\bf 10}\, &\to\, {\bf 6} + {\bf 4}
\end{align}
so that $\{\hat{E}_{IJ}\} \,=\, \{\hat{E}_{ij},\,\hat{E}_{i5}\}$, with $i,j=1,\ldots,4$.

For vanishing Romans mass, $m=0$, we can easily construct a generalised parallelisation that realises the \emph{ISO}$(4)$ algebra.
 We choose the frame
\be\label{parall_S3_meq0}
\hat{E}_{IJ} \,=\, \begin{cases} \;\ \hat{E}_{ij} \,=\, v_{ij} + \rho_{ij} + \iota_{v_{ij}}\!\rg{B} \ ,\\[1mm]
\;\ \hat{E}_{i5} \,=\, y_i + \kappa_i - y_i \!\rg{B}\ , \end{cases} 
\ee
where $v_{ij}$ are the $\SO(4)$ Killing vectors and
\begin{align}
\rho_{ij} \,&=\ \rg{*}(R^2 \dd y_i\wedge \dd y_j) \,=\, R\, \epsilon_{ijkl}\, y^k\mathrm{d}y^l \ ,\nn \\
\kappa_{i} \,&=\  \rg{*}(R\, \dd y_i) \,=\, \frac{R^2}{2}\,\epsilon_{ijkl}\, y^j \mathrm{d}y^k \wedge \mathrm{d}y^l\ .
\end{align}
Here, we have twisted the frame by the $B$ field,\footnote{The twist by $B$ acts on a vector $\check V$ of the untwisted generalised tangent bundle $\check E$ on $M_3$ as
\begin{equation*}
V \,=\, \rme^{-B}\cdot \check V \,=\, \check v + (\check \lambda + \iota_{\check v} B) + \check \omega_0 + (\check \omega_2 -\check \omega_0 B)\ .
\end{equation*}
}
chosen in such a way that
\be\label{HproptoVol_3}
\rg{H} \ =  \dd \!\rg{B_{}} \ =\, \frac{2}{R} \rg{\rm vol}_3\ , 
\ee
which, again recalling~\eqref{eq:contrvol}, implies
\begin{equation}
\label{hcn}
\iota_{v_{ij}}\!\! \rg{H} \ =\, \mathrm{d}\rho_{ij}\ .
\end{equation}
This frame is globally defined and orthonormal; hence it defines a generalised parallelisation.
Recalling appendix~\ref{app:constrcoo} and relation~\eqref{hcn}, one can check that the Dorfman derivative with $m=0$ yields
\begin{align}
L_{\hat{E}_{ij}}\hat{E}_{kl} \,&=\, 2R^{-1}\big(\delta_{i[k}\hat{E}_{l]j} - \delta_{j[k}\hat{E}_{l]i}  \big)\ ,\nn\\[1mm]
L_{\hat{E}_{ij}}\hat{E}_{k5} \,&=\, -2R^{-1}\delta_{k[i}\hat{E}_{j]5} \ ,\nn \\[1mm]
L_{\hat{E}_{i5}}\hat{E}_{kl} \,&=\, 2R^{-1}\delta_{i[k}\hat{E}_{l]5}\ , \nn\\[1mm]
L_{\hat{E}_{i5}}\hat{E}_{k5} \,&=\, 0\ , \label{ISO4-alg}
\end{align}
and the relation~\eqref{LeibnizParall} is satisfied, with structure constants
\be\label{XofISO4}
X_{[II'][JJ']}{}^{[KK']} \,=\, 2\, \delta_{[I}^{[K}Y_{I'][J}\delta_{J']}^{K']}\ , \qquad  Y_{II'} = \frac{2}{R}\, {\rm diag}(1,1,1,1,0)\ .
\ee
Note that, as  the Dorfman derivative is antisymmetric on this frame, it realises a Lie algebra (rather than just a Leibniz algebra), which in this case is the $\ISO(4)\simeq\CSO(4,0,1)$ algebra. 

A consistent truncation of massless type IIA supergravity to maximal $D=7$ supergravity with gauge group $\ISO(4)$ has been known for some time. This can be obtained starting from the well-known reduction of eleven-dimensional supergravity on $S^4$, which yields maximal $D=7$, $\SO(5)$ supergravity~\cite{Nastase:1999kf}, and implementing the limiting procedure of \cite{Hull:1988jw}. In the limit, $S^4$ degenerates into $\mathbb R \times S^3$; correspondingly, the $\SO(5)$ gauge group of the seven-dimensional theory is contracted to $\ISO(4)$.\footnote{This is analogous to the way the $\ISO(7)$ reduction of massless IIA supergravity on $S^6$ is obtained from the $\SO(8)$ reduction of eleven-dimensional supergravity on $S^7$.}
The bosonic part of this $S^3$ reduction was worked out in detail in~\cite{Cvetic:2000ah} (where the $\SO(4)$ subgroup of the gauge group was emphasised). A discussion of the resulting maximal supergravity can be found in \cite{Samtleben:2005bp}. In seven dimensions, the embedding tensor determining the gauging transforms in the ${\bf 15} + {\bf 40'}$ representation of the global symmetry group $\SL(5)$ \cite{Samtleben:2005bp}. For the $\ISO(4)$ gauging, its non-vanishing components are solely in the {\bf 15}, and match those in~\eqref{XofISO4} obtained from the parallelisation. In addition to the metric, the fourteen $\SL(5)/\SO(5)$ scalars and the ten $\ISO(4)$ gauge vectors, the bosonic field content of the seven-dimensional theory is made of a massless two-form and four massive self-dual three-forms.
The scalar potential does not admit stationary points, and the most symmetric ground state solution is a domain wall, describing the near-horizon geometry of NS5-branes.

We would now like to see whether the frame~\eqref{parall_S3_meq0}  gives a generalised parallelisation also for $m\neq0$. In this case the problems
appear even before considering the action of the massive Dorfman derivative. Indeed the frame~\eqref{parall_S3_meq0} requires the existence of
a non trivial field strength $H$, while we know from ~\eqref{eq:H-exact} that  for $m\neq 0$ $H$ is exact. 
 
\subsection{Massive algebras on $S^3$ and $S^4$}\label{massive_algebras}

In the previous sections we saw that,  contrary to the case of $S^6$,  the massless frames for $S^3$ and 
$S^4$  do not lead to good parallelisations when the Romans mass is turned on. 
In this section, we provide some understanding of why the frame on $S^6$  is the only one that satisfies a good algebra also in the massive Dorfman derivative. We also explore the possibility of finding other parallelisations that  do satisfy an algebra of the desired type.
For $S^3$ we derive a no-go theorem showing that, under mild assumptions, one cannot find a frame which gives rise to a maximally supersymmetric consistent truncation with gauge group $\SO(4)$ (or larger).

Given a $d$-dimensional sphere $S^d$ with a non-zero flux for a $d$-form field-strength, one can build a $\GL(d+1)$ generalised tangent bundle,
 which is isomorphic to  $T\oplus \Lambda^{d-2}T^*$. Since this admits a global generalised frame, the sphere is generalised parallelisable \cite{spheres}. This generalised frame is  a $\GL(d+1)$ rotation of the coordinate frame. 
For spheres, the $\GL(d+1)$ generalised tangent bundle is always a sub-bundle of the full  $E_{d+1(d+1)}\times\bbR^+$ bundle  and, in fact, it is possible to decompose the whole generalised tangent bundle into representations of  the $\GL(d+1)$ subgroup. Moreover, all the parts of the parallelisations of the bundle $E$  are related to the corresponding coordinate frames by the same $\GL(d+1)$ transformation.

In the previous sections we constructed the frame $\hE_A$ and the respective Leibniz algebra for type IIA on $S^d$, $d=3,4,6$. We consider now 
the effect of adding the Romans mass  to the massless Dorfman derivative. As the given frame on $S^d$ already satisfies a Leibniz algebra for the massless Dorfman derivative with constant structure constants $X_{AB}{}^C$, the structure constants of the same frame with the massive Dorfman derivative will be $X_{AB}{}^C + Y_{AB}{}^C$, where
\begin{equation}
	Y_{AB}{}^C = \hE_A{}^M \hE_B{}^N E^C{}_P\, \underline{m}_{MN}{}^P \, , 
\end{equation}
are the frame components of the Romans mass map $\underline{m}_{MN}{}^P$ defined in section~\ref{sec:massive_genLie}.
 The frame $\hE_A$ will thus give a generalised Leibniz parallelisation in the massive Dorfman derivative if the additional coefficients $Y_{AB}{}^C$ are constant.
 
A natural way for this to happen would be if the components $Y_{AB}{}^C$ are equal to the components $\underline{m}_{MN}{}^P$, which are constant by definition.
This would mean that the frame $\hE_A{}^M$ must lie in the  stabiliser group of the Romans mass, namely in the subgroup of $E_{d+1(d+1)} \times\bbR^+$ that leaves $\underline{m}_{MN}{}^P$ invariant. The stabiliser can be determined by combining~\eqref{IIAadjvecCompact} and~\eqref{eq:commAdjIIA} with %
\begin{equation}
	(R\cdot \underline{m}) (V) = [ R, \underline{m}(V) ] - \underline{m} (R\cdot V) \, ,
\end{equation}
where $R$ is an element of the adjoint of $E_{d+1(d+1)}\times\bbR^+$, see \eqref{section_adjoint}. For instance, in six dimensions we find  that $R \cdot \underline{m} = 0$ for $R$ of the form\footnote{For lower-dimensional spheres it is enough to truncate to the relevant potentials.}
\begin{equation}\label{eq:stabiliser}
R \, = \, l + \varphi + r  + \beta  +  \tilde B    + \Gamma_5  + C \, ,
\end{equation} 
where  $l = -\varphi$ and $\Gamma_5$ is a five-vector, while $C=C_1+ C_3 + C_5$.  The stabiliser group is the semi-direct product of a Lie group $G$
with a nilpotent group $G'$.   The Lie algebra $\mathfrak{g}$ of $G$ is 
generated by  $r$, $\Gamma_5$, $C_5$ and $l = -\varphi$ in~\eqref{eq:stabiliser}. The Lie algebra of $G'$ is $\mathfrak{g}' = \mathfrak{g}'_1 \oplus \mathfrak{g}'_2$ where $\mathfrak{g}_1$ and $\mathfrak{g}_2$ are generated by
 $\beta$ and $C_3$, and $C_1$ and $\tilde{B}$, respectively.  The algebra $\mathfrak{g}'$ is graded so that the commutator of two  $\mathfrak{g}_1$  elements is in $\mathfrak{g}_2$ and all other commutators vanish. 
 The stabiliser groups of $\underline{m}$ for the dimensions of interest in this paper  are summarised in table~\ref{tab:stab}. In the table, $\rep{R}_1$ and $\rep{R}_2$ denote the representations of  $G$ in which $\mathfrak{g}'_1$ and $\mathfrak{g}'_2$  transform. 
%

\begin{table}[h]
\centering
\begin{tabular}{rccc}
   $d$ & $G$ & $\rep{R}_1$ & $\rep{R}_2$ \\ 
   \hline
   $6$ & $\GL(7)$ & $\rep{35}$ &  $\rep{7'}$ \\
   $5$ & $\SL(5)\times\SL(2)\times\bbR^+$ & $(\rep{10},\rep{2})_{+1}$ 
   	& $(\rep{5},\rep{1})_{+2}$ \\
   $\leq 4$ & $\GL(d) \times \bbR^+$ & $(\Lambda^2 T)_{+1} \oplus (\Lambda^3 T^*)_{+1}$ 
   	& $(T^*)_{+2}$
\end{tabular}
\caption{Constituents of the stabiliser group of $\underline{m}_{MN}{}^P$.}
\label{tab:stab}
\end{table}

It is noteworthy that only for $d=6$ the group $G$  coincides with  $GL(d+1)$. 
Since the frame $\hE_A{}^M$ is an element of $\GL(d+1)$, we see that for $S^6$ the frame does lie in the relevant stabiliser group.\footnote{Note that for  $d=6$  the full stabiliser group is isomorphic to the geometric subgroup of $E_{7(7)}\times\bbR^+$ for M-theory.} Hence the massless frame remains a good Leibniz parallelisation when the Romans mass is switched on. However, for $d \leq 5$ it does not, and this provides a partial explanation for why these frames do not give Leibniz parallelisations in massive IIA. By this reasoning, one is not surprised that $S^6$ is the only case which works in massive IIA without modifying the frame.

However, the above argument does not rule out the possibility that there are alternative Leibniz generalised parallelisations of the lower-dimensional spheres in 
the massive IIA. In what follows, we  explore this possibility focusing on the case of $S^3$, for simplicity. As noted before, in massive type IIA $H_3$ must be trivial in cohomology. As $S^3$ has only a non-trivial 3-cycle, this means that there can be no cohomologically non-trivial field strengths. We thus assume that the background field configuration has non-zero Romans mass and  all other fields are zero. This implies that the generalised tangent space has no twisting and is just given by the direct sum 
\begin{equation}
	E \,=\, \check{E} \,=\, T \oplus T^* \oplus \Lambda^0 T^* \oplus \Lambda^2 T^* \, .
\end{equation}
Suppose now that there exists a  generalised Leibniz parallelisation $\hE_A$  that gives an $SO(4)$ algebra
\begin{align}\label{massive_algebra_S3}
\phantom{i}L_{\hat{E}_{ij}}\hat{E}_{kl} \,&=\, 2R^{-1}\big(\delta_{i[k}\hat{E}_{l]j} -  \delta_{j[k}\hat{E}_{l]i} \big)\ ,\nn\\[1mm]
\phantom{i}L_{\hat{E}_{ij}}\hat{E}_{k5} \,&=\, -2R^{-1}\delta_{k[i}\hat{E}_{j]5} \ , \nn\\[1mm]
\phantom{i}L_{\hat{E}_{i5}}\hat{E}_{kl} \,&=0\ , \nn\\[1mm]
\phantom{i}L_{\hat{E}_{i5}}\hat{E}_{k5} \,&= 0\ ,
\end{align}
where $L$ is the  massive Dorfman derivative.  This  implies that the generalised metric $G^{-1} = \delta^{AB} \hE_A \otimes \hE_B$ is preserved by the Dorfman derivative so that the $\hE_A$ are generalised Killing vectors~\cite{Grana:2008yw, Lee:2015xga}. Thus the gauge transformations of the background fields generated by the $\hE_A$ all vanish. As we have no gauge fields, this leads to the conditions
\begin{equation}
\label{eq:gen-Killing}
	\mathcal{L}_{v_A} \!\rg{g}\ = 0\, , \hs{30pt} \dd \lambda_A = 0\,,
	\hs{30pt} \dd \omega_A - m \lambda_A = 0\, ,
\end{equation}
which imply that the Dorfman derivative reduces to the Lie derivative term only
\begin{equation}
	L_{\hE_A} \equiv \mathcal{L}_{v_A}  \, . 
\end{equation}
As the vector parts of the $\hE_{ij}$ satisfy the $\SO(4)$ algebra, these must be the $S^3$ Killing vectors (up to an overall constant automorphism), and we have that
\begin{equation}
	L_{\hE_{ij}} \,\equiv\, \mathcal{L}_{v_{ij}} 
\end{equation}
is the action of the $\SO(4)$ isometry group. The second of~\eqref{massive_algebra_S3} then says that the $\hE_{k 5}$ components of the frame transform in the vector representation. This implies that
\begin{equation}
	\hE_{i5} = a_1 k_i + a_2 y_i + a_3 \dd y_i + a_4 \rg{*}\! \dd y_i
\end{equation}
for some real coefficients $a_n$, where $y^i$, with $i=1, \ldots, 4$, are the constrained coordinates on $\bbR^4$, while $k_i$ are the standard conformal Killing vectors on the sphere (cf.~appendix~\ref{app:constrcoo}). As $L_{\hE_{i5}} \hE_{j5} = 0$ we have $a_1=0$ and~\eqref{eq:gen-Killing} gives us $a_2 = ma_3$ and $a_4=0$. 
 One can then see that
 \be\label{parall_S3_m}
\hat{E}_A \,\equiv\, \hat{E}_{IJ} \,=\, \begin{cases} \ \;\hat{E}_{ij} \,=\, v_{ij} + R^2 \,\dd y_i \wedge \dd y_j  \\[1mm]
\;\ \hat{E}_{i5} \,=\, R\,(m\,y_i + \dd y_i)\ , \end{cases}  
\ee
where $R$ is the radius of $S^3$, is the unique frame 
giving a parallelisation of the generalised tangent bundle on $S^3$ which satisfies the $SO(4)$ algebra  \eqref{massive_algebra_S3}.\footnote{In appendix  \ref{IIBonS3} we show that in type IIB it is possible to find
a parallelisation for the generalised tangent bundle on $S^3$ that satisfies the same Leibniz algebra~\eqref{massive_algebra_S3}.} 
If $mR=1$, the frame is also orthonormal in the generalised metric.
However, the frame~\eqref{parall_S3_m} fails to be in the $\SL(5,\bbR)\times\bbR^+$ generalised frame bundle. We recall from~\cite{Coimbra:2011ky} that the generalised frame bundle is defined to be those frames which are related to the coordinate frame by an $E_{d+1(d+1)}\times\bbR^+$ transformation. In the $\SL(5,\bbR)\times\bbR^+$ case, this means that there must also be a parallelisation $\hE_I$ of the bundle $W \simeq (\det T)^{-1/2} \otimes (T + \det T)$, discussed in~\cite{spheres}, such that
\begin{equation}
	\hE_{IJ} \,=\, \hE_I \wedge \hE_J \,.
\end{equation}
It is simple to show that our frame~\eqref{parall_S3_m} is not of this form, and is thus outside of the generalised frame bundle. This means that one cannot use it to describe a consistent truncation of supergravity. For example, the Scherk--Schwarz twist of this frame does not define a generalised metric which can be parameterised in terms of supergravity fields, and as such it does not provide a scalar ansatz for such a reduction via eq.~\eqref{invG_from_parall}.

Having ruled out the possibility of the algebra~\eqref{massive_algebra_S3}, one could still wonder if there are other frame algebras containing $\SO(4)$ which could fare better. The obvious alternatives would be the $ISO(4)$ algebra~\eqref{ISO4-alg} or $\SO(5)$. The latter is immediately excluded as it is impossible for the vector parts $v_A$ of the frame to generate an $\SO(5)$ isometry group in three dimensions. The following argument will show that the $ISO(4)$ algebra is also excluded.


For the $\hE_{ij}$ parts of the frame, we can use the same generalised Killing vector arguments as above to deduce that  $L_{\hE_{ij}} \equiv \mathcal{L}_{v_{ij}}$, so we can again decompose the frames into $\SO(4)$ representations. This decomposition implies that the one-form part of $\hE_i$ is closed, and, together with the generalised Killing vector condition, that the one-form part of $\hE_{ij}$ vanishes. From~\eqref{ISO4-alg} we have the constraint  $L_{\hE_{i5}} \hE_{j5} = 0$ which implies that $L_{\hE_{i5}} \equiv - (\dd \omega_{2,i}) \cdot$ is the adjoint action of $\dd \omega_{2,i} \in \Lambda^3 T^* \subset \adj$, where $\omega_{2,i}$ is the two-form part of $\hat E_{i5}$.
However, this contradicts another of the hypothesised algebra relations $L_{\hat{E}_{i5}}\hat{E}_{kl} = 2R^{-1}\delta_{i[k}\hat{E}_{l]5}$ as the image of $\dd \omega_{2,i} \in \adj$ is contained in $\Lambda^2 T^* \subset E$, while $\hE_{i5}$ must feature one-form parts in order for $\hE_{IJ}$ to give a parallelisation.

We have thus shown that the most likely frame algebras featuring $\SO(4)$ in the gauge group cannot be realised in massive type IIA parallelisations. While these arguments do not systematically rule out all possibilities, they are highly suggestive that there is no maximally supersymmetric consistent truncation of massive type IIA on $S^3$ with gauge group $\SO(4)$ (or larger).

It seems that a similar conclusion can be reached for the $S^4$ case. We note that \eqref{parall_S3_m}, augmented by an additional piece $\hE = \vol_4$, also yields a Leibniz parallelisation of the type IIA generalised tangent bundle on $S^4$, satisfying the $\SO(5)$ algebra. However, again one can prove this is not an $\SO(5,5)\times {\mathbb R}^+$ frame. 
%
One can construct an $\SO(5,5)\times\bbR^+$ covariant projection acting on four generalised vectors $E^4 \ra \Lambda^4 T^*$. This is done by taking the projections to the bundle $N$ of the two pairs of generalised vectors and then contracting the resulting sections of $N$, which transform in the vector representation of $\SO(5,5)$, using the $\SO(5,5)$ invariant metric. Due to the $\bbR^+$ weights, the inner product is in fact a volume form and transforms under $\bbR^+$, but it is $\SO(5,5)$ invariant. By explicit computation, one can check that the components of this quartic $\SO(5,5)$ invariant on $E$ are not preserved, or rescaled, when one moves to the frame \eqref{parall_S3_m} combined with $\hE = \vol_4$, showing that this frame is not an $\SO(5,5)\times\bbR^+$ frame.


\subsection{$S^2$ parallelisation and $D=8$, $\SO(3)$ supergravity}

We conclude our set of examples by considering type IIA supergravity on the two-sphere $S^2$. 
Again, we will see that while it is easy to define a  generalised Leibniz parallelisation for $m=0$, in the massive case the most likely frame does not work.

On a two-dimensional manifold, the U-duality group is $\SL(3)\times \SL(2)$, and the
generalised tangent bundle reads
\be
E \, \simeq\, T \oplus T^* \oplus \mathbb{R} \oplus \Lambda^2 T^*\, ,
\ee
which factorises as
\be\label{2dEfactorised}
	E \,\simeq\, (\mathbb{R} \oplus \det T^* ) \otimes ( T \oplus \mathbb{R})\, =\, U \otimes W \,,
\ee
where $U$ transforms as an $\SL(2)$ doublet and $W$ as an $\SL(3)$ triplet.

An  $\SL(3)\times \SL(2)$ frame is specified by $\{\hat E_{i\alpha}\}$, where $i=1,2,3$ is an $\SL(3)$ index while $\alpha =\pm$ is an $\SL(2)$ index.
According to the factorisation~\eqref{2dEfactorised}, it can be written as 
\be\label{factorSL3SL2}
	\hat{E}_{i \alpha} = \hat{E}_\alpha \otimes \hat{E}_i\,,
\ee
where $\hat{E}_\alpha$ is a frame for $U$ and $\hat{E}_i$ is a frame for $W$. This guarantees that the scalar matrix $\mathcal{M}^{i\alpha,j\beta}$ defined by the generalised Scherk--Schwarz ansatz parameterises the seven-dimensional coset $\frac{\SL(3)}{\SO(3)}\times \frac{\SL(2)}{\SO(2)}$, as expected for maximal supergravity in eight dimensions.

For vanishing Romans mass, a generalised Leibniz parallelisation on $S^2$ is given by
\be\label{S2parallel}
\begin{cases}
\ \ \hat E_{i+} \!\!\!&=\, v_i + y_i + \iota_{v_i} \!\rg{C_1}\ ,
 \\[1mm]
\ \ \hat E_{i-} \!\!\!&=\, \dd y_i + y_i \rg{\rm vol}_2 -\, \dd y_i\, \wedge \rg{C_1}\ ,
\end{cases}
\ee
where $v_i = \frac 12\epsilon_{i}{}^{jk}v_{jk}$ are the $\SO(3)$ Killing vectors and $\rg{\rm vol}_2$ is the volume on the round $S^2$ of unitary radius. Notice that (before twisting by $\rg{C_1}$) the factorisation condition~\eqref{factorSL3SL2} is satisfied by taking
\begin{align}
	\hat{E}_i \,&=\, v_i + y_i\,, \nn\\[1mm]
	\hat{E}_\alpha \,&=\, { 1 \choose {\rm vol}_2 }_\alpha\,.
\end{align}
Moreover, choosing the two-form flux as
\be
\rg{F_2} \ = \, \dd\! \rg{C_1}\ =\, \frac{1}{R} \rg{\rm vol}_2\ , 
\ee
so that $\iota_{v_i}\dd \!\rg{C_1} = c\,R\,\dd y_i$, the massless Dorfman derivative yields
\bea\label{S2Leibniz}
L_{\hat E_{i+}}\hat E_{j+} &=& - \tfrac{1}{R}\epsilon_{ij}{}^k \hat E_{k+} \ ,\qquad L_{\hat{E}_{i+}} \hat{E}_{j-} \ =\ - \tfrac{1}{R}\epsilon_{ij}{}^k \hat{E}_{k+}\ , \nn \\ [1mm]
L_{\hat{E}_{i-}}\hat{E}_{j+} &=& 0 \ ,\;\qquad\qquad\qquad L_{\hat{E}_{i-}} \hat{E}_{j-} \ =\ 0 \, ,
\eea
which is a Leibniz algebra leading to an $\SO(3)$ gauge algebra.

Hence we have an $\SL(3)\times \SL(2)$ Leibniz parallelisation with associated $\SO(3)$ gauge algebra. 
This can be used to define a generalised Scherk--Schwarz reduction of massless type IIA supergravity on $S^2$, down to to maximal supergravity in eight dimensions with gauge group $\SO(3)$. 
As pointed out in \cite{Boonstra:1998mp}, this consistent reduction on $S^2$ is the same as the conventional Scherk--Schwarz reduction of eleven-dimensional supergravity on the group manifold $\SU(2)\simeq S^3$, presented long ago in~\cite{Salam:1984ft}.
The explicit truncation ansatz for the metric, dilaton and RR two-form on $S^2$ can be found in~\cite[sect.$\:$6]{Cvetic:2000dm}, and its relation with the $S^3$ reduction of eleven-dimensional supergravity is explained in~\cite{Cvetic:2003jy}.

  When the Romans mass is switched on, the frame \eqref{S2parallel} fails to satisfy an algebra under the Dorfman derivative with $m\neq 0$. One could consider the alternative generalised frame
\be\label{alternateS2frame}
\begin{cases}
\ \ \hat E_{i+} \!\!\!&=\, v_i + y_i \!\rg{\rm vol}_2\,,
 \\[1mm]
\ \ \hat E_{i-} \!\!\!&=\, \dd y_i + y_i \,,
\end{cases}
\ee
which compared to \eqref{S2parallel} has the role of the $\mathbb{R}$ and $\Lambda^2 T^*$ terms exchanged, and is not twisted by $\rg{C_1}$.
This frame is still globally defined, orthonormal and  can easily be checked to satisfy the $\SO(3)$ algebra under the massive Dorfman derivative for $mR=1$. However, it cannot be put in the form \eqref{factorSL3SL2}, so it is not an acceptable $\SL(3)\times \SL(2)$ frame. This means that a Scherk--Schwarz reduction based on \eqref{alternateS2frame} would not define a generalised metric of the type given by the supergravity degrees of freedom~\eqref{genmetr}, so it would not make sense to define an ansatz like \eqref{invG_from_parall}. The $S^2$ case is thus on the same footing as $S^3$ and $S^4$, that is it does not seem to allow for a consistent truncation of massive type IIA supergravity preserving maximal supersymmetry.

\section{Conclusions}\label{sec:conclusions}

In this paper we derived  the exceptional generalised geometry formalism for type IIA supergravity on a manifold $M_d$ of dimension $d\leq 6$, 
completing and complementing the work in~\cite{Hull:2007zu,Grana:2009im} and in particular showing how to include the Romans mass in the
formalism. The Romans mass defines a deformation of the massless generalised Lie derivative which generates the internal diffeomorphisms and gauge transformations of the supergravity fields. 

We then applied this formalism to the  construction of  generalised Scherk--Schwarz reductions of type IIA supergravity. These reductions are based on the existence of a generalised Leibniz parallelisation of the $E_{d+1(d+1)}\times \mathbb{R}^+$ generalised tangent bundle on $M_d$,  and are conjectured to yield consistent truncations down to ($10-d$)-dimensional maximal gauged supergravities.
The Leibniz algebra satisfied by the generalised parallelising frame directly determines the embedding tensor of the lower-dimensional theory, and thus completely specifies it.
While the truncation ansatz for the lower-dimensional scalar fields was already discussed in~\cite{spheres}, our derivation of the ansatz for the fields with one or two external legs from the generalised parallelisation is new in generalised geometry. For these latter objects, the corresponding ans\"atze for Scherk--Schwarz reductions have appeared in recent work studying dimensional reductions of the exceptional field theory description of eleven-dimensional and type IIB supergravity in~\cite{Hohm:2014qga,Baguet:2015xha,Baguet:2015sma}. Here, however,  we gave explicit expressions in terms of type IIA supergravity fields, directly truncating the supergravity, without enlarging the dimension of the space-time or introducing additional degrees of freedom.
 We also gave a partial proof of the consistency of the generalised Scherk--Schwarz truncations by showing that the bosonic gauge transformations reduce consistently and yield the gauge transformations of maximal gauged supergravity.

We applied our construction to concrete examples, and found generalised Leibniz parallelisations on $d$-dimensional spheres and hyperbolic 
spaces. In particular, we obtained a generalised parallelisation on $S^6$ satisfying the $\ISO(7)$ algebra, and spelled out the corresponding truncation ansatz as obtained from the generalised Scherk--Schwarz prescription. As recently described in \cite{Guarino:2015jca,Guarino:2015vca}, the Romans mass introduces a magnetic gauging of the $\ISO(7)$ translations in the truncated four-dimensional theory, yielding a symplectic deformation~\cite{Dall'Agata:2014ita} of the type first found in~\cite{Dall'Agata:2012bb} for the $\SO(8)$ gauging. We found the same phenomenon for type IIA supergravity on the six-dimensional hyperboloids $H^{p,7-p}$: on these spaces one can define a consistent truncation down to $\ISO(p,7-p)$ supergravity in four-dimensions; switching the Romans mass on leads to the symplectically-deformed $\ISO(p,7-p)$ gauging described in~\cite{Dall'Agata:2014ita}.
We also obtained generalised Leibniz parallelisations on $S^4$, $S^3$ and $S^2$ for vanishing Romans mass, reproducing the Leibniz algebra of known consistent truncations of massless type IIA supergravity on these manifolds. When the Romans mass is switched on, these parallelisations no more satisfy a Leibniz algebra. We offered an explanation of why this is the case by showing that the frame lies in the stabiliser group of the Romans mass only for the parallelisation on $S^6$. For massive type IIA on $S^3$ we presented a no-go result indicating that a consistent truncation including the $\SO(4)$ algebra does not exist. It would be interesting to see whether similar no-go theorems can be proved for the $S^4$ and $S^2$ cases. 

In this paper we focused on consistent truncations that preserve maximal supersymmetry.  In the last few years a  vast literature was devoted to the construction of consistent truncations with less than maximal supersymmetry, which are interesting per se and for applications to the AdS/CFT correspondence. An approach similar to the one of this paper, but employing a  non-trivial $G$-structure on the generalised tangent bundle rather than the identity structure associated with a  parallelisation, may help clarifying the general structure of such consistent truncations preserving a fraction of supersymmetry.

It is also noteworthy that there is an alternative massive type IIA theory \cite{Howe:1997qt} which can be obtained from eleven-dimensional supergravity by gauging a combination of the $\GL(1)$ global symmetry and the trombone symmetry of the equations of motion.\footnote{We thank Paul Richmond for bringing this case to our attention.} As such, this theory does not have an action. In~\cite{Tsimpis:2005vu} it was argued by superspace methods that this theory and Romans' original massive theory exhaust all possibilities. It is natural to ask how this deformation appears in our formalism. To be diffeomorphism invariant any deformation parameters must appear as $\GL(6)$ singlets with zero $\bbR^+$ weight. There are precisely two such singlets in the $\rep{912_{-1}}$ representation of $E_{7(7)}\times\bbR^+$, one of which we have already identified as the Romans mass deformation. There is also a singlet in the $\rep{56_{-1}}$ representation, which is another part of the generalised torsion~\cite{Coimbra:2011ky}, and which could also be used to deform the Dorfman derivative. When performing generalised Scherk-Schwarz reductions, this additional $\rep{56_{-1}}$ part of the embedding tensor is generated by gauging the trombone symmetry~\cite{LeDiffon:2008sh}, and the resulting theory does not have an action. It is natural to conjecture that deforming the Dorfman derivative by switching on a combination of the second singlet in  $\rep{912_{-1}}$ and the singlet in $\rep{56_{-1}}$ would give the relevant gauge algebra for the theory described in \cite{Howe:1997qt}. 
One could also try to argue that there are no other singlet deformations by considering the closure of the gauge algebra, thus corroborating the result of \cite{Tsimpis:2005vu}.

The formalism developed in the first part of this paper may also be applied to investigate problems different from consistent truncations, for instance the study of marginal deformations of superconformal gauge theories with a (massive) type IIA dual.
We hope to come back to these interesting directions in the near future.

\medskip

{\bf Note added:} On completion of this work, we became aware of~\cite{Ciceri:2016dmd}, which provides an analogous construction of the massive generalised Lie derivative in the context of exceptional field theory and reproduces massive type IIA supergravity upon imposing the section condition.

\section*{Acknowledgements}
We would like to thank Emanuel Malek, Ruben Minasian and Paul Richmond for useful discussions. D.C. is supported by an European Commission Marie Curie Fellowship under the contract PIEF-GA-2013-627243. O.d.F. is supported by the	ILP	LABEX (under reference ANR-10-LABX-63) through French	state funds managed	by the ANR within the Investissements	d'Avenir programme under reference ANR-11-IDEX-0004-02. The work of C.S-C has been supported by the ANR grant 12-BF05-003-002 and a grant from the Foundational Questions Institute (FQXi) Fund, a donor advised fund of the Silicon Valley Community Foundation on the basis of proposal FQXi-RFP3-1321 (this grant was administered by Theiss Research). DW is supported by the STFC Consolidated Grant ST/L00044X/1, the EPSRC Programme Grant EP/K034456/1 ``New Geometric Structures from String Theory'' and the EPSRC Standard Grant EP/N007158/1 ``Geometry for String Model Building''.


\appendix

\section{Notation and conventions}\label{app:notation}

The indices used in this paper are:
\begin{align}
\mu, \nu  \ &: \ \text{external spacetime indices} \, ,\nn\\
m,n \ &:\ \text{curved indices on the internal manifold $M_d$} \ ,\nn\\
a,b \ &:\ \text{frame indices on $M_d$} \, ,\nn\\
i,j \ &:\ \text{indices for the embedding coordinates of $S^d$ in $\mathbb R^{d+1}$ (or $H^{p,q}$ in $\mathbb{R}^{p+q}$)} \, ,\nn\\
I,J \ &:\ \text{$\SL(d+2,\mathbb R)$ indices}\, ,\nn \\
M,N \ &: \ \text{curved indices for the $E_{d+1(d+1)}\times \mathbb{R^+}$ generalised tangent space on $M_d$} \, ,\nn\\
A,B \ &: \ \text{frame indices for the $E_{d+1(d+1)}\times \mathbb{R^+}$ generalised tangent space on $M_d$} \, .\nn
\end{align}

\vskip 3mm

Our tensor conventions are the same as in~\cite{Coimbra:2012af}. We collect here the ones relevant for our computations. 
On a $d$-dimensional manifold $M_d$, given a form $\lambda \in \Lambda^p T^*$ and a poly-vector $w \in \Lambda^q T$, 
\be
\lambda \,=\, \frac{1}{p!} \lambda_{m_1\ldots m_p} \dd x^{m_1} \wedge \cdots \wedge \dd x^{m_p}\ ,\qquad w \,=\, \frac{1}{q!} w^{m_1 \ldots m_q} \frac{\partial}{\partial x^{m_1}}\wedge \cdots \wedge\frac{\partial}{\partial x^{m_q}} \,,
\ee
we define the contraction
\begin{align}\label{deflrcorner}
(w \,\lrcorner\,\lambda)_{m_1\ldots m_{p-q}} \,&=\, \frac{1}{q!}w^{n_1 \ldots n_q}\lambda_{n_1\ldots n_q m_1 \ldots m_{p-q}} \qquad \qquad \text{if}\ q\leq p\,,\nn\\[1mm]
(w \,\lrcorner\,\lambda)^{m_1\ldots m_{q-p}} \,&=\, \frac{1}{p!}w^{m_1 \ldots m_{q-p}n_1 \ldots n_p}\lambda_{n_1\ldots n_p} \qquad \qquad \text{if}\ p< q\,.
\end{align}
The contraction of a vector $v\in T$ with a form $\lambda$ is also denoted by $\iota_v \lambda \equiv v\,\lrcorner \,\lambda$.

The contraction of a poly-vector $w$ with a tensor $\tau \in T^*\otimes \Lambda^d T^*$ is defined as
\be
(w\,\lrcorner\, \tau)_{m_1\ldots m_{d-q+1}} \,=\, \frac{1}{(q-1)!} w^{n_1\ldots n_q} \tau_{n_1,\,n_2\ldots n_q m_1\ldots m_{d-q+1}} \,.
\ee

Moreover, for $\lambda \in \Lambda^{p}T^*$ and $\mu \in \Lambda^{d-p+1}T^*$, we define the ``$j$-operator'' giving  $j \lambda \wedge \mu \in T^* \otimes \Lambda^d T^*$ as: 
\be
\left(j \lambda \wedge \mu\right)_{m,\,m_1\ldots m_d} \,=\, \frac{d!}{(p-1)!(d-p+1)!}\,\lambda_{m[m_1\ldots m_{p-1}}\mu_{m_p\ldots m_d]}\ .
\ee 
This is the same as $j\lambda \wedge \mu = \diff x^m \otimes (\iota_m \lambda \wedge \mu)$. Upon exchanging $\lambda$ and $\mu$ one has
\be 
j \lambda \wedge \mu \,=\, (-1)^{p(d-p+1)+1}\, j\mu \wedge \lambda\ .
\ee

For the Hodge star we take
\be
(*\lambda)_{m_1\cdots m_{d-p}} \,=\, \frac{1}{p!}\sqrt{g}\,\epsilon_{m_1\cdots m_{d-p}}{}^{n_1\ldots n_p}\lambda_{n_1\ldots n_p} \ ,
\ee
with $\epsilon_{12\ldots d} = +1$.

The action of a $\mathfrak{gl}(d)$ element $r \in T \otimes T^*$ on a vector $v\in T $ and on a $p$-form is defined as
\be
(r \cdot v)^m = r^m{}_n v^n\,, \qquad (r\cdot \lambda)_{m_1\ldots m_p} = -p\, r^{n}{}_{[m_1} \lambda_{|n| m_2\ldots m_p]}  \,. 
\ee

\section{IIA exceptional generalised geometry from M-theory}
\label{fromMthToIIA}

In this section we derive the exceptional generalised geometry for type IIA supergravity on a six-dimensional manifold $M_6$ by dimensional reduction of the M-theory exceptional generalised geometry on a seven-dimensional space $M_7$.

\subsection{M-theory exceptional generalised geometry}\label{Mth_excep_geom}

The M-theory exceptional generalised geometry was constructed in \cite{Hull:2007zu, Pacheco:2008ps, Coimbra:2011ky,Coimbra:2012af}. While we refer to these papers for a detailed discussion, here we briefly summarise the main structures that are needed to derive their type IIA counterpart. 
We use the same notation and conventions as~\cite{Coimbra:2011ky,Coimbra:2012af}. 

In M-theory compactified on a seven-dimensional manifold $M_7$, the fibres of the generalised tangent bundle $E$ transform in the ${\bf 56_1}$ representation of the $E_{7(7)}\times\mathbb R^+$ structure group. Under $\GL(7)$, $E$ decomposes as
\begin{equation}
\label{gense7}
   E \ \simeq\ \check E \,\equiv\, TM_7 \oplus \Lambda^2T^*M_7 \oplus \Lambda^5T^*M_7
       \oplus (T^*M_7\otimes\Lambda^7T^*M_7)\, .
\end{equation} 
A section can be written as 
 \begin{equation}
 V \ =\ v +  \omega +  \sigma +  \tau\, ,
\end{equation}
where at each point on $M_7$,  $v \in TM_7$ is an ordinary vector, $\omega\in T^*M_7$, $\sigma \in \Lambda^5T^*M_7$ and $\tau \in (T^*\otimes \Lambda^7 T^*)M_7$.  

The adjoint bundle $\adj$ decomposes under $\GL(7)$ as
\begin{equation}
\adj \ =\  \bbR	\oplus (TM_7\otimes T^*M_7) \oplus \Lambda^3 T^*M_7
		\oplus \Lambda^6 T^*M_7 \oplus \Lambda^3 TM_7 \oplus \Lambda^6 TM_7
  \ ,
\end{equation}
with sections transforming in the ${\bf 133_0}+{\bf 1_0}$ representation of $E_{7(7)}\times \mathbb{R}^+$ given by
\begin{equation}
\label{eq:Ggeom-M}
R \ = \  l +   r +   a +  \tilde a  +  \alpha +  \tilde\alpha \ ,
\end{equation}
where $l \in \bbR$ gives the shift of the warp factor,  $r \in End(T M_7)$, $a \in \Lambda^3 T^*M_7$ is related to the three-form potential of M-theory, $\tilde a \in \Lambda^6T^*M_7$ to its dual, while 
$\alpha \in \Lambda^3 TM_7$ and $ \tilde\alpha \in \Lambda^6 TM_7$ are a three- and a six-vector.

The adjoint action of the $E_{7(7)}\times \mathbb{R}^+$ algebra on a generalised vector is denoted as $V' = R \cdot V$ and reads:
\begin{align}\label{Mth_adjoint_act}
v' \,&=\, l \,  v +  r  \cdot v  + \alpha  \,\lrcorner\,\,  \omega- \tilde{\alpha} \,\lrcorner\, \sigma\ , \nn \\[1mm]
\omega' \,&=\,  l \, \omega +   r \cdot   \omega +v   \,\lrcorner\, a+  \alpha \,\lrcorner\, \sigma + \tilde{\alpha} \,\lrcorner\, \tau\ , \nn \\[1mm]
\sigma' \,&=\,  l \,\sigma +  r \cdot  \sigma  +v   \,\lrcorner\, \tilde a  + a \wedge \omega+ \alpha \,\lrcorner\, \tau \ ,\nn \\[1mm]
\tau'\,&=\,  l\, \tau+  r \cdot  \tau - j \tilde a \wedge \omega+ j a \wedge \sigma  \ .
\end{align}
The $E_{7(7)}$ subalgebra is given by $ \frac 12 {\rm tr}(r)=l$.
The adjoint commutator $R''= [R , R']$ is
\begin{align}\label{comm_Mth_adj}
l'' \,&=\, \tfrac{1}{3} (\alpha \,\lrcorner\, a' - \alpha' \,\lrcorner\, a) + \tfrac{2}{3}  (\tilde \alpha' \,\lrcorner\, \tilde a - \tilde \alpha \,\lrcorner\, \tilde a')\ , \nn \\[1mm]
r'' \,&=\, [ r, r'] + j \alpha \,\lrcorner\, j a'  -  j \alpha' \,\lrcorner\, j  a  -  \tfrac{1}{3}  (\alpha \,\lrcorner\, a' - \alpha' \,\lrcorner\, a) \mathbbm{1}  \nn \\[1mm]
&   \ \quad +  j \tilde \alpha' \,\lrcorner\, j \tilde a  - j \tilde \alpha \,\lrcorner\, j \tilde a'  - \tfrac{2}{3}   (\tilde \alpha' \,\lrcorner\, \tilde a - \tilde \alpha \,\lrcorner\, \tilde a')\mathbbm{1} \ , \nn \\[1mm]
a'' \,&=\, r \cdot a' - r' \cdot a + \alpha' \,\lrcorner\, \tilde a - \alpha \,\lrcorner\,  \tilde a' \ ,\nn \\[1mm]
\tilde a'' \,&=\, r \cdot \tilde a' - r' \cdot \tilde a - a \wedge  a' \ ,\nn \\
\alpha'' \,&=\, r \cdot \alpha' - r' \cdot \alpha +\tilde \alpha' \,\lrcorner\, a - \tilde \alpha \,\lrcorner\, a' \ ,\nn \\[1mm]
\tilde \alpha'' \,&=\, r \cdot  \tilde \alpha' - r' \cdot \tilde \alpha - \alpha \wedge  \alpha'\ .
\end{align}

The generalised tangent bundle $E$ is actually twisted to take into account the non-trivial gauge potentials of M-theory, and this is why it is only isomorphic to $\check E$ in~\eqref{gense7}.  Given a section $\check{V}$ of the untwisted tangent bundle
$\check E$, a section $V$ of $E$ is defined as
\be\label{twist_Mth}
 V \,=\, \rme^{A + \tilde A}  \cdot \check{V} \, , 
\ee 
where $A + \tilde A$ is an element of the adjoint bundle.  The patching condition on the overlaps $U_{\alpha} \cap U_{\beta}$ is 
\be
V_{(\alpha)} \,=\, \rme^{\dd \Lambda_{(\alpha \beta)} + \dd \tilde \Lambda_ {(\alpha \beta)}} \cdot V_{(\beta)} \, , 
\ee
where $\Lambda_{(\alpha \beta)}$ and $ \tilde\Lambda_ {(\alpha \beta)}$ are a two- and five-form, respectively. This corresponds to gauge-transforming the three- and six-form potentials in~\eqref{twist_Mth} as 
\begin{align}
A_{(\alpha)} \,&=\, A_{(\beta)} + \dd \Lambda_{(\alpha \beta)}\ , \nn \\
\tilde A_{(\alpha)} \,&=\, \tilde A_{(\beta)} + \dd \tilde \Lambda_{(\alpha \beta)}  -\frac{1}{2}   \dd \Lambda_{(\alpha \beta)}  \wedge A_{(\beta)} \ .
\end{align}
The respective gauge-invariant field-strengths reproduce the supergravity ones:
\begin{align}
F \,&=\, \dd A \, ,  \nn \\
\tilde F \,&=\, \dd \tilde A - \frac{1}{2} A \wedge F \, .
\end{align}

The Dorfman derivative is constructed as a generalisation of the Lie derivative.  The Lie derivative  between two ordinary vectors $v$ and $v'$ of $TM_7$ 
can be written in components as a $\mathfrak{gl}(7)$ action 
\be
( \mathcal{L}_v v')^m \,=\,  v^n \,\partial_n v'^{\,m} - (\partial \times v)^m{}_n \,v'^{\,n}  \, ,
\ee
where the symbol $\times$ denotes the product of the fundamental and dual representation of $\GL(7)$. Similarly, indicating by $V^M$ the components of $V$ in a standard coordinate basis, and embedding the standard derivative operator 
as a section of the dual generalised tangent bundle $E^*$, one can define the Dorfman derivative as
\be 
\label{eq:Liedefgapp}
(L_V V')^M \,=\,  V^N \partial_N  V'^M - (\partial \times_{\rm ad} V)^M{}_N V'^N \, , 
\ee
where $ \times_{\rm ad}$ is the projection onto the adjoint bundle,
\be
 \times_{\rm ad} \, : \, E^* \otimes E \rightarrow  {\rm ad}\, . 
\ee
In the $\GL(7)$  decomposition, \eqref{eq:Liedefgapp} becomes
\begin{equation}\label{dorf7}
L_{V} V' = \mathcal{L}_{v} v' + \left(\mathcal{L}_{v}  \omega^{\prime} -\iota_{v^\prime}\mathrm{d} \omega\right) + \left(\mathcal{L}_{v} \sigma' -\iota_{v^\prime}\mathrm{d}\sigma -  \omega^{\prime}\wedge \mathrm{d} \omega\right) + \left(\mathcal{L}_{v} \tau^{\prime} - j \sigma^{\prime}\wedge \mathrm{d} \omega - j  \omega' \wedge\mathrm{d}\sigma \right) .
\end{equation}
Note that this is not antisymmetric under the exchange of $V$ and $V'$. 

Another object we will need is the bundle $N$ first introduced in~\cite{Coimbra:2011ky}. This is a sub-bundle of the symmetric product of two generalised tangent bundles, $N \subset S^2 E$, and can be expressed as
\begin{align}\label{MthNbundle}
N \,&\simeq\, T^*M_7 \oplus \Lambda^4 T^*M_7 \oplus (T^*M_7 \otimes \Lambda^6T^*M_7)\nn\\[1mm]
&\;\quad \oplus (\Lambda^3T^*M_7\otimes \Lambda^7T^*M)\oplus (\Lambda^6T^*M_7 \otimes \Lambda^7T^*M_7)\,.
\end{align}
Formally, $N$ can be described via a series of exact sequences
\begin{equation}
\label{eq:N-sequences}
\begin{aligned}
   0 \longrightarrow \Lambda^4 T^*M_7 \longrightarrow N' &
      \longrightarrow T^* \longrightarrow 0\, , \\[1mm]
   0 \longrightarrow T^*M_7\otimes \Lambda^6T^*M_7 \longrightarrow N'' & 
      \longrightarrow N' \longrightarrow 0 \,, \\[1mm]
   0 \longrightarrow \Lambda^7 T^*M_7\otimes \Lambda^3 T^*M_7
      \longrightarrow N''' &
      \longrightarrow N'' \longrightarrow 0\,, \\[1mm]
   0 \longrightarrow \Lambda^7 T^*M\otimes \Lambda^6 T^*M
      \longrightarrow N &
      \longrightarrow N''' \longrightarrow 0 \,.
\end{aligned}
\end{equation}
Under $E_{7(7)}\times \mathbb{R}^+$, sections of $N$ transform in the ${\bf 133_{2}}$ representation. Their expression in terms of the symmetric product of generalised vectors can be found in~\cite{Coimbra:2011ky}. 

The simplest of the intermediate bundles appearing in~\eqref{eq:N-sequences} is $N'$, whose type IIA counterpart will be relevant for the scopes of this paper. This can be expressed as
\be\label{MthN'}
N' \,\simeq \, T^*M_7 \oplus \Lambda^4 T^*M_7\,.
\ee
Given a basis $\{\hat E_A\}$, $A = 1,\ldots, 56$, for the generalised tangent bundle $E$, a section $S$ of $N'$ has the form
\be
S \,=\, S^{AB}\hat E_A \otimes_{N'} \hat E_B \,,
\ee
where $S^{AB}$ are functions on the manifold and the map $\otimes_{N'}: E \otimes E \to N'$ is defined by
\be\label{N'prod_Mth}
V \otimes_{N'} V' \,=\, (v \,\lrcorner\, \omega' + v'\,\lrcorner\,\omega) + (v \,\lrcorner\, \sigma' + v' \,\lrcorner\, \sigma - \omega \wedge \omega')\,.
\ee
We make this definition as it is the result of taking the $E_{7(7)}\times\bbR^+$ covariant projection of $V\otimes V'$ onto $N$ (from~\cite{Coimbra:2011ky}) and then projecting onto $N'$ using the natural mappings in~\eqref{eq:N-sequences}. We stress that the sections of $N'$ themselves do not transform in a definite representation of $E_{7(7)}\times \mathbb{R}^+$.

\subsection{Reduction to type IIA}\label{app:reductionIIA}

We can now proceed and reduce the structures above to type IIA supergravity (in string frame) on a six-dimensional manifold $M_6$. 
Decomposing the $E_{7(7)}\times \bbR^+$ generalised tangent bundle $E$ under the $\GL(6)$ structure group of $M_6$, we obtain 
\begin{equation}
\label{app:gentb}
	E \ \simeq\ T \oplus T^* \oplus \Lambda^5 T^*\oplus (T^* \otimes \Lambda^6 T^*)
		\oplus \Lambda^{\rm even}T^*\ ,
\end{equation}
where $\Lambda^{\rm even}T^*=\mathbb{R} \oplus \Lambda^2 T^* \oplus \Lambda^4 T^* \oplus \Lambda^6 T^*$ and each term in the direct sum is now on $M_6$.  A section, transforming again in the fundamental of $E_{7(7)}\times \bbR^+$, can be written as
\begin{equation}	
\label{app:genvec}	
	V \ =\ v + \lambda + \sigma + \tau + \omega\ ,
\end{equation}
where $\omega=\omega_0 + \omega_2 + \omega_4 + \omega_6$ is a poly-form in $ \Lambda^{\text{even}} T^*$.

The $\GL(6)$ decomposition of the adjoint bundle is
\begin{equation}
	\adj F \ =\ \bbR_\Delta \oplus \bbR_\phi 
		\oplus (T\otimes T^*) \oplus \Lambda^2 T \oplus \Lambda^2 T^*
		\oplus \Lambda^6 T \oplus \Lambda^6 T^*
		\oplus \Lambda^{\text{odd}} T \oplus \Lambda^{\text{odd}} T^*\ ,
\end{equation}
with section
\begin{equation}
\label{Edd_adjoint}
R \ = \ l + \varphi + r + \beta + b + \tilde \beta + \tilde b + \alpha + a\ ,
\end{equation}
where $\alpha = \alpha_1 + \alpha_3 + \alpha_5 \in \Lambda^{\text{odd}} T$ and  $a= a_1 + a_3 + a_5 \in \Lambda^{\text{odd}} T^*$ are antisymmetric poly-vectors and poly-forms,
respectively.  

Let us now derive the action of the adjoint of $E_{7(7)}\times \bbR^+$ on a generalised vector and the commutators of two adjoints in type IIA language. 
 Denoting by $z$ the coordinate along the seventh direction, a type IIA generalised vector is related to an M-theory generalised vector as
\bea
\label{redrules}
v_{\rm M}  \,&=&\, v + \omega_0 \partial_z\, , \nn \\[1mm]
\omega_{\rm M}  \,&=&\, \omega_2 -  \lambda \wedge \mathrm{d}z  \, , \nn \\[1mm]
\sigma_{\rm M}  \,&=&\, \sigma + \omega_4 \wedge \mathrm{d}z\, , \nn \\[1mm]
\tau_{\rm M}  \,&=&\, \tau \wedge \mathrm{d}z + \mathrm{d}z \otimes (\omega_6 \wedge \mathrm{d}z)\ ,
\eea
where $\tau = \tau_1 \otimes \tau_6$ and  the subscript M denotes the M-theory quantities defined in section~\ref{Mth_excep_geom}. Similarly, the M-theory adjoint~\eqref{eq:Ggeom-M} decomposes as
\begin{equation}
\label{redrulesad}
\begin{array}{l}
l_{\rm M}  \ =\ l - \tfrac13 \varphi\\[1mm]
a_{\rm M} \ =\ a_3 + b \wedge   \mathrm{d}z  \\[1mm]
\tilde a_{\rm M}  \ = \ \tilde b + a_5   \wedge   \mathrm{d}z  \\[1mm]
\alpha_{\rm M}  \ =\ \alpha_3 + \beta \wedge \partial_z   \\[1mm] 
\tilde \alpha_{\rm M} \ =\  \tilde \beta + \alpha_5 \wedge \partial_z 
\end{array}
\qquad \qquad 
r_{\rm M}   = \begin{pmatrix} 
r + \tfrac13 \varphi \,\id \,&\, -  \alpha_1 \\[1mm] 
a_1 \,&\, -\tfrac23 \varphi
\end{pmatrix}
\end{equation}
where the identification $l_{\rm M} = l - \tfrac13 \varphi$ follows from the relation between the M-theory and IIA warp factors  $\Delta_{\text{M}} = \Delta_{\text{IIA}} - \tfrac13 \phi$.

Decomposing the M-theory adjoint action given in~\eqref{Mth_adjoint_act} yields the IIA adjoint action  on a generalised vector. Denoting this by $V' = R\cdot V$, we have
\begin{align}
\label{IIAadjvecCompact}
v' \,&=\, l v +  r \cdot v - [ \alpha \,\lrcorner\, s(\omega)]_{-1} - \beta \,\lrcorner\, \lambda - \tilde \beta \,\lrcorner\, \sigma \ , \nn \\[1mm]
\lambda' \,&=\, l \lambda  +  r \cdot \lambda  - v \,\lrcorner\, b - [\alpha\,\lrcorner\, s(\omega)]_1 - \tilde \beta \,\lrcorner\, \tau \ ,\nn \\[1mm]
\sigma' \,&=\,  (l-2\varphi) \sigma  +  r \cdot \sigma +  v \,\lrcorner\, \tilde b - [\omega\wedge s(a)]_5 - \beta \,\lrcorner\, \tau \ ,\nn \\[1mm]
\tau' \,&=\,  (l-2\varphi) \tau  +  r \cdot \tau + j a \wedge s(\omega) + j \tilde b \wedge \lambda - j b \wedge \sigma \ ,\nn \\[1mm]
\omega' \,&=\, (l-\varphi) \omega + r \cdot \omega + b\wedge \omega +  v \,\lrcorner\, a + \lambda \wedge a + \beta \,\lrcorner\, \omega + \alpha \,\lrcorner\, \sigma  +\alpha\,\lrcorner \,\tau \ ,\quad
\end{align}
where $s$ is the sign operator $s(\omega_n) = (-1)^{[n/2]} \omega_n$ for $\omega_n \in \Lambda^n T^*$, and $[\ldots]_p$
denotes the form of degree $p$ in the formal sum inside the parenthesis (by $-1$ we mean we pick the vector component). The $E_{7(7)}$ subalgebra is specified by $\frac 12{\rm tr}(r) = l - \varphi$. In particular, the $O(6,6) \subset \E7$ action is generated by $r, b$ and~$\beta$, also setting $\varphi = -\frac 12{\rm tr}(r)$ and all other generators to zero. 

Reducing the M-theory commutator \eqref{comm_Mth_adj} with the decomposition \eqref{redrulesad} we find that 
the IIA adjoint commutator $R'' = [R,R']$ reads 
\bea\label{eq:commAdjIIA}
l'' \!&=&\!   -\tfrac{1}{2} (\alpha_1 \,\lrcorner\, a_1' -  \alpha_1' \,\lrcorner\, a_1)+\tfrac{1}{2} (\alpha_3 \,\lrcorner\, a_3' -  \alpha_3' \,\lrcorner\, a_3)   - \tfrac{1}{2} (\alpha_5 \,\lrcorner\, a_5' -  \alpha_5' \,\lrcorner\, a_5) + (\tilde \beta' \,\lrcorner\,  \tilde b - \tilde \beta \,\lrcorner\, \tilde b') \nn \\
\phi'' \!&=&\! \tfrac{3}{2} (\alpha_1' \,\lrcorner\, a_1 -  \alpha_1 \,\lrcorner\, a_1') + \tfrac{1}{2} (\alpha_3 \,\lrcorner\, a_3' -  \alpha_3' \,\lrcorner\, a_3)  - \tfrac{1}{2}
 (\alpha_5' \,\lrcorner\, a_5 -  \alpha_5 \,\lrcorner\, a_5') \nn \\
\!&&\!   - ( \beta \,\lrcorner\,  b' - \beta' \,\lrcorner\,  b) + (\tilde \beta' \,\lrcorner\,  \tilde b - \tilde \beta \,\lrcorner\, \tilde b') \nn \\ 
r'' \!&=&\! [ r, r'] + j \alpha_1' \,\lrcorner\,  j a_1 -  j  \alpha_1\,\lrcorner\, j a_1'  + j\alpha_3 \,\lrcorner\, j a_3' -  j\alpha_3' \,\lrcorner\, ja_3  - j\alpha_5 \,\lrcorner\, j a_5'   +j \alpha_5' \,\lrcorner\,  j a_5 \nn \\ 
\!&&\!+ j \beta \,\lrcorner\,  j b'  -  j \beta' \,\lrcorner\, j b - j \tilde \beta \,\lrcorner\, j \tilde b'  +  j  \tilde \beta' \,\lrcorner\, j \tilde b + \tfrac{1}{2}\mathbbm{1} (\alpha_1' \,\lrcorner\, a_1 - \alpha_1 \,\lrcorner\,  a_1') + \tfrac{1}{2} \mathbbm{1}  (\alpha_3' \,\lrcorner\, a_3 - \alpha_3 \,\lrcorner\,  a_3') \nn \\
\!&&\! + \tfrac{1}{2} \mathbbm{1}  (\alpha_5' \,\lrcorner\, a_5 - \alpha_5 \,\lrcorner\,  a_5') + \mathbbm{1}(\tilde \beta \,\lrcorner\, \tilde b' - \tilde\beta' \,\lrcorner\, \tilde b) \nn \\
b'' \!&=&\! r \cdot b'-  r' \cdot b  +\alpha_1 \,\lrcorner\, a_3'  - \alpha_1' \,\lrcorner\, a_3 - \alpha_3 \,\lrcorner\, a_5' + \alpha_3' \,\lrcorner\, a_5  \nn \\
\tilde b'' \!&=&\! r  \cdot \tilde b' - r' \cdot \tilde b  -2\varphi \tilde b' + 2\varphi'\tilde b  + a_1 \wedge a_5' - a_1' \wedge a_5 - a_3 \wedge a_3' \nn \\
a'' \!&=&\!  r \cdot a' - r' \cdot a - \varphi a'  + \varphi' a  + b\wedge  a' - b'\wedge a   + \beta \,\lrcorner\, a' - \beta' \lrcorner\, a -\alpha \,\lrcorner\, \tilde b'  + \alpha' \lrcorner\, \tilde b  \nn \\
\beta'' \!&=&\! r \cdot \beta' - r' \cdot \beta + \alpha_3' \,\lrcorner\, a_1 - \alpha_3  \,\lrcorner\,  a_1'  - \alpha_5' \,\lrcorner\, a_3 + \alpha_5 \,\lrcorner\,  a_3'   \nn \\
\tilde \beta'' \!&=&\!  r \cdot \tilde \beta' - r' \cdot \tilde \beta + 2\varphi \tilde \beta' - 2\varphi' \tilde \beta   + \alpha_1 \wedge \alpha_5' - \alpha_3 \wedge \alpha_3' + \alpha_5\wedge \alpha_1'\nn \\
\alpha'' \!&=&\!  r \cdot \alpha'  - r' \cdot \alpha +\varphi \alpha'  - \varphi' \alpha + \beta\wedge\alpha' -\beta'\wedge\alpha  - \alpha \,\lrcorner\, b'  + \alpha' \lrcorner\, b -  \tilde \beta  \,\lrcorner\, a' + \tilde \beta'  \lrcorner\, a  \ .
\eea

We next obtain the explicit expression for the Dorfman derivative between two type IIA generalised vectors $V$ and $V'$. By plugging \eqref{redrules} into \eqref{dorf7} we find:
\begin{align}\label{dorf6_expanded}
L_V V' =&\ \mathcal{L}_v v' + \left(\mathcal{L}_v \lambda' - \iota_{v^\prime} \mathrm{d}\lambda\right) + \left(\iota_v \mathrm{d}\omega_0' - \iota_{v^\prime} \mathrm{d}\omega_0\right) \nn \\[1mm]
&\, + \left(\mathcal{L}_v \omega_2^\prime - \iota_{v^\prime}\mathrm{d}\omega_2 - \lambda' \wedge \mathrm{d}\omega_0 +  \omega_0' \mathrm{d}\lambda\right) \nn \\[1mm]
&\, + \left( \mathcal{L}_v \omega_4^\prime - \iota_{v^\prime}\mathrm{d}\omega_4 - \lambda' \wedge \mathrm{d}\omega_2 +  \omega_2^\prime \wedge \mathrm{d}\lambda\right) \nn \\[1mm]
&\, + \left(\mathcal{L}_v \omega_6^\prime  -  \lambda' \wedge \mathrm{d}\omega_4 +  \omega_4' \wedge \mathrm{d}\lambda\right) \nn \\[1mm]
&\, + \left( \mathcal{L}_v \sigma' - \iota_{v^\prime}\mathrm{d}\sigma + \omega_0^\prime \mathrm{d}\omega_4 - \omega_2^\prime \wedge \mathrm{d}\omega_2 + \omega_4' \wedge \mathrm{d}\omega_0\right)  \\[1mm]
&\, + \left(\mathcal{L}_v \tau' + j \sigma' \wedge \mathrm{d}\lambda + \lambda^\prime \otimes \mathrm{d}\sigma + \mathrm{d}\omega_0 \otimes \omega_6^\prime + j \omega_4^\prime \wedge \mathrm{d}\omega_2 
 - j \omega_2^\prime \wedge \mathrm{d}\omega_4 
\right) \nn \, . 
\end{align}
This expression can be cast in the more compact form given in \eqref{dorf6}.

As in M-theory, the presence in type IIA of non-trivial gauge potentials leads to the definition of a twisted generalised tangent bundle whose sections
are related to  \eqref{app:genvec}  by the  twist \eqref{eq:twistC_short}.  In order to derive the explicit form of the twist we need to 
exponentiate the  $E_{7(7)}$ adjoint action on a generalised vector  \eqref{IIAadjvecCompact} with  $l=\varphi = r = \beta =\tilde\beta= \alpha= 0$. This corresponds to exponentiating a nilpotent subalgebra of the $E_{7(7)}$ algebra, comprising precisely the form potentials of type IIA supergravity.  We find that the series expansion
\be
V' \ =\ \rme^{R}\cdot V \ \equiv\ V + R \cdot V + \tfrac12 R \cdot(R \cdot V) + \ldots \ee 
truncates at fifth order, and is given by
\bea\label{eq:exp_adj}
v'\!&=&\! v \ ,\nn \\ [1mm]
\lambda' \!&=&\! \lambda - \iota_v b \ ,\nn \\ [1mm]
\sigma' \!&=&\! \sigma + \iota_v \tilde b - \left[\calB^{(1)} \wedge s(a)\wedge \omega + \calB^{(2)}\wedge s(a) \wedge \iota_v a \right]_5 + a_1 \wedge a_3 \wedge \left(\lambda -\tfrac 13 \iota_v b \right) \ , \nn \\ [1mm] 
\tau' \!&=&\! \tau   + j \tilde b \wedge \left(\lambda - \tfrac 12 \iota_v b \right) - j s(a) \wedge \left(\calB^{(1)}\wedge\omega + \calB^{(2)}\wedge (\iota_v a +  \lambda \wedge a)+ \calB^{(3)}\wedge a\wedge \iota_v b \right)  \nn \\ [1mm]
\!&&\!\!\! - j b \wedge \left(\sigma + \tfrac 12 \iota_v \tilde b - \calB^{(2)}\wedge s(a) \wedge \omega - \calB^{(3)}\wedge s(a)\wedge \iota_v a + \tfrac 13 a_1 \wedge a_3 \wedge \left(\lambda-\tfrac 14 \iota_v b\right)\right),\nn \\ [1mm]
\omega' \!&=&\! \rme^b \wedge\omega + \calB^{(1)}\wedge (\iota_v a + \lambda \wedge a) +  \calB^{(2)}\wedge a\wedge \iota_v b \ ,
\eea
where we introduced the shorthand notation
\begin{align}
\calB^{(1)} \ &= \ \frac{\rme^b-1}{b} \ = \ 1 + \tfrac12 b + \tfrac{1}{3!}b \wedge b + \ldots \ ,\nn \\
\calB^{(2)} \ &= \ \frac{\rme^b-1-b}{b\wedge b} \ = \ \tfrac12 + \tfrac{1}{3!} b + \tfrac{1}{4!}b\wedge b + \ldots \ ,\nn \\
\calB^{(3)} \ &= \ \frac{\rme^b-1-b- \tfrac 12 b \wedge b}{b\wedge b\wedge b} \ = \ \tfrac{1}{3!} + \tfrac{1}{4!}b + \tfrac{1}{5!}b \wedge b + \ldots\ .
\end{align}

We can also reduce to type IIA the bundle $N \subset S^2 E$ given in \eqref{MthNbundle}. In terms of bundles on $M_6$, we obtain 
\be
N \,\simeq\, \mathbb{R}  \oplus \Lambda^4T^*  \oplus \Lambda^{\rm odd}T^*  \oplus \Lambda^6T^*  \oplus (T^* \otimes \Lambda^5 T^*) \oplus (\Lambda^2T^*  \oplus \Lambda^6T^*  \oplus \Lambda^{\rm odd}T^*)\otimes \Lambda^6T^* .
\ee
The full $N$ bundle in type IIA is described as a similar set of exact sequences to those in M-theory~\eqref{eq:N-sequences}. Again, these provide us with a natural projection onto a smaller bundle $N'$, which is isomorphic to
\be
N' \,\simeq \, \mathbb{R} \oplus \Lambda^4T^* \oplus \Lambda^{\rm odd} T^* \,
\ee
(note that this also includes $\Lambda^5 T^*$ and is thus not just the reduction of the M-theory $N'$ bundle given in~\eqref{MthN'}).
Given a basis $\{\hat E_A\}$, $A = 1,\ldots, 56$, for the generalised tangent bundle $E$, a section $S$ of $N'$ has the form
\be
S \,=\, S^{AB}\hat E_A \otimes_{N'} \hat E_B \,,
\ee
where $S^{AB}$ are functions on the manifold and the map 
\be
\otimes_{N'}: E \otimes E \to N'
\ee 
is defined as
\begin{align}
V \otimes_{N'} V' \,&=\, v\,\lrcorner\, \lambda' + v' \lrcorner\, \lambda\nn\\[1mm]
\,&\ +  v \,\lrcorner \,\sigma' + v' \,\lrcorner \,\sigma + [\omega \wedge s(\omega')]_4 \nn\\[1mm]
\,&\ + v \,\lrcorner\, \omega' + \lambda\wedge \omega' + v' \lrcorner\, \omega + \lambda'\wedge \omega\,.
\end{align}
As for~\eqref{N'prod_Mth}, this is the $E_{7(7)}\times\bbR^+$ covariant projection to $N$ further projected onto $N'$.

\subsection{The split frame}\label{splitfr_MtoIIA}

As discussed in section \ref{gen_frame_metric}, a convenient way to compute the generalised metric is starting from the conformal split frame, namely a specific choice of frame
on the generalised tangent bundle \eqref{IIAtangentbundle}.  Here we derive the type IIA conformal split frame by reducing the M-theory one given in~\cite{Coimbra:2011ky}. The latter reads
\begin{align}
\label{eq:geom-basis}
  \mathcal{E}_{{\rm M} \, \hat{a}} \,&=\, \rme^{\Delta_{\rm M}} \Big( \hat{e}_{\hat{a}} + i_{\hat{e}_{\hat{a}}} A
      + i_{\hat{e}_{\hat{a}}} \tilde A
      + \tfrac{1}{2} A \wedge i_{\hat{e}_{\hat{a}}} A 
      + j A \wedge i_{\hat{e}_{\hat{a}}}\tilde{A}
      + \tfrac{1}{6}j A \wedge A \wedge i_{\hat{e}_{\hat{a}}} A \Big)\, , \nn \\[1mm]
 \mathcal{E}_{\rm M}^{\hat{a}_1\hat{a}_2} \,&=\, \rme^{\Delta_{\rm M}} \left( e^{\hat{a}_1\hat{a}_2} +  A \wedge e^{\hat{a}_1\hat{a}_2}  
      - j \tilde{A}\wedge e^{\hat{a}_1\hat{a}_2}
      + \tfrac{1}{2}j A \wedge A \wedge e^{\hat{a}_1\hat{a}_2} \right)\, , \nn\\[1mm]
   \mathcal{E}_{\rm M}^{\hat{a}_1\dots \hat{a}_5} \,&=\, \rme^{\Delta_{\rm M}} \left( e^{\hat{a}_1\dots \hat{a}_5} 
      + j A \wedge e^{\hat{a}_1\dots \hat{a}_5} \right)\, , \nn \\[1mm]
 \mathcal{E}_{\rm M}^{\hat{a},\hat{a}_1\dots \hat{a}_7} \,&=\, \rme^{\Delta_{\rm M}}\, e^{\hat{a},\hat{a}_1\dots \hat{a}_7}\, ,
\end{align}
where $\Delta_{\rm M}$ is the M-theory warp factor and $A$ and $\tilde A$ are the three- and six-form potentials of M-theory.   $\hat{e}_{\hat{a}}$  is 
a  frame for $T M_7$, $e_{\hat{a}} $ is the dual one and  $e^{\hat{a}_1\ldots \hat{a}_p}= e^{\hat{a}_1}\wedge \cdots \wedge e^{\hat{a}_p}$,  and $e^{\hat{a},\hat{a}_1\ldots \hat{a}_7} = 
e^{\hat{a}}\otimes e^{\hat{a}_1\ldots \hat{a}_7}$.  The index $\hat{a}$ goes from 1 to 7 and,  not to clutter the notation,  we omitted  the subscript M on   $\hat{e}_{\hat{a}} $ and  $e^{\hat{a}} $.

In reducing to type IIA, we decompose the M-theory potentials as 
\begin{align}
A \,&=\, C_3 - B \wedge \mathrm{d}z\, ,\nn\\[1mm]
\tilde{A} \,&=\, \tilde{B} - \tfrac{1}{2}C_5 \wedge C_1 + (C_5 - \tfrac{1}{2}B \wedge C_3) \wedge \mathrm{d}z\, , 
\end{align}
where $z$ denotes again the circle direction along which we are reducing, and $B$, $\tilde{B}$ and $C_k$ are the IIA potentials.  As already recalled, the IIA and M-theory warp factors are related by
\be
\Delta_{\rm M} \,=\, \Delta - \phi/3 \ . 
\ee
To reduce the split frame \eqref{eq:geom-basis}, we also need to  decompose the seven-dimensional indices as $\hat{a} = (a, 7)$, with $a=1, \ldots, 6$, and
write the seven-dimensional  frames as 
\be
\hat{e}_{{\rm M} \, \hat{a}} \,=\, \begin{cases} \rme^{\phi/3}\left(\hat{e}_a + C_a \partial_z\right)\\[1mm]
 \rme^{-2\phi/3}\partial_z
\end{cases}
\qquad\qquad
e_{\rm M}^{\hat{a}} \,=\, \begin{cases}  \rme^{-\phi/3}e^a \\[1mm]
 \rme^{2\phi/3}\left(\mathrm{d}z - C_1\right)\ ,
\end{cases}
\ee
where $\hat{e}_a$ and $e^a$ are basis for the IIA frame and dual frame bundle, respectively, while $C_a$ denotes the components of the one-form $C_1$. 
The reduction gives
\begin{equation}
\{\hat{E}_A\} = \{\hat \calE_a\} \cup \{\calE^a\} \cup \{\calE^{a_1 \ldots a_5} \} \cup \{\calE^{a,a_1\ldots a_6}\} \cup \{\calE\} \cup \{\calE^{a_1a_2}\} \cup \{\calE^{a_1\ldots a_4}\} \cup \{\calE^{a_1 \ldots a_6}\}\, ,
\end{equation} 
with
\begin{equation}
\label{splitframe_explicit}
\begin{aligned}
\hat{\mathcal{E}}_a \,&=\, \rme^{\Delta} \left(\hat e_a + \iota_{\hat e_a}B + \rme^{-B}\wedge\iota_{\hat e_a}(C_1 + C_3 + C_5) + \iota_{\hat e_a}\tilde{B} + j\tilde{B} \wedge \iota_{\hat e_a}B \right.\\
& \phantom{= \rme^{\Delta}}  - \tfrac{1}{2} C_1 \wedge \iota_{\hat e_a}C_5 +\tfrac{1}{2} C_3 \wedge \iota_{\hat e_a} C_3 -\tfrac{1}{2} C_5 \wedge \iota_{\hat e_a} C_1 - \tfrac{1}{2} j C_5 \wedge \iota_{\hat e_a}C_3  \\[1mm]
& \phantom{= \rme^{\Delta}} \left. + \tfrac{1}{2}j B\wedge C_3 \wedge \iota_{\hat e_a}C_3  -\tfrac{1}{2}j B\wedge C_5 \wedge \iota_{\hat e_a}C_1  -\tfrac{1}{2}j B\wedge C_1 \wedge \iota_{\hat e_a}C_5 \right) , \\[2mm]
\mathcal{E}^a \,&=\, \rme^{\Delta} \big(e^a - \rme^{-B}\wedge (C_1+ C_3 + C_5) \wedge e^a  + j\tilde{B} \wedge e^a   - C_3 \wedge C_1 \wedge e^a   \\[1mm]
& \phantom{=\rme^{\Delta}} - j B \wedge C_3 \wedge C_1\wedge e^a + \tfrac{1}{2}j C_1 \wedge C_5\wedge e^a - \tfrac{1}{2} j C_3 \wedge C_3\wedge e^a + \tfrac{1}{2}j C_5 \wedge C_1\wedge e^a \big)\, , \\[2mm]
\mathcal{E}^{a_1 \ldots a_5} \,&=\, \rme^{\Delta-2\phi}\left(e^{a_1 \ldots a_5} + jB \wedge e^{a_1 \ldots a_5}  \right)\, , \\[2mm]
\mathcal{E}^{a,a_1\ldots a_6} \,&=\, \rme^{\Delta-2\phi}\left(e^{a,a_1 \ldots a_6} \right)\, , \\[2mm]
\mathcal{E} \,&=\,  \rme^{\Delta-\phi} \left( \rme^{-B} - C_5 - j B\wedge C_5 \right)\ , \\[2mm]
\mathcal{E}^{a_1 a_2} \,&=\,  \rme^{\Delta-\phi} \left( \rme^{-B}\wedge e^{a_1 a_2} + C_3\wedge e^{a_1 a_2}  - jC_5 \wedge e^{a_1 a_2} + jB \wedge C_3\wedge e^{a_1 a_2} \right)\, , \\[2mm]
\mathcal{E}^{a_1 \ldots a_4} \,&=\, \rme^{\Delta-\phi}\left(\rme^{-B}\wedge e^{a_1 \ldots a_4} - C_1 \wedge e^{a_1 \ldots a_4} + jC_3 \wedge e^{a_1 \ldots a_4} -jB\wedge C_1 \wedge e^{a_1\ldots a_4} \right) , \\[2mm]
\mathcal{E}^{a_1 \ldots a_6} \,&=\, \rme^{\Delta-\phi}\left(e^{a_1 \ldots a_6} - j C_1 \wedge e^{a_1\ldots a_6} \right)\, .
\end{aligned}
\end{equation}
These expressions can be summarised in the twist given in~\eqref{twist_splitfr}.


\section{Twisted bundle and  gauge transformations}
\label{app:gauge-patching}

In this appendix, we show how one can derive the patching conditions \eqref{patching} of the generalised tangent bundle starting from the supergravity gauge transformations. The key requirement will be that the generalised vector generates the  diffeomorphism and gauge transformations that act on the supergravity fields. 
We include the Romans mass in our computation, the massless case simply follows by setting $m=0$.

We start imposing that  in each patch $U$ of the manifold $M_6$, a generalised vector $V$ generates a diffeomorphism and gauge transformation of the type IIA supergravity potentials:
\begin{align}\label{eq:gauge-trans-by-V-E1}
	\delta_{V} B \,&=\, \mathcal{L}_v B - \dd \lambda \,, \nn\\[1mm]
	\delta_{V} C \,&=\, \mathcal{L}_v C -  \rme^B \wedge (\dd \omega - m \lambda)  \,, \nn\\[1mm]
	\delta_{V} \tB \,&=\,  \mathcal{L}_v \tB - (\dd \sigma + m \,\omega_6) 
		- \tfrac12 \big[\rme^B \wedge (\dd \omega - m \lambda) \wedge s(C) \big]_6 \,,
\end{align}
where all the fields are defined on $U$.
In these expressions, the diffeomorphism is generated by the ordinary Lie derivative $\mathcal{L}$ along a vector $v$, while the remaining terms correspond to the supergravity gauge transformation.

We next require that the generalised diffeomorphism~\eqref{eq:gauge-trans-by-V-E1} be globally well-defined. This means that on the intersection of a patch $U_\alpha$ with another patch $U_\beta$, the new field configuration defined by~\eqref{eq:gauge-trans-by-V-E1} is patched in the same way as the original one, so as to preserve the global structure (which cannot be changed by an infinitesimal transformation). 
The patching conditions for the gauge potentials on $U_{\alpha} \cap U_{\beta}$ are
given by the gauge transformation of the supergravity fields. At the linearised level, these read
\begin{align}
\label{eq:lin-gauge-E1}
	B_{(\alpha)} &= B_{(\beta)} + \dd \Lambda_{(\alpha\beta)} \, ,\nn\\[1mm]
	C_{(\alpha)} &= C_{(\beta)} + \rme^{B_{(\beta)} } \wedge (\dd \Omega_{(\alpha\beta)} -m \Lambda_{(\alpha\beta)})\, , \nn\\[1mm]
	\tB_{(\alpha)} &= \tB_{(\beta)} + \dd \tilde \Lambda_{(\alpha\beta)} + m \Omega_{6(\alpha\beta)} 
		  + \tfrac12 \left[\rme^{B_{(\beta)}}\wedge (\dd \Omega_{(\alpha\beta)} -m\Lambda_{(\alpha\beta)})\wedge s(C_{(\beta)}) \right]_6 \, ,
\end{align}
where the labels $(\alpha)$ and $(\beta)$ indicate fields on $U_\alpha$ and $U_\beta$, respectively, while $(\alpha\beta)$ denotes a field defined just on $U_{\alpha} \cap U_{\beta}$. Note that these gauge transformations have the opposite signs with respect to those in~\eqref{eq:gauge-trans-by-V-E1}, as that equation describes an active transformation which shifts the field configuration to a physically equivalent one; contrastingly, equation~\eqref{eq:lin-gauge-E1} describes a patching relation needed to define the fields on the whole manifold, similar to coordinate invariance in general relativity, which is a passive transformation.
From \eqref{eq:lin-gauge-E1} we construct the corresponding finite transformation.
Its form is not uniquely determined, since it depends on the order one chooses for the exponentiation of the infinitesimal transformations. We choose to exponentiate first the action of the RR transformation with parameter $\Omega$, then the NSNS transformation by
$ \Lambda$ and finally the one by $\tilde \Lambda$. This gives:
\begin{align}\label{eq:finite-gauge-E1}
	B_{(\alpha)} \,&=\, B_{(\beta)}  + \dd \Lambda_{(\alpha\beta)} \ ,\nn\\[1mm]
	C_{(\alpha)} \,&=\, C_{(\beta)}  + \rme^{B_{(\beta)}  + \dd \Lambda_{(\alpha\beta)}} \wedge \dd \Omega_{(\alpha\beta)} 
		- m \, \rme^{B_{(\beta)} } \wedge\Lambda_{(\alpha\beta)}  \wedge \left( \tfrac{\rme^{\dd \Lambda}-1}{\dd \Lambda} \right)_{(\alpha\beta)}  \ , \nn\\[1mm]
	\tB_{(\alpha)} \,&=\, \tB_{(\beta)}  + \dd \tilde \Lambda_{(\alpha\beta)} + m \Omega_{6 \, (\alpha\beta)} 
		+ \tfrac12\, m \, \Lambda_{(\alpha\beta)} \wedge \Big[  \rme^{-B_{(\beta)} } \wedge \left(\tfrac{ \rme^{-\dd \Lambda}-1}{\dd \Lambda} \right)_{(\alpha\beta)}
			 \wedge C_{(\beta)}  \Big]_5 \nn\\[1mm]
		&\quad\;  + \tfrac12 \Big[ 
			\dd \Omega_{(\alpha\beta)}  \wedge  \rme^{B_{(\beta)} + \dd \Lambda_{(\alpha\beta)}} \wedge  s( C_{(\beta)}  )
			- m\, \dd \Omega_{(\alpha\beta)}  \wedge \Lambda_{(\alpha\beta)} \wedge   \left( \tfrac{\rme^{\dd \Lambda}-1}{\dd \Lambda} 
				   \right)_{(\alpha\beta)} \Big]_6 \,,
\end{align}
where we used the shorthand notation
\be
\tfrac{\rme^{\pm\dd \Lambda} - 1}{\dd \Lambda} = \pm 1 + \tfrac{1}{2} \dd \Lambda \pm  \tfrac{1}{3!} \dd \Lambda \wedge \dd \Lambda\, + \dots \,.
\ee

Imposing that the new field configurations \eqref{eq:gauge-trans-by-V-E1} in two overlapping patches $U_\alpha$ and $U_\beta$ are still related in the intersection $U_\alpha \cap U_\beta$ by the transformation \eqref{eq:finite-gauge-E1}, and working to first order in the components of $V$, we obtain
\begin{align}\label{eq:well-defined}
 \delta_{V_{(\alpha)}} B_{(\alpha)} &=   \delta_{V_{(\beta)}}  B_{(\beta)}\,,  \nn\\[1mm]
 \delta_{V_{(\alpha)}} C_{(\alpha)} &=   \delta_{V_{(\beta)}} C_{(\beta)}+  \rme^{ B_{(\beta)}  +\dd \Lambda} \wedge \delta_{V_{(\beta)}}  B_{(\beta)}  \wedge \dd \Omega 
- m \, \rme^{  B_{(\beta)} } \wedge \delta_{V_{(\beta)}}  B_{(\beta)} \wedge   \Lambda   \wedge \big( \tfrac{\rme^{\dd \Lambda}-1}{\dd \Lambda} \big) , \nn\\[1mm]
\delta_{V_{(\alpha)}} \tB_{(\alpha)}  & =\,  \delta_{V_{(\beta)}} \tB_{(\beta)}    
		+ \tfrac12 m \, \Lambda \wedge \Big[  \rme^{-B_{(\beta)} } \wedge \big(\tfrac{ \rme^{-\dd \Lambda}-1}{\dd \Lambda} \big)
			 \wedge ( \delta_{V_{(\beta)}} C_{(\beta)} - \delta_{V_{(\beta)}} B_{(\beta)} \wedge C_{(\beta)})  \Big]_5 \nn\\[1mm]
		&\quad\;  + \tfrac12 \Big[ 
			\dd \Omega  \wedge  \rme^{B_{(\beta)} + \dd \Lambda} \wedge\Big(  s( \delta_{V_{(\beta)}} C_{(\beta)}) + \delta_{V_{(\beta)}} B_{(\beta)} \wedge s(C_{(\beta)})\Big)   \Big]_6 \,,
\end{align}
where for ease of notation we are omitting the label $(\alpha\beta)$ on $\Lambda$, $\tilde\Lambda$ and $\Omega$.
This equation can be solved to give relations between the components of $V_{(\alpha)}$ and $V_{(\beta)}$. Also requiring that these relations are linear in $V_{(\alpha)}$ and $V_{(\beta)}$, we obtain the following patching rules for the generalised vector: 
\begin{align}
v_{(\alpha)}&= v_{(\beta)} \ ,\nn \\[1mm]
\lambda_{(\alpha)} &= \lambda_{(\beta)} + \iota_{v_{(\beta)}} \dd \Lambda \ ,\nn \\[1mm]
\sigma_{(\alpha)} &= \sigma_{(\beta)} +\iota_{v_{(\beta)}}(\dd\tilde\Lambda+m\Omega_6)  + \dd\Omega_{0} \wedge \dd\Omega_{2} \wedge (\lambda_{(\beta)} + \iota_{ v_{(\beta)}}\dd\Lambda) \nn\\[1mm] 
& \quad - \big[ s(\dd\Omega)\wedge \big( \rme^{-\dd\Lambda} \wedge\omega_{(\beta)} + m \big(\tfrac{\rme^{-\dd\Lambda}-1}{\dd\Lambda}\big)\wedge (\iota_{v_{(\beta)}} \Lambda + \lambda_{(\beta)} \wedge \Lambda)
 \big) + \tfrac{1}{2}s(\dd\Omega)\wedge \iota_{v_{(\beta)}}\dd\Omega \big]_5\nn \\[1mm]
&\quad - \big[m \big(\tfrac{\rme^{-\dd\Lambda}-1}{\dd\Lambda}\big) \wedge\Lambda\wedge \omega_{(\beta)} + m^2 \big(\tfrac{\rme^{-\dd\Lambda}-1+ \dd\Lambda}{\dd\Lambda\wedge \dd\Lambda}
\big)\wedge \Lambda \wedge \iota_{v_{(\beta)}} \Lambda \big]_5  \ , \nn \\[1mm] 
\omega_{(\alpha)} &= \rme^{-\dd\Lambda} \wedge\omega_{(\beta)} + \iota_{v_{(\beta)}}\dd\Omega +  ( \lambda_{(\beta)} + \iota_{ v_{(\beta)}}\dd\Lambda)\wedge \dd\Omega\nn \\[1mm] 
&\quad + m \big(\tfrac{\rme^{-\dd\Lambda}-1}{\dd\Lambda}\big)\wedge (\iota_{v_{(\beta)}} \Lambda + \lambda_{(\beta)} \wedge \Lambda) + m \big(\tfrac{\rme^{-\dd\Lambda}-1+ \dd\Lambda}{\dd\Lambda\wedge \dd\Lambda}
\big)\wedge \Lambda\wedge \iota_{v_{(\beta)}} \dd\Lambda \ .
\end{align}
Setting $m=0$, these terms match precisely those following from eq.~\eqref{patching} for the patching of the twisted generalised tangent space relevant to massless type IIA. Keeping \hbox{$m \neq 0$}, we recover the corresponding terms of equation~\eqref{patching_m}. 

Note however that by this procedure,  one can construct the full twisted bundle $E$ only for compactifications on manifolds $M_d$ of dimension $d \leq 5$. 
 Indeed  one can directly deduce the patching of the differential form parts of the generalised vector (which form a section of the bundle $E'''$ in \eqref{eq:twistE}), but not the dual graviton charge, as there is no known treatment of the (non-linear) gauge transformations of the dual graviton field in an arbitrary background. One can nevertheless infer the transformation of the $\tau$ component of the generalised vector by insisting that the patching is an $E_{d+1 (d+1)}$ adjoint action. 
In particular, for $m=0$ this yields:
\begin{align}
\tau_{(\alpha)} \,&=\,  \tau_{(\beta)} + j \dd\Lambda\wedge  \sigma_{(\beta)} + j\dd\tilde\Lambda\wedge (\lambda_{(\beta)} + \iota_{ v_{(\beta)}}\dd\Lambda) \nn\\[1mm]
\,&\, \quad -js(\dd\Omega)\wedge\big(\rme^{-\dd\Lambda}\wedge\omega_{(\beta)} + \tfrac12\iota_{ v_{(\beta)}}\dd\Omega + \tfrac12 (\lambda_{(\beta)} + \iota_{ v_{(\beta)}}\dd\Lambda) \wedge \dd\Omega \big)\, .
\end{align}
%

\section{Exceptional generalised tangent bundle as an extension of $O(d,d)$ generalised geometry}
\label{app:Odd}

In the formulae for exceptional generalised geometry for (massive) type IIA, one can identify combinations of terms familiar from Hitchin's generalised geometry~\cite{Hitchin:2004ut,Gualtieri:2003dx}. We devote this appendix to showing how the exceptional generalised tangent space can be formulated as an extension of that introduced by Hitchin, by $O(d,d)\times\bbR^+$ tensor bundles. This clarifies how exceptional geometry constructions like the Dorfman derivative~\eqref{dorf6m}, are built out of objects and operators naturally associated to these $O(d,d)\times\bbR^+$ generalised geometric bundles.

Recall that Hitchin's generalised tangent space~\cite{Hitchin:2004ut,Gualtieri:2003dx}, which we denote by $E'$, has the structure of an extension
\begin{equation}
\label{eq:HitchinE}
   0 \longrightarrow T^* \stackrel{\pi^*}{\longrightarrow} \;  E' 
      \stackrel{\pi}{\longrightarrow} T \longrightarrow 0 \,. 
\end{equation}
The supergravity $B$-field provides a splitting of the sequence and thus an isomorphism
\begin{equation}
\label{eq:E'-isom}
	E' \,\cong\, T \oplus T^*\,.
\end{equation}
As in~\cite{Coimbra:2011nw}, we will view $E'$ as an $O(d,d)\times\bbR^+$ vector bundle with zero $\bbR^+$-weight. We normalise the $\bbR^+$ weight by fixing the line bundle $L \cong \Lambda^d T^*$ to have unit weight. The spinor bundles associated to $E'$ with weight $\frac12$, denoted $S^\pm(E')_{\frac12}$, can then be represented as local polyforms
\begin{equation}
\label{eq:SE'-isom}
	S^\pm(E')_{\frac{1}{2}} \,\cong\, \Lambda^{\rm even/odd} T^*\,,
\end{equation}
while (in six dimensions) there is also an isomorphism
\begin{equation}
\label{eq:E'L-isom}
	E' \otimes L \,\cong\, \Lambda^5 T^* \oplus (T^* \otimes \Lambda^6 T^*)\,.
\end{equation}
The bundles $S^\pm(E')_{\frac12}$ and $E' \otimes L$ are themselves naturally formed from extensions, and the isomorphisms~\eqref{eq:SE'-isom} and~\eqref{eq:E'L-isom} are also provided by the supergravity $B$ field.

The (massive) type IIA exceptional generalised tangent space $E$ then fits into the exact sequences
\begin{equation}
\label{eq:Odd-E}
\begin{aligned}
   0 \longrightarrow S^+(E')_{\frac12} \longrightarrow \; &E''  
      \longrightarrow E' \longrightarrow 0 \,, \\
   0 \longrightarrow E' \otimes L \longrightarrow \; &E 
      \longrightarrow E'' \longrightarrow 0 \,.
\end{aligned}
\end{equation}
These give us a mapping
\begin{equation}
\label{eq:Odd-anchor}
\begin{aligned}
	\pi' : E &\ra \; E'  \\
	V &\mapsto X = v + \lambda\,,
\end{aligned}
\end{equation}
which serves as an analogue of the anchor map when viewing the exceptional generalised tangent space $E$ as an extension of $E'$. 

Some useful $O(d,d)\times\bbR^+$ covariant maps can be defined as follows. First, given a section $\tilde{b} \in L$, one has the mapping
\begin{equation}
\label{eq:density-map}
\begin{aligned}
	\tilde{b} : E' &\ra \; E' \otimes L \\
	v + \lambda &\mapsto  i_v \tilde{b}  - \lambda \otimes \tilde{b}
\end{aligned}
\end{equation}
There is also a natural derivative
\begin{equation}
\label{eq:d-sigma-map}
\begin{aligned}
	\der : E' \otimes L &\ra \;  L \\
	\tilde{X} = \sigma + \tau &\mapsto \langle \der , \tilde{X} \rangle = \dd \sigma ,
\end{aligned}
\end{equation}
which is the analogue of the (covariant) divergence of a vector density in Riemannian geometry, and a covariant pairing of spinors of opposite chirality
\begin{equation}
\label{eq:Odd-spinor-bilinear}
\begin{aligned}
	\langle (\dots), \Gamma^{(1)} (\dots) \rangle : 
		S^+ (E')_{\frac{1}{2}} \otimes S^- (E')_{\frac{1}{2}} 
		&\ra \;  E' \otimes L \\
	\langle \omega , \Gamma^{(1)} \theta \rangle 
		= - [s(\omega) \wedge \theta ]_5 - [j s(\omega)\wedge \theta ]_{1,6}\,.
\end{aligned}
\end{equation}
The supergravity fields\footnote{In this appendix we use the $A$-basis for the RR fields (see footnote~\ref{Abasis_foot}) as we wish for the $B$ field to appear purely in the twisting of the $O(d,d)$ bundles in~\eqref{eq:Odd-E} and not in defining the isomorphism~\eqref{eq:Odd-E-isom}.} $A$ and $\tilde{B}$ are naturally collections of local sections of $S^- (E')_{\frac{1}{2}}$ and $L$ respectively, patched by the relevant supergravity gauge transformations. These provide splittings of the sequences~\eqref{eq:Odd-E} and thus an isomorphism 
\begin{equation}
\label{eq:Odd-E-isom}
\begin{aligned}
	E &\cong E' \oplus S^+(E')_{\frac12} \oplus (E' \otimes L) \\
	V &\mapsto \check{X} + \check{\omega}  + \check{\tilde{X}}
\end{aligned}
\end{equation}
which is given explicitly in terms of the maps above as
\begin{equation}
\label{eq:Odd-twist}
\begin{aligned}
	\check{X} &= X \\
	\check{\omega} &= \omega - X \cdot A \\
	\check{\tilde{X}} &= \tilde{X} - \tilde{B} \cdot X 
		- \langle \omega - \tfrac12 X \cdot A, \Gamma^{(1)} A \rangle
\end{aligned}
\end{equation}
where $X\cdot A$ is the Clifford product. 

Let us now show how to write the massless type IIA Dorfman derivative~\eqref{dorf6} in terms of natural operations in $O(d,d)\times\bbR^+$ generalised geometry. Denote by $X = \pi' (V) = v + \lambda$ and $X' = \pi'(V') = v' + \lambda'$ the projections of the generalised vectors $V$ and $V'$ onto $E'$ using the mapping~\eqref{eq:Odd-anchor}. The vector and one-form parts of \eqref{dorf6} correspond to the $O(d,d)$ Dorfman derivative $L_X X' = \mathcal{L}_v v' + \left(\mathcal{L}_v \lambda' - \iota_{v^\prime} \mathrm{d}\lambda\right)$, so that one has $\pi' (L_V V') = L_{\pi' (V)} \pi' (V')$. This is reminiscent of the situation for the usual anchor map $\pi : E \ra TM$, which satisfies $\pi (L_V V') = \mathcal{L}_{\pi(V)} \pi(V')$ so that the Dorfman derivative descends to the Lie derivative. Here, the mapping $\pi'$ preserves the Dorfman derivative structure. We remark that the map $\pi'$ and the Dorfman derivatives can be viewed as providing a generalisation of the notion of an algebroid, where one replaces the tangent bundle with Hitchin's generalised tangent bundle.

The poly-forms $\omega$ and $\omega'$ are local sections of the $O(6,6)$ spinor bundle $S^+(E')_{\frac{1}{2}}$, and these are treated in an $O(6,6)$-covariant way in~\eqref{dorf6}. Indeed, $L_X \omega' = (\mathcal{L}_v + \dd \lambda \wedge)\omega'$ is a spinorial Lie derivative in $O(d,d)$ generalised geometry, while $(\iota_v +  \lambda\wedge)\dd\omega$ is the Clifford action of the $O(6,6)$ generalised vector $X$ on $\dd \omega$. 

In six dimensions, the last two parts $\sigma$ and $\tau$ form a local section $\tilde{X}$ of $E' \otimes L$ as in~\eqref{eq:E'L-isom}. We see that the $O(d,d)\times\bbR^+$ Dorfman derivative $L_X \tilde{X}' = \mathcal{L}_v \sigma' + \mathcal{L}_v \tau' + j \sigma' \wedge \dd \lambda$ accounts for some of the terms involving these in $L_V V'$. From~\eqref{eq:d-sigma-map}, one can write $\dd \sigma = \langle \der , X \rangle$, a section of $L$, which can act on $X'$ via the map~\eqref{eq:density-map}, to give $\langle \der , X \rangle (X') = i_{v'} \dd \sigma - \lambda'\otimes\dd\sigma$. Finally, the exterior derivative gives the natural $O(d,d)\times\bbR^+$ Dirac operator $S^+(E')_{\frac{1}{2}} \ra S^-(E')_{\frac{1}{2}}$ and the pairing between $\omega'$ and $\dd\omega$ in the first and second lines of \eqref{dorf6} is the $O(6,6)\times\bbR^+$ invariant given in~\eqref{eq:Odd-spinor-bilinear}. 

Putting all of this together, we can write the Dorfman derivative in terms of $O(d,d)\times\bbR^+$ objects as
\begin{equation}
\label{eq:Odd-Edd-Dorfman}
\begin{aligned}
	L_V V' = L_X X' + (L_X \omega' - X' \cdot \dd \omega)
		 + (L_X \tilde{X}' - \langle \der , \tilde{X} \rangle (X') 
			- \langle \omega' , \Gamma^{(1)} \dd \omega \rangle)\,.
\end{aligned}
\end{equation}
This can be easily enhanced to include the mass terms in~\eqref{dorf6m}. The mass can be viewed as a local section of the spinor bundle $S^+(E')_{\frac12}\cong \Lambda^{\rm even} T^*$ and we can write the massive version of~\eqref{eq:Odd-Edd-Dorfman} as
\begin{equation}
\label{eq:massive-Odd-Edd-Dorfman}
\begin{aligned}
	L_V V' &= L_X X' + \Big[ L_X \omega' - X' \cdot (\dd\omega - X\cdot m)\Big] \\
		 & \qquad 
		 + \Big[ L_X \tilde{X}' - \big(\langle \der , \tilde{X} \rangle + \langle \omega , m \rangle\big)(X') 
			- \langle \omega' , \Gamma^{(1)} (\dd \omega - X\cdot m)\rangle \Big]\,.
\end{aligned}
\end{equation}
Finally, we remark that the projected generalised metric appearing in~\eqref{eq:GB-metric} is formalised by the construction of this appendix as $\mathcal{H}^{-1} \in S^2(E')$, which is the image of the exceptional generalised metric $G^{-1} \in S^2(E)$ in the anchor-like mapping $\pi' : E \ra E'$ from~\eqref{eq:Odd-anchor}. This is much like the first line of~\eqref{invG_comp_1}, where $\rme^{2\Delta} g^{-1} \in S^2(TM)$ is seen to be the image of $G^{-1}$ in the anchor map $\pi : E \ra TM$.

\section{$S^d$ in constrained coordinates}
\label{app:constrcoo}

In the following we provide some useful formulae for the embedding coordinate description of the round sphere $S^d$, mostly taken from~\cite{spheres}. These are needed to study the parallelisations of the exceptional tangent bundle presented in the main text.

We parameterise $\mathbb{R}^{d+1}$ in Cartesian coordinates as $x^i = r\, y^i$, $i=1,\ldots d+1$, with 
\be
\delta_{ij}\,y^iy^j \,=\, 1\ .
\ee 
Then the $d$-dimensional sphere $S^d$ of radius $R$ is obtained by fixing $r=R$. The standard metric and volume form on $\mathbb{R}^{d+1}$ induce the following round metric and \hbox{volume form on $S^d$:}
\be\label{roundSdmetric}
\rg{g} \, = \, R^2 \, \delta_{ij}\dd y^i \dd y^j \ ,
\ee
\be
\rg{\rm vol}_{d} \, = \, \frac{R^d}{d!} \,\epsilon_{i_1 \ldots i_{d+1}} y^{i_1} \dd y^{i_2}\wedge \cdots \wedge \dd y^{i_{d+1}}\ .
\ee
The Killing vector fields generating the $SO(n+1)$ isometries can be written as 
\begin{equation}
v_{ij} \,=\, R^{-1}\left(y_i k_j - y_j k_i \right)\ ,
\end{equation}
where $k_i$ are conformal Killing vectors, satisfying 
\begin{align}
\mathcal{L}_{k_i}\! \rg{g} \,&=\, - 2 y^i \rg{g}\ , \\[1mm]
k_i (y_j) \,&\equiv\, \iota_{k_i}\dd y_j \, = \, \delta_{ij} - y_i y_j\ .
\end{align}
The index on the coordinates $y^i$ is lowered using the $\mathbb{R}^{d+1}$ metric $\delta_{ij}$.
The Killing vectors $v_{ij}$ generate the $\mathfrak{so}(n+1)$ algebra,
\be\label{algebra_Killing_v}
\calL_{ v_{ij}} v_{kl} \, =\, R^{-1}\left(\delta_{ik}v_{lj} - \delta_{il}v_{kj} - \delta_{jk}v_{li} + \delta_{jl} v_{ki} \right)\ ,
\ee
while the constrained coordinates $y_k$ and their differentials $\dd y_k$ transform in the fundamental representation of $SO(n+1)$ under the Lie derivative, 
\begin{align}\label{Lie_on_y}
\mathcal{L}_{v_{ij}}y_k \,&\equiv\, \iota_{v_{ij}}\dd y_k \,=\, R^{-1}\left( y_i \delta_{jk} -  y_j\delta_{ik}\right)\ , \nn \\[1mm]
\mathcal{L}_{v_{ij}} \mathrm{d} y_k \, &=\,  R^{-1}\left(\mathrm{d}y_i \delta_{jk} - \mathrm{d}y_j\delta_{ik}\right) \ .
\end{align}
The $(d-1)$-form
\be
\kappa_i \, = \, - \rg{*}(R\,\dd y_i) \, =\,  \frac{R^{d-1}}{(d-1)!}\,\epsilon_{ij_1\ldots j_d}\, y^{j_1}\dd y^{j_2}\wedge \cdots \wedge \dd y^{j_d}
\ee
transforms under $\mathcal{L}_{v_{ij}}$ exactly as $\dd y_k$ (since $\calL_{v_{ij}}$ preserves the round metric~\eqref{roundSdmetric}, it commutes with the Hodge star):
\be
\mathcal{L}_{v_{ij}} \kappa_k \,=\, R^{-1}\left(\kappa_i \delta_{jk} - \kappa_j\delta_{ik}\right) \ .
\ee

We also introduce the forms
\begin{align}
\omega_{ij} \,&=\, R^2\, \dd y_i \wedge \dd y_j \ ,\nn \\[1mm]
\rho_{ij} \,&=\ \rg{*}\omega_{ij}  \,=\,   \frac{R^{d-2}}{(d-2)!} \,\epsilon_{ijk_1\ldots k_{d-1}} y^{k_1} \dd y^{k_2} \wedge \cdots \wedge \dd y^{k_{d-1}}\ ,\nn \\[1mm]
\tau_{ij} \,&=\, R\left(y_i\mathrm{d}y_j  - y_j\mathrm{d}y_i\right) \otimes \rg{\rm vol}_d\ ,
\end{align}
which transform in the adjoint representation of $\SO(d+1)$ under $\calL_{v_{ij}}$. Namely,
\be\label{Lie_on_omega}
\calL_{v_{ij}} \omega_{kl} \, = \, R^{-1}\left(\delta_{ik}\omega_{lj} - \delta_{il}\omega_{kj} - \delta_{jk}\omega_{li} + \delta_{jl} \omega_{ki} \right) \ ,
\ee
and similarly for the others, with the same overall factor $R^{-1}$.

Furthermore, one can show the relations
\begin{align}
\iota_{v_{ij}}\! \rg{\rm vol}_d \,&=\, \frac{R}{d-1} \mathrm{d}\rho_{ij}\, , \label{eq:contrvol}\\[1mm]
\dd\kappa_i \,&=\, \frac{d}{R}\, y_i \rg{\rm vol}_d\ , \\[1mm]
\dd \,\iota_{v_{ij}}\kappa_k \,&=\, - \dd\,(y_k \rho_{ij}) \ ,
\end{align}
which are proven by making use of the trivial identity $y_{[i_1}\epsilon_{i_2\ldots i_{d+2}]}=0$.

When computing the norm of our generalised frames, we will need the following ``squares'' of the forms defined above:
\begin{align}\label{contractions_sphere}
v_{ij} \,\lrcorner\, v_{kl} 
\,&=\, y_iy_k \delta_{jl} - y_jy_k \delta_{il} - y_iy_l \delta_{jk} +y_jy_l \delta_{ik}\ ,\nn\\[1mm]
\omega_{ij} \,\lrcorner\, \omega_{kl} \,=\, \rho_{ij}\,\lrcorner\,\rho_{kl}
\,&=\, \delta_{ik}\delta_{jl}- \delta_{il}\delta_{jk} - (y_iy_k \delta_{jl} - y_jy_k \delta_{il} - y_iy_l \delta_{jk} +y_jy_l \delta_{ik})\ , \nn \\[1mm]
\tau_{ij}\,\lrcorner\,\tau_{kl} \,&=\, y_iy_k \delta_{jl} - y_jy_k \delta_{il} - y_iy_l \delta_{jk} +y_jy_l \delta_{ik}\ , \nn \\[1mm]
\kappa_i \,\lrcorner\, \kappa_j \,=\, R^2\, \dd y_i \,\lrcorner\, \dd y_j \,&=\,  \delta_{ij} - y_iy_j \ .
\end{align}
Here, the round metric $\rg{g}$ and its inverse are used to lower/raise the indices; for instance, $\omega_{ij} \,\lrcorner\, \omega_{kl} \equiv \frac{1}{2} \rg{g}{}^{\!mp}\rg{g}{}^{\!nq}(\omega_{ij})_{mn}(\omega_{kl})_{pq}$, and so on.

\section{Type IIB parallelisation on $S^3$}\label{IIBonS3}

In this appendix, we present a parallelisation of the type IIB generalised tangent bundle on $S^3$ which satisfies an $\SO(4)$ gauge algebra.
 A consistent truncation of type IIB supergravity on $S^3$ down to $\SO(4)$ maximal supergravity in seven dimensions has recently been worked out in~\cite{Malek:2015hma} adopting an exceptional field theory approach. This was related to the $S^3$ reduction of massless type IIA by an external automorphism of $\SL(4)$ exchanging the ${\bf 10}\subset{\bf 15}$ and the ${\bf 10'}\subset {\bf 40'}$ representations. 
 Here we show that this type IIB truncation can also be understood in terms of generalised parallelisations.

The type IIB generalised tangent bundle $E$ on a three-dimensional manifold $M_3$ is
\begin{equation}
\label{gb}
E\,\simeq\, T \oplus T^* \oplus T^* \oplus \Lambda^3 T^*\, ,
\end{equation}
and has structure group $E_{4(4)} \times \mathbb R^+\simeq \SL\left(5,\mathbb{R}\right)\times \mathbb{R}^+$.
A generalised vector transforms in the ${\bf 10_1}$ representation and can be written as
\begin{equation*}
V \,=\, v + \lambda + \rho + \zeta\ ,
\end{equation*}
where $v\in T$, $\lambda \in T^*$, $\rho \in T^*$, and $\zeta\in \Lambda^3 T^*$.
The relevant Dorfman derivative can be obtained by truncating to three dimensions the type IIB, five-dimensional Dorfman derivative given in~\cite{spheres,Ashmore:2015joa}. This yields
\begin{equation}
\label{dorfb3}
L_V V' \,=\, \mathcal{L}_v v' + \left(\mathcal{L}_v \lambda' - \iota_{v^\prime} \mathrm{d}\lambda \right) + \left(\mathcal{L}_v \rho' - \iota_{v^\prime} \mathrm{d}\rho \right) + \left(\mathcal{L}_v \zeta' + \mathrm{d}\lambda \wedge \rho'\right)\ .
\end{equation}

As in the type IIA example discussed in section~\ref{S3and7dsugra}, we decompose the generalised frame $\hat{E}_{IJ}$, $I,J =1,\ldots, 5$, under $\SL\left(4,\mathbb{R}\right)$ as $\{E_{ij} , E_{i5}\}$, with  $i=1,\ldots,4$.
Then we define a generalised parallelisation on $S^3$ as
\begin{equation}
\label{glpb}
\hat{E}_{IJ} \,=\, \begin{cases}\  \hat{E}_{ij} = v_{ij} + \rho_{ij} + \iota_{v_{ij}}\!\! \rg{B}\, ,\\[1mm]
\;\ \hat{E}_{i5} = R\,\dd y_i + y_i \rg{\rm vol}_3 + R\,\dd y_i\, \wedge\! \rg{B}\, , \end{cases}
\end{equation}
with
\be
\rho_{ij} \,=\, \rg{*}(R^2 \dd y_i \wedge\dd y_j)   \,=\, R\, \epsilon_{ijkl}\, y^k\mathrm{d}y^l \, .
\ee
Here, $\hat{E}_{ij}$ parallelises $T \oplus T^*$, that is the part of the generalised tangent bundle common to type IIA, while $\hat{E}_i$ is a parallelisation on the complement $T^* \oplus \Lambda^3 T^*$.
As in section~\ref{S3and7dsugra}, the background two-form potential $\rg{B}$ is chosen such that 
\be
\rg{H} \ =\, \dd\! \rg{B}\, \ = \, \frac{2}{R}\rg{\rm vol}_3\ 
\ee
(we could also have twisted by a background RR two-form potential $\rg{C_2}\,$).

Evaluating the Dorfman derivative on the frame~\eqref{glpb}, we obtain
\begin{align}
\phantom{i}L_{\hat{E}_{ij}}\hat{E}_{kl} \,&=\, 2R^{-1}\big(\delta_{i[k}\hat{E}_{l]j} -  \delta_{j[k}\hat{E}_{l]i} \big)\ ,\nn\\[1mm]
\phantom{i}L_{\hat{E}_{ij}}\hat{E}_{k5} \,&=\, -2R^{-1}\delta_{k[i}\hat{E}_{j]5} \ , \nn\\[1mm]
\phantom{i}L_{\hat{E}_{i5}}\hat{E}_{kl} \,&=0\ , \nn\\[1mm]
\phantom{i}L_{\hat{E}_{i5}}\hat{E}_{k5} \,&= 0\ ,
\end{align}
which corresponds to an $SO(4)$ frame algebra.\footnote{This is the same algebra satisfied by the massive IIA generalised parallelisation on $S^3$ discussed in section~\ref{massive_algebras} (cf.~eq.~\eqref{massive_algebra_S3}) -- however in that case the parallelisation fails to be an $\SL(5)$ frame.} This is consistent with the $\SO(4)$ gauging of $D=7$ maximal supergravity originally discussed in~\cite{Samtleben:2005bp}. To see this, it is convenient to dualise $\hat E_{ij}$ to $\widetilde{E}_{ij} = \frac{1}{2} \epsilon_{ij}{}^{kl}\hat{E}_{kl}$. Also renaming $\widetilde{E}_{i5}= \hat{E}_{i5}$, the frame algebra now reads
\be
L_{\widetilde{E}_{II'}} \widetilde{E}_{JJ'} \, =\, -X_{[II'][JJ']}{}^{[KK']} \widetilde{E}_{KK'}\ ,
\ee
with
\be\label{strconst_m_S3}
X_{[II'][JJ']}{}^{[KK']} = -4 \, \epsilon_{5II'L[J}w^{L[K}\delta_{J']}^{K']}\ ,\qquad
w^{IJ} \,=\, \frac{1}{2R} \,{\rm diag}\big(1,1,1,1,0\big)\ ,
\ee
which matches the embedding tensor given in~\cite{Samtleben:2005bp} for the $\SO(4)$ gauging. 

\bibliographystyle{JHEP}
\bibliography{Bibliography}

\providecommand{\href}[2]{#2}\begingroup\raggedright\begin{thebibliography}{10}

\bibitem{deWit:1986oxb}
B.~de~Wit and H.~Nicolai, \emph{{The Consistency of the $S^7$ Truncation in
  $d=11$ Supergravity}},
  \href{http://dx.doi.org/10.1016/0550-3213(87)90253-7}{\emph{Nucl. Phys.} {\bf
  B281} (1987) 211}.

\bibitem{Nastase:1999kf}
H.~Nastase, D.~Vaman and P.~van Nieuwenhuizen, \emph{{Consistency of the $AdS_7
  \times S_4$ reduction and the origin of selfduality in odd dimensions}},
  \href{http://dx.doi.org/10.1016/S0550-3213(00)00193-0}{\emph{Nucl. Phys.}
  {\bf B581} (2000) 179--239}, [\href{http://arxiv.org/abs/hep-th/9911238}{{\tt
  hep-th/9911238}}].

\bibitem{Cvetic:2000nc}
M.~Cvetic, H.~Lu, C.~N. Pope, A.~Sadrzadeh and T.~A. Tran, \emph{{Consistent
  SO(6) reduction of type IIB supergravity on $S^5$}},
  \href{http://dx.doi.org/10.1016/S0550-3213(00)00372-2}{\emph{Nucl. Phys.}
  {\bf B586} (2000) 275--286}, [\href{http://arxiv.org/abs/hep-th/0003103}{{\tt
  hep-th/0003103}}].

\bibitem{Riccioni:2007au}
F.~Riccioni and P.~C. West, \emph{{The E(11) origin of all maximal
  supergravities}},
  \href{http://dx.doi.org/10.1088/1126-6708/2007/07/063}{\emph{JHEP} {\bf 07}
  (2007) 063}, [\href{http://arxiv.org/abs/0705.0752}{{\tt 0705.0752}}].

\bibitem{Berman:2012uy}
D.~S. Berman, E.~T. Musaev and D.~C. Thompson, \emph{{Duality Invariant
  M-theory: Gauged supergravities and Scherk-Schwarz reductions}},
  \href{http://dx.doi.org/10.1007/JHEP10(2012)174}{\emph{JHEP} {\bf 10} (2012)
  174}, [\href{http://arxiv.org/abs/1208.0020}{{\tt 1208.0020}}].

\bibitem{Aldazabal:2013mya}
G.~Aldazabal, M.~Gra\~na, D.~Marqu\'es and J.~A. Rosabal, \emph{{Extended
  geometry and gauged maximal supergravity}},
  \href{http://dx.doi.org/10.1007/JHEP06(2013)046}{\emph{JHEP} {\bf 06} (2013)
  046}, [\href{http://arxiv.org/abs/1302.5419}{{\tt 1302.5419}}].

\bibitem{Godazgar:2013dma}
H.~Godazgar, M.~Godazgar and H.~Nicolai, \emph{{Generalised geometry from the
  ground up}}, \href{http://dx.doi.org/10.1007/JHEP02(2014)075}{\emph{JHEP}
  {\bf 02} (2014) 075}, [\href{http://arxiv.org/abs/1307.8295}{{\tt
  1307.8295}}].

\bibitem{spheres}
K.~Lee, C.~Strickland-Constable and D.~Waldram, \emph{{Spheres, generalised
  parallelisability and consistent truncations}},
  \href{http://arxiv.org/abs/1401.3360}{{\tt 1401.3360}}.

\bibitem{Hohm:2014qga}
O.~Hohm and H.~Samtleben, \emph{{Consistent Kaluza-Klein Truncations via
  Exceptional Field Theory}},
  \href{http://dx.doi.org/10.1007/JHEP01(2015)131}{\emph{JHEP} {\bf 01} (2015)
  131}, [\href{http://arxiv.org/abs/1410.8145}{{\tt 1410.8145}}].

\bibitem{Ciceri:2014wya}
F.~Ciceri, B.~de~Wit and O.~Varela, \emph{{IIB supergravity and the E$_{6(6)}$
  covariant vector-tensor hierarchy}},
  \href{http://dx.doi.org/10.1007/JHEP04(2015)094}{\emph{JHEP} {\bf 04} (2015)
  094}, [\href{http://arxiv.org/abs/1412.8297}{{\tt 1412.8297}}].

\bibitem{Baguet:2015xha}
A.~Baguet, O.~Hohm and H.~Samtleben, \emph{{E$_{6(6)}$ Exceptional Field
  Theory: Review and Embedding of Type IIB}}, {\emph{PoS} {\bf Corfu2014}
  (2015) 133}, [\href{http://arxiv.org/abs/1506.01065}{{\tt 1506.01065}}].

\bibitem{Baguet:2015sma}
A.~Baguet, O.~Hohm and H.~Samtleben, \emph{{Consistent Type IIB Reductions to
  Maximal 5D Supergravity}},
  \href{http://dx.doi.org/10.1103/PhysRevD.92.065004}{\emph{Phys. Rev.} {\bf
  D92} (2015) 065004}, [\href{http://arxiv.org/abs/1506.01385}{{\tt
  1506.01385}}].

\bibitem{Malek:2015hma}
E.~Malek and H.~Samtleben, \emph{{Dualising consistent IIA/IIB truncations}},
  \href{http://dx.doi.org/10.1007/JHEP12(2015)029}{\emph{JHEP} {\bf 12} (2015)
  029}, [\href{http://arxiv.org/abs/1510.03433}{{\tt 1510.03433}}].

\bibitem{Hull:2007zu}
C.~M. Hull, \emph{{Generalised Geometry for M-Theory}},
  \href{http://dx.doi.org/10.1088/1126-6708/2007/07/079}{\emph{JHEP} {\bf 07}
  (2007) 079}, [\href{http://arxiv.org/abs/hep-th/0701203}{{\tt
  hep-th/0701203}}].

\bibitem{Pacheco:2008ps}
P.~P. Pacheco and D.~Waldram, \emph{{M-theory, exceptional generalised geometry
  and superpotentials}},
  \href{http://dx.doi.org/10.1088/1126-6708/2008/09/123}{\emph{JHEP} {\bf 09}
  (2008) 123}, [\href{http://arxiv.org/abs/0804.1362}{{\tt 0804.1362}}].

\bibitem{Scherk:1979zr}
J.~Scherk and J.~H. Schwarz, \emph{{How to Get Masses from Extra Dimensions}},
  \href{http://dx.doi.org/10.1016/0550-3213(79)90592-3}{\emph{Nucl. Phys.} {\bf
  B153} (1979) 61--88}.

\bibitem{Grana:2009im}
M.~Grana, J.~Louis, A.~Sim and D.~Waldram, \emph{{$E_{7(7)}$ formulation of
  $N=2$ backgrounds}},
  \href{http://dx.doi.org/10.1088/1126-6708/2009/07/104}{\emph{JHEP} {\bf 07}
  (2009) 104}, [\href{http://arxiv.org/abs/0904.2333}{{\tt 0904.2333}}].

\bibitem{Coimbra:2011ky}
A.~Coimbra, C.~Strickland-Constable and D.~Waldram, \emph{{$E_{d(d)} \times
  \mathbb{R}^+$ generalised geometry, connections and M theory}},
  \href{http://dx.doi.org/10.1007/JHEP02(2014)054}{\emph{JHEP} {\bf 02} (2014)
  054}, [\href{http://arxiv.org/abs/1112.3989}{{\tt 1112.3989}}].

\bibitem{Dall'Agata:2012bb}
G.~Dall'Agata, G.~Inverso and M.~Trigiante, \emph{{Evidence for a family of
  SO(8) gauged supergravity theories}},
  \href{http://dx.doi.org/10.1103/PhysRevLett.109.201301}{\emph{Phys. Rev.
  Lett.} {\bf 109} (2012) 201301}, [\href{http://arxiv.org/abs/1209.0760}{{\tt
  1209.0760}}].

\bibitem{Dall'Agata:2014ita}
G.~Dall'Agata, G.~Inverso and A.~Marrani, \emph{{Symplectic Deformations of
  Gauged Maximal Supergravity}},
  \href{http://dx.doi.org/10.1007/JHEP07(2014)133}{\emph{JHEP} {\bf 07} (2014)
  133}, [\href{http://arxiv.org/abs/1405.2437}{{\tt 1405.2437}}].

\bibitem{Guarino:2015jca}
A.~Guarino, D.~L. Jafferis and O.~Varela, \emph{{String Theory Origin of Dyonic
  N=8 Supergravity and Its Chern-Simons Duals}},
  \href{http://dx.doi.org/10.1103/PhysRevLett.115.091601}{\emph{Phys. Rev.
  Lett.} {\bf 115} (2015) 091601}, [\href{http://arxiv.org/abs/1504.08009}{{\tt
  1504.08009}}].

\bibitem{Guarino:2015qaa}
A.~Guarino and O.~Varela, \emph{{Dyonic ISO(7) supergravity and the duality
  hierarchy}}, \href{http://dx.doi.org/10.1007/JHEP02(2016)079}{\emph{JHEP}
  {\bf 02} (2016) 079}, [\href{http://arxiv.org/abs/1508.04432}{{\tt
  1508.04432}}].

\bibitem{Guarino:2015vca}
A.~Guarino and O.~Varela, \emph{{Consistent $ \mathcal{N}=8 $ truncation of
  massive IIA on S$^{6}$}},
  \href{http://dx.doi.org/10.1007/JHEP12(2015)020}{\emph{JHEP} {\bf 12} (2015)
  020}, [\href{http://arxiv.org/abs/1509.02526}{{\tt 1509.02526}}].

\bibitem{Hull:1988jw}
C.~M. Hull and N.~P. Warner, \emph{{Noncompact Gaugings From Higher
  Dimensions}},
  \href{http://dx.doi.org/10.1088/0264-9381/5/12/005}{\emph{Class. Quant.
  Grav.} {\bf 5} (1988) 1517}.

\bibitem{Lee:2015xga}
K.~Lee, C.~Strickland-Constable and D.~Waldram, \emph{{New gaugings and
  non-geometry}},  \href{http://arxiv.org/abs/1506.03457}{{\tt 1506.03457}}.

\bibitem{Bergshoeff:2001pv}
E.~Bergshoeff, R.~Kallosh, T.~Ortin, D.~Roest and A.~Van~Proeyen, \emph{{New
  formulations of D = 10 supersymmetry and D8 - O8 domain walls}},
  \href{http://dx.doi.org/10.1088/0264-9381/18/17/303}{\emph{Class.Quant.Grav.}
  {\bf 18} (2001) 3359--3382}, [\href{http://arxiv.org/abs/hep-th/0103233}{{\tt
  hep-th/0103233}}].

\bibitem{Romans:1985tz}
L.~J. Romans, \emph{{Massive N=2a Supergravity in Ten-Dimensions}},
  \href{http://dx.doi.org/10.1016/0370-2693(86)90375-8}{\emph{Phys. Lett.} {\bf
  B169} (1986) 374}.

\bibitem{Bergshoeff:1997ak}
E.~Bergshoeff, Y.~Lozano and T.~Ortin, \emph{{Massive branes}},
  \href{http://dx.doi.org/10.1016/S0550-3213(98)00045-5}{\emph{Nucl. Phys.}
  {\bf B518} (1998) 363--423}, [\href{http://arxiv.org/abs/hep-th/9712115}{{\tt
  hep-th/9712115}}].

\bibitem{Bergshoeff:2006qw}
E.~A. Bergshoeff, M.~de~Roo, S.~F. Kerstan, T.~Ortin and F.~Riccioni,
  \emph{{IIA ten-forms and the gauge algebras of maximal supergravity
  theories}},
  \href{http://dx.doi.org/10.1088/1126-6708/2006/07/018}{\emph{JHEP} {\bf 07}
  (2006) 018}, [\href{http://arxiv.org/abs/hep-th/0602280}{{\tt
  hep-th/0602280}}].

\bibitem{Hitchin:2004ut}
N.~Hitchin, \emph{{Generalized Calabi-Yau manifolds}},
  \href{http://dx.doi.org/10.1093/qjmath/54.3.281}{\emph{Quart. J. Math.} {\bf
  54} (2003) 281--308}, [\href{http://arxiv.org/abs/math/0209099}{{\tt
  math/0209099}}].

\bibitem{Gualtieri:2003dx}
M.~Gualtieri, \emph{{Generalized complex geometry}}.
\newblock PhD thesis, Oxford U., 2003.
\newblock \href{http://arxiv.org/abs/math/0401221}{{\tt math/0401221}}.

\bibitem{Coimbra:2011nw}
A.~Coimbra, C.~Strickland-Constable and D.~Waldram, \emph{{Supergravity as
  Generalised Geometry I: Type II Theories}},
  \href{http://dx.doi.org/10.1007/JHEP11(2011)091}{\emph{JHEP} {\bf 11} (2011)
  091}, [\href{http://arxiv.org/abs/1107.1733}{{\tt 1107.1733}}].

\bibitem{Coimbra:2012af}
A.~Coimbra, C.~Strickland-Constable and D.~Waldram, \emph{{Supergravity as
  Generalised Geometry II: $E_{d(d)} \times \mathbb{R}^+$ and M theory}},
  \href{http://dx.doi.org/10.1007/JHEP03(2014)019}{\emph{JHEP} {\bf 03} (2014)
  019}, [\href{http://arxiv.org/abs/1212.1586}{{\tt 1212.1586}}].

\bibitem{Hohm:2011cp}
O.~Hohm and S.~K. Kwak, \emph{{Massive Type II in Double Field Theory}},
  \href{http://dx.doi.org/10.1007/JHEP11(2011)086}{\emph{JHEP} {\bf 11} (2011)
  086}, [\href{http://arxiv.org/abs/1108.4937}{{\tt 1108.4937}}].

\bibitem{Kaloper:1999yr}
N.~Kaloper and R.~C. Myers, \emph{{The Odd story of massive supergravity}},
  \href{http://dx.doi.org/10.1088/1126-6708/1999/05/010}{\emph{JHEP} {\bf 05}
  (1999) 010}, [\href{http://arxiv.org/abs/hep-th/9901045}{{\tt
  hep-th/9901045}}].

\bibitem{Dall'Agata:2005ff}
G.~Dall'Agata and S.~Ferrara, \emph{{Gauged supergravity algebras from twisted
  tori compactifications with fluxes}},
  \href{http://dx.doi.org/10.1016/j.nuclphysb.2005.03.039}{\emph{Nucl. Phys.}
  {\bf B717} (2005) 223--245}, [\href{http://arxiv.org/abs/hep-th/0502066}{{\tt
  hep-th/0502066}}].

\bibitem{D'Auria:2005er}
R.~D'Auria, S.~Ferrara and M.~Trigiante, \emph{{E(7(7)) symmetry and dual gauge
  algebra of M-theory on a twisted seven-torus}},
  \href{http://dx.doi.org/10.1016/j.nuclphysb.2005.10.020}{\emph{Nucl. Phys.}
  {\bf B732} (2006) 389--400}, [\href{http://arxiv.org/abs/hep-th/0504108}{{\tt
  hep-th/0504108}}].

\bibitem{Hull:2005hk}
C.~M. Hull and R.~A. Reid-Edwards, \emph{{Flux compactifications of string
  theory on twisted tori}},
  \href{http://dx.doi.org/10.1002/prop.200900076}{\emph{Fortsch. Phys.} {\bf
  57} (2009) 862--894}, [\href{http://arxiv.org/abs/hep-th/0503114}{{\tt
  hep-th/0503114}}].

\bibitem{Hull:2006tp}
C.~M. Hull and R.~A. Reid-Edwards, \emph{{Flux compactifications of M-theory on
  twisted Tori}},
  \href{http://dx.doi.org/10.1088/1126-6708/2006/10/086}{\emph{JHEP} {\bf 10}
  (2006) 086}, [\href{http://arxiv.org/abs/hep-th/0603094}{{\tt
  hep-th/0603094}}].

\bibitem{Samtleben:2008pe}
H.~Samtleben, \emph{{Lectures on Gauged Supergravity and Flux
  Compactifications}},
  \href{http://dx.doi.org/10.1088/0264-9381/25/21/214002}{\emph{Class. Quant.
  Grav.} {\bf 25} (2008) 214002}, [\href{http://arxiv.org/abs/0808.4076}{{\tt
  0808.4076}}].

\bibitem{Baguet:2015iou}
A.~Baguet, C.~N. Pope and H.~Samtleben, \emph{{Consistent Pauli reduction on
  group manifolds}},
  \href{http://dx.doi.org/10.1016/j.physletb.2015.11.062}{\emph{Phys. Lett.}
  {\bf B752} (2016) 278--284}, [\href{http://arxiv.org/abs/1510.08926}{{\tt
  1510.08926}}].

\bibitem{deWit:2007kvg}
B.~de~Wit, H.~Samtleben and M.~Trigiante, \emph{{The Maximal D=4
  supergravities}},
  \href{http://dx.doi.org/10.1088/1126-6708/2007/06/049}{\emph{JHEP} {\bf 06}
  (2007) 049}, [\href{http://arxiv.org/abs/0705.2101}{{\tt 0705.2101}}].

\bibitem{Boonstra:1998mp}
H.~Boonstra, K.~Skenderis and P.~Townsend, \emph{{The domain wall / QFT
  correspondence}},
  \href{http://dx.doi.org/10.1088/1126-6708/1999/01/003}{\emph{JHEP} {\bf 9901}
  (1999) 003}, [\href{http://arxiv.org/abs/hep-th/9807137}{{\tt
  hep-th/9807137}}].

\bibitem{Hull:1984yy}
C.~M. Hull, \emph{{A New Gauging of $N=8$ Supergravity}},
  \href{http://dx.doi.org/10.1103/PhysRevD.30.760}{\emph{Phys. Rev.} {\bf D30}
  (1984) 760}.

\bibitem{DallAgata:2011aa}
G.~Dall'Agata and G.~Inverso, \emph{{On the Vacua of N = 8 Gauged Supergravity
  in 4 Dimensions}},
  \href{http://dx.doi.org/10.1016/j.nuclphysb.2012.01.023}{\emph{Nucl. Phys.}
  {\bf B859} (2012) 70--95}, [\href{http://arxiv.org/abs/1112.3345}{{\tt
  1112.3345}}].

\bibitem{Gallerati:2014xra}
A.~Gallerati, H.~Samtleben and M.~Trigiante, \emph{{The $ \mathcal{N}>2 $
  supersymmetric AdS vacua in maximal supergravity}},
  \href{http://dx.doi.org/10.1007/JHEP12(2014)174}{\emph{JHEP} {\bf 12} (2014)
  174}, [\href{http://arxiv.org/abs/1410.0711}{{\tt 1410.0711}}].

\bibitem{Varela:2015uca}
O.~Varela, \emph{{AdS$_{4}$ solutions of massive IIA from dyonic ISO(7)
  supergravity}}, \href{http://dx.doi.org/10.1007/JHEP03(2016)071}{\emph{JHEP}
  {\bf 03} (2016) 071}, [\href{http://arxiv.org/abs/1509.07117}{{\tt
  1509.07117}}].

\bibitem{Cassani:2009ck}
D.~Cassani and A.-K. Kashani-Poor, \emph{{Exploiting N=2 in consistent coset
  reductions of type IIA}},
  \href{http://dx.doi.org/10.1016/j.nuclphysb.2009.03.011}{\emph{Nucl. Phys.}
  {\bf B817} (2009) 25--57}, [\href{http://arxiv.org/abs/0901.4251}{{\tt
  0901.4251}}].

\bibitem{Hull:1984vg}
C.~M. Hull, \emph{{Noncompact Gaugings of $N=8$ Supergravity}},
  \href{http://dx.doi.org/10.1016/0370-2693(84)91131-6}{\emph{Phys. Lett.} {\bf
  B142} (1984) 39}.

\bibitem{Hull:1984qz}
C.~M. Hull, \emph{{More Gaugings of $N=8$ Supergravity}},
  \href{http://dx.doi.org/10.1016/0370-2693(84)90091-1}{\emph{Phys. Lett.} {\bf
  B148} (1984) 297--300}.

\bibitem{Cowdall:1998rs}
P.~M. Cowdall, \emph{{On gauged maximal supergravity in six-dimensions}},
  \href{http://dx.doi.org/10.1088/1126-6708/1999/06/018}{\emph{JHEP} {\bf 06}
  (1999) 018}, [\href{http://arxiv.org/abs/hep-th/9810041}{{\tt
  hep-th/9810041}}].

\bibitem{Cvetic:2000ah}
M.~Cvetic, H.~Lu, C.~Pope, A.~Sadrzadeh and T.~A. Tran, \emph{{$S^3$ and $S^4$
  reductions of type IIA supergravity}},
  \href{http://dx.doi.org/10.1016/S0550-3213(00)00466-1}{\emph{Nucl.Phys.} {\bf
  B590} (2000) 233--251}, [\href{http://arxiv.org/abs/hep-th/0005137}{{\tt
  hep-th/0005137}}].

\bibitem{Bergshoeff:2007ef}
E.~Bergshoeff, H.~Samtleben and E.~Sezgin, \emph{{The Gaugings of Maximal D=6
  Supergravity}},
  \href{http://dx.doi.org/10.1088/1126-6708/2008/03/068}{\emph{JHEP} {\bf 0803}
  (2008) 068}, [\href{http://arxiv.org/abs/0712.4277}{{\tt 0712.4277}}].

\bibitem{Samtleben:2005bp}
H.~Samtleben and M.~Weidner, \emph{{The Maximal D=7 supergravities}},
  \href{http://dx.doi.org/10.1016/j.nuclphysb.2005.07.028}{\emph{Nucl.Phys.}
  {\bf B725} (2005) 383--419}, [\href{http://arxiv.org/abs/hep-th/0506237}{{\tt
  hep-th/0506237}}].

\bibitem{Grana:2008yw}
M.~Grana, R.~Minasian, M.~Petrini and D.~Waldram, \emph{{T-duality, Generalized
  Geometry and Non-Geometric Backgrounds}},
  \href{http://dx.doi.org/10.1088/1126-6708/2009/04/075}{\emph{JHEP} {\bf 04}
  (2009) 075}, [\href{http://arxiv.org/abs/0807.4527}{{\tt 0807.4527}}].

\bibitem{Salam:1984ft}
A.~Salam and E.~Sezgin, \emph{{$d = 8$ supergravity}},
  \href{http://dx.doi.org/10.1016/0550-3213(85)90613-3}{\emph{Nucl. Phys.} {\bf
  B258} (1985) 284}.

\bibitem{Cvetic:2000dm}
M.~Cvetic, H.~Lu and C.~N. Pope, \emph{{Consistent Kaluza-Klein sphere
  reductions}}, \href{http://dx.doi.org/10.1103/PhysRevD.62.064028}{\emph{Phys.
  Rev.} {\bf D62} (2000) 064028},
  [\href{http://arxiv.org/abs/hep-th/0003286}{{\tt hep-th/0003286}}].

\bibitem{Cvetic:2003jy}
M.~Cvetic, G.~W. Gibbons, H.~Lu and C.~N. Pope, \emph{{Consistent group and
  coset reductions of the bosonic string}},
  \href{http://dx.doi.org/10.1088/0264-9381/20/23/013}{\emph{Class. Quant.
  Grav.} {\bf 20} (2003) 5161--5194},
  [\href{http://arxiv.org/abs/hep-th/0306043}{{\tt hep-th/0306043}}].

\bibitem{Howe:1997qt}
P.~S. Howe, N.~D. Lambert and P.~C. West, \emph{{A New massive type IIA
  supergravity from compactification}},
  \href{http://dx.doi.org/10.1016/S0370-2693(97)01199-4}{\emph{Phys. Lett.}
  {\bf B416} (1998) 303--308}, [\href{http://arxiv.org/abs/hep-th/9707139}{{\tt
  hep-th/9707139}}].

\bibitem{Tsimpis:2005vu}
D.~Tsimpis, \emph{{Massive IIA supergravities}},
  \href{http://dx.doi.org/10.1088/1126-6708/2005/10/057}{\emph{JHEP} {\bf 10}
  (2005) 057}, [\href{http://arxiv.org/abs/hep-th/0508214}{{\tt
  hep-th/0508214}}].

\bibitem{LeDiffon:2008sh}
A.~Le~Diffon and H.~Samtleben, \emph{{Supergravities without an Action: Gauging
  the Trombone}},
  \href{http://dx.doi.org/10.1016/j.nuclphysb.2008.11.010}{\emph{Nucl. Phys.}
  {\bf B811} (2009) 1--35}, [\href{http://arxiv.org/abs/0809.5180}{{\tt
  0809.5180}}].

\bibitem{Ciceri:2016dmd}
F.~Ciceri, A.~Guarino and G.~Inverso, \emph{{The exceptional story of massive
  IIA supergravity}},  \href{http://arxiv.org/abs/1604.08602}{{\tt
  1604.08602}}.

\bibitem{Ashmore:2015joa}
A.~Ashmore and D.~Waldram, \emph{{Exceptional Calabi--Yau spaces: the geometry
  of $\mathcal{N}=2$ backgrounds with flux}},
  \href{http://arxiv.org/abs/1510.00022}{{\tt 1510.00022}}.

\end{thebibliography}\endgroup

\end{document}